\documentclass[prb,aps,twocolumn,superscriptaddress,longbibliography,notitlepage]{revtex4-1}
\pdfoutput=1
\usepackage[colorlinks,linkcolor=darkblue,citecolor=darkblue,urlcolor=darkblue]{hyperref}
\usepackage{amsmath}
\usepackage{xcolor}
\usepackage{verbatim}
\usepackage{graphicx}
\usepackage{esint}
\usepackage{float}
\usepackage{bm}
\usepackage{bbm}
\definecolor{darkblue}{rgb}{0,0,0.6}
\usepackage{tcolorbox}
\usepackage{amssymb}

\newcommand{\rev}[1]{\textcolor{black}{{#1}}}

\begin{document}

\title{Violation of the fluctuation-dissipation theorem and effective temperatures in spin ice}

\author{Valentin Raban}

\affiliation{Universit\'e de Lyon, ENS de Lyon, CNRS, Laboratoire de Physique, F-69342 Lyon, France} 

\affiliation{Laboratoire Charles Coulomb (L2C), Universit\'e de Montpellier, CNRS,
 34095 Montpellier, France}

\author{Ludovic Berthier}

\affiliation{Laboratoire Charles Coulomb (L2C), Universit\'e de Montpellier, CNRS,
 34095 Montpellier, France}

\affiliation{Yusuf Hamied Department of Chemistry, University of Cambridge, Lensfield Road, Cambridge CB2 1EW, United Kingdom}

\author{Peter C. W. Holdsworth}

\affiliation{Universit\'e de Lyon, ENS de Lyon, CNRS, Laboratoire de Physique, F-69342 Lyon, France} 

\date{\today}

\begin{abstract}
We present numerical tests of the fluctuation-dissipation theorem (FDT) in the dumbbell model of spin ice with parameters suitable for dysprosium titanate. The tests are made for local spin variables, magnetic monopole density, and energy. We are able to achieve local equilibrium in which the FDT is satisfied down to $T=0.4$~K below which the system completely freezes. Non-equilibrium dynamics, together with violation of the FDT, are nonetheless observed following a thermal quench into the non-contractable monopole pair regime. Despite FDT violation, an approximate linear response regime allows for the identification of effective non-equilibrium temperatures which are different for each variable. The spin variable appears hotter than the heat reservoir, the monopole concentration responds with a lower effective temperature while the energy has a negative effective temperature. Results are discussed in the context of the monopole picture of spin ice and compared to the structure of FDT violations in other glassy materials. Prospectives for future experiments are reviewed.
\end{abstract}

\maketitle

\section{Introduction}

Spin ice materials and models~\cite{spinicebook2021} share a number of characteristics with spin glasses~\cite{mezard1987spin} and other glassy materials~\cite{berthier2011theoretical}. They show slow dynamics at temperatures below the interaction energy scale \cite{Matsuhira2001,Snyder2004,Jaubert2009,Quilliam2011,Matsuhira2011,Yaraskavitch2012,Revell2012,Takatsu2013,eyvazov2018common}, ergodicity breaking between field cooled (FC) and zero-field cooled (ZFC) protocols \cite{Snyder2004} and, under certain conditions appear to show stretched exponential slowing down characteristic of supercooled liquids~\cite{kassner2015supercooled}. It is therefore interesting to borrow tools developed in the field of spin and structural glasses to analyse spin ice in order to understand better the similarities and differences between these different types of frustrated materials undergoing some kind of ergodicity breaking.

Glassy systems, once inside the ergodicity breaking time frame, typically  violate the fluctuation-dissipation theorem~\cite{chandler1987introduction} (FDT) with spontaneous fluctuations and response functions no longer related to one another with a prefactor corresponding to the temperature of the thermal bath~\cite{bouchaud1998out,cugliandolo1993analytical,cugliandolo2011effective,cugliandolo1997energy,crisanti2003violation,kurchan2005and}. Many glassy materials are seen to `age' as they evolve through a hierarchical structure of metastable states~\cite{struik1977physical}. Despite this, integrated response functions often remain proportional to their conjugate correlation functions~\cite{bouchaud1998out,cugliandolo1993analytical} allowing for the definition of an effective temperature~\cite{cugliandolo1997energy,cugliandolo2011effective} replacing that of the thermal bath and associated with the explored non-equilibrium states. As a consequence the study of effective temperatures and of the so-called fluctuation-dissipation ratio has proved a rich and powerful tool to study glassy materials and their non-equilibrium aging dynamics at low temperatures~\cite{cugliandolo2011effective,crisanti2003violation}. FDT violations have been studied in many disordered materials with slow dynamics, such as spin glasses~\cite{marinari1998violation,barrat2001real,ricci2003measuring,herisson2002fluctuation}, and structural glasses~\cite{PhysRevLett.79.3660,barrat1999fluctuation,grigera1999observation,bellon2001violation,berthier2007efficient}. The same tools have been applied to scores of non-equilibrium dynamics, like coarsening~\cite{barrat1998monte,berthier1999response}, non-equilibrium critical dynamics~\cite{godreche2000response,henkel2001aging,berthier2001nonequilibrium,mayer2003fluctuation,calabrese2005ageing}, sheared complex fluids~\cite{barrat2000fluctuation,berthier2002nonequilibrium,berthier2002shearing}, and active matter~\cite{loi2008effective,levis2015single}. While FDT violations can of course be generically expected in non-equilibrium regimes, the specific structure of the measured FDT violations in glassy materials has helped distinguishing different types of aging dynamics~\cite{cugliandolo2011effective,crisanti2003violation,kurchan2005and}. 

In this paper we present a numerical study the fluctuation dissipation theorem, its satisfaction and violation within the monopole picture of spin ice. We show that annealing below the FC-ZFC ergodicity threshold \cite{Snyder2004}, the FDT remains satisfied, a result that we argue is consistent with the absence of a hierarchical structure in phase space. However, on making a rapid quench in temperature, which traps a finite concentration of `non-contractable pairs' of monopole quasi-particle excitations~\cite{castelnovo2010thermal}, we find distinct FDT violations, with different effective temperatures for energy, spin and monopole degrees of freedom. In particular, the energy shows a negative effective temperature reminiscent of kinetically constrained models of glass forming materials~\cite{mayer2006activated,leonard2007non,jack2006fluctuation,garriga2009negative}, where aging dynamics is similarly controlled by localised defects with thermally activated dynamics.  

We explicitly limit ourselves to the monopole approximation~\cite{Ryzhkin2005,castelnovo2008magnetic}, that is, to the dumbbell model \cite{molller2006needle} in which the magnetic moments of spin ice materials, localised on the nodes of the corner-sharing pyrochlore lattice are replaced by needles,  extended to touch at the centres of the tetrahedra, forming a diamond lattice of vertices, see Fig.~\ref{Dumbbell}. The vertices carry monopole charges interacting via Coulomb's law ~\cite{castelnovo2008magnetic}. The monopole picture has been hugely successful in providing a theoretical framework for such a complex frustrated system, giving a quantitative description of the thermodynamics and a good qualitative description of both the static and dynamic magnetic properties of spin ice materials~\cite{castelnovo2021modelling}. To this picture we highlight two further simplifications which should be separated from the use of the monopole Hamiltonian. Firstly we limit ourselves to periodic boundary conditions, and secondly we model the real dynamics using Metropolis dynamics. This choice corresponds, at the microscopic level, to modelling spin (or in our case needle) flips with a single tunnelling rate, independently of the local environment \cite{Jaubert11}. 

This simplest of starting points allows for the identification of the intrinsic mechanisms that can drive FDT violations. These should be contrasted with extrinsic mechanisms or corrections to this simplest of monopole pictures which, as a consequence, will be identifiable in future work. Such effects have already been identified in the modelling of observed slow dynamics at low temperature where the influence of the rapidly falling monopole density is strongly enhanced by the presence of defects, of open boundaries, and of the temperature dependence of the microscopic tunnelling process \cite{Revell2012,Takatsu2013,Tomasello2019}. 

The rest of the paper is organised as follows. In Sec.~\ref{sec:dumbbell} we review the dumbbell model and the numerical techniques used to study dynamical correlation and response functions. In Sec.~\ref{sec:FDT} we review the FDT adapted to spin ice models. In Sec.~\ref{sec:eq} we show numerical data revisiting the slow equilibrium dynamics with FDT satisfaction as the monopole concentration falls at low temperature. In Sec.~\ref{sec:noneq} we turn to non-equilibrium dynamics in the non-contractable pair regime in which our numerical data shows FDT violations. Some conclusions and perspectives are given in Sec.~\ref{sec:conclusion}. 

\section{The monopole picture and the dumbbell model}

\label{sec:dumbbell}

\begin{figure}
\centering{\includegraphics[scale=0.25]{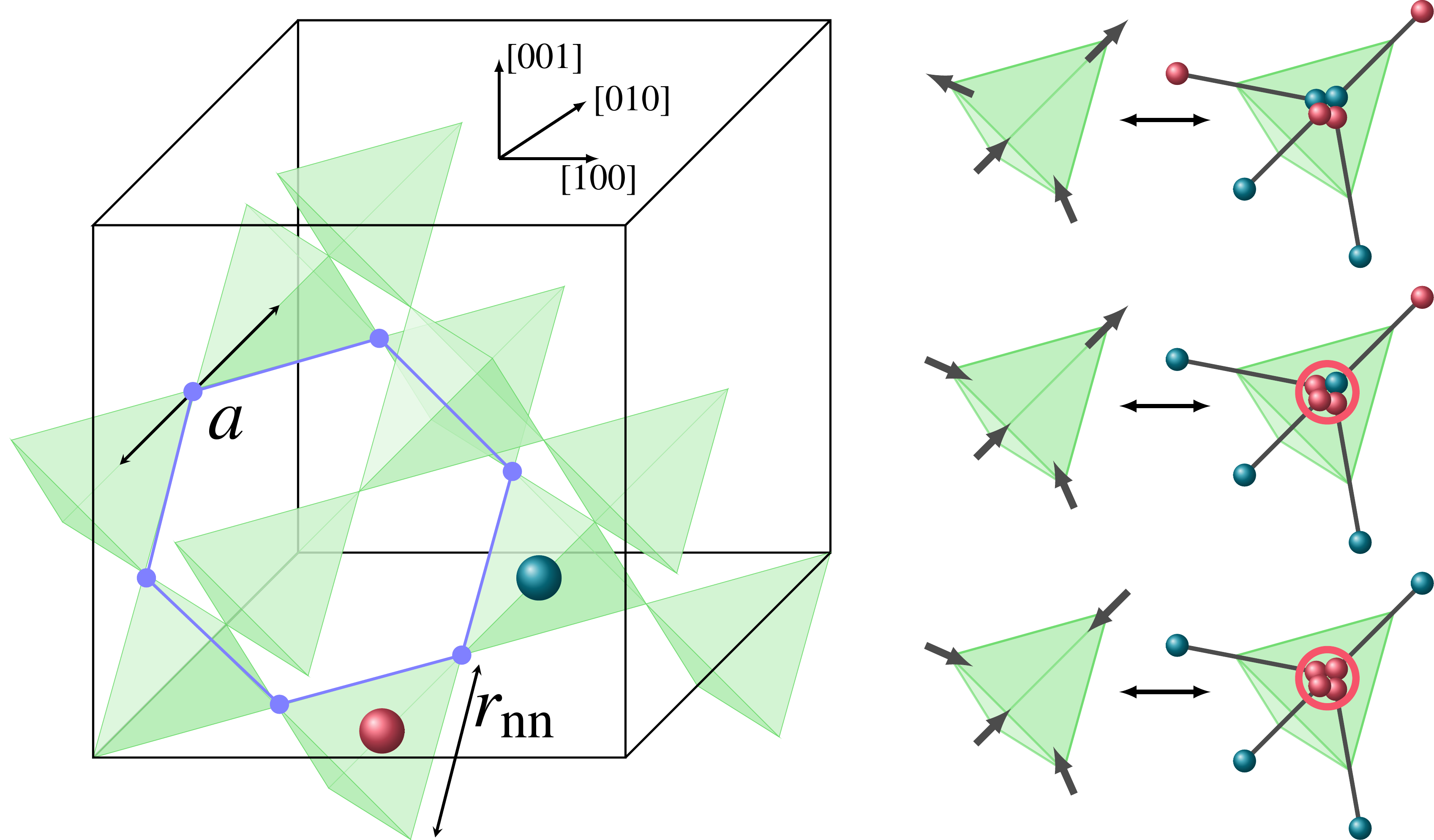}}
\caption{From spins to dumbbells. Left: Pyrochlore lattice of corner sharing tetrahedra. Tetrahedron centres form a diamond lattice. Blue and red spheres illustrate monopoles of charge $\pm Q$.
  Right: The point dipoles are extended to needles touching at the diamond lattice sites. The needles carry magnetic flux and and dumbbelles of charge $q=\pm m/a$. In a 2in-2out configuration (top) the vertex is charge neutral. A 3in-1out (3out-1in) configuration carries a monopole charge $Q=2m/a$ ($-Q=-2m/a$) (centre). A 4in (4out) configuration carries a double monopole charge $2Q=4m/a$ ($-2Q=-4m/a$) (bottom).}
\label{Dumbbell}
\end{figure}

Spin ice materials \cite{harris1997geometrical,bramwell2001spin} such as dysprosium or holmium titanate are Ising-like frustrated ferromagnets with magnetic degrees of freedom on the sites of the corner sharing pyrochlore lattice, with moments pointing along the body-centred diagonal axes that join at the centres of the tetrahedra, which themselves form a diamond lattice of vertices, see Fig.~\ref{Dumbbell}. Interactions, both exchange and dipolar, are on the one Kelvin scale and are captured by the dipolar spin ice model \cite{denHertog2000,melko2004monte} (DSI). The lowest energy states, with spins on each tetrahedron satisfying the Bernal-Fowler ice-rules~\cite{doi:10.1063/1.1749327} of two spins in and two out in each tetrahedron, form a low-energy extensive band containing the Pauling entropy of approximately $\frac{1}{2}\ln{\left(\frac{3}{2}\right)}$ per spin. The band width of these states is well below the energy scale of interactions, as the long-range part of the dipole interactions are almost self-screened throughout the band \cite{denHertog2000,isakov2005projective,melko2001long}. The separation of energy scales allows for the interpretation of the magnetic moments as elements of an emergent magnetostatic field with the low-energy states constrained by a divergence-free condition \cite{brooks2014magnetic,lhotel2020fragmentation}. Excitations above the low-energy band are topological defects to the emergent field \cite{Ryzhkin2005} which, being dressed by the real magnetic fields of the dipoles, carry a magnetic charge \cite{castelnovo2008magnetic}. These are the magnetic monopoles that have been much studied over the last decade. 

The monopole approximation for spin ice is captured by 
the dumbbell model~\cite{castelnovo2008magnetic,Kaiser2018,castelnovo2021modelling} in which the point-like dipole moments of the DSI are replaced by magnetic needles which touch at the centres of the tetrahedra. The magnetic flux is channeled down the needles, so that they
 carry dumbbells of magnetic charge $\pm m/a$ where $m$ is the magnetic moment and $a$ the distance between tetrahedron centres. Configurations with two spins in and two out are by construction charge neutral and the total charge on site $i$ is $Q_i=0, \pm Q, \pm 2Q$, with $Q=2m/a$ the monopole charge. Also by construction in this model, the ensemble of Pauling states are exactly degenerate, so that the model violates the third law of thermodynamics. The Hamiltonian for excitations above the extensively degenerate ground state is 
\begin{equation} 
{\cal H}_{\rm db}=\frac{u}{2} \sum_{i \neq j} \left( \frac{a}{r_{ij}} \right) \hat{n}_i \hat{n}_j - \mu \sum_i \hat{n}_i^2 ,
\label{H-dumb}
\end{equation}
where $\hat{n}_i=Q_i/Q=0,\pm 1, \pm 2$ is a site occupation variable, $r_{ij}$ the distance between sites $i$ and $j$, $u=\frac{\mu_0 Q^2}{4\pi a}$ is the Coulomb energy scale, $\mu_0$ is the permeability of free space and $\mu<0$ the chemical potential~\cite{Jaubert11,Castelnovo2011,raban2019multiple}. 

Throughout the paper we follow standard notation for spin ice simulations and refer to a dimensionless length $L$, measured in cubic units. Each cubic cell contains 16 spins (dumbbells) so that the number of tetrahedra (monopole sites) is $N_0=8L^3$. We index diamond lattice sites hosting the monopoles with $i,j$ and and the pyrochlore lattice sites hosting the spins by $\kappa,\lambda$. Quantitative measures refer to the spin ice material Dy$_2$Ti$_2$O$_7$ (DTO) for which we take the the diamond lattice constant to be  $\;a=4.36$ \AA, the nearest neighbour spin distance $r_{nn}=\sqrt{\frac{2}{3}}a=3.56$ \AA,  and cube length  $a_c=\frac{4a}{\sqrt{3}} \approx 10.1$~\AA~(see Fig.~\ref{Dumbbell})~\cite{zhou2012parameters}. We use the Kelvin energy scale, fixing the Boltzmann constant to unity.  Taking an estimate for the magnetic moment $m=9.87$ $\mu_B$~\cite{Yavorskii2008} yields $u=2.88$ K and $\mu=-4.35$~K.

\rev{Dynamics are simulated using a Metropolis Monte Carlo algorithm for dumbbell flips between their discrete orientations. Through the dumbbell flips the charges associated with the monopole quasi-particles move, are created or are destroyed in nearest neighbour pairs. The long range nature of the Coulomb interaction is dealt with using the Ewald summation method. Time steps are recorded as Monte Carlo moves per dumbbell (MCS). Spin configurations at each step can be recovered by replacing each dumbbell by the corresponding spin. }

The emergent monopole picture can also be cast in the language of a lattice field which can be decomposed into `longitudinal', `transverse' and `harmonic' components via a lattice Helmholtz decomposition~\cite{brooks2014magnetic,bramwell2017harmonic,lhotel2020fragmentation}. The monopoles can be reconstituted from the longitudinal part ($m$), with the leftover, defined by the constraints of fixed spin density, shared between the other two components. The divergence-free transverse component ($d$) has characteristic dipolar correlations~\cite{PhysRevLett.91.167004,Isakov2004}. With periodic boundaries and in zero field, the harmonic term can be set to zero to an excellent approximation, so that for each microstate  the spin at site $\kappa$ can be written
\begin{equation}
S_{\kappa}=S_{\kappa}^m+S_{\kappa}^d.
\end{equation}
Vector fields can be built at each site by multiplying the components by the local unit vectors connecting the diamond lattice sites with appropriate sign convention~\cite{lhotel2020fragmentation}. \rev{The closed loops of the transverse component provide the residual entropy as the temperature falls to zero while the energy (of the monopoles) is carried uniquely by the longitudinal component.} In Fourier space, the two components fall parallel and perpendicular to the propagation vector $\vec q$ defined within the first Brillouin zone. This apparent fragmentation of the magnetic moments into two orthogonal parts generates a form of spin-charge separation in which magnetic monopoles and transverse spin components  develop largely independent fluctuations and response functions. 

\section{The fluctuation dissipation theorem out of  equilibrium}

\label{sec:FDT}

\subsection{The fluctuation-dissipation theorem}

The fluctuation dissipation theorem (FDT) relates the equilibrium fluctuations of an observable $A$ (in this case extensive), to the linear response to an applied field $f$ conjugate to $A$, such that the Hamiltonian at finite $f$ is $H(f)=H(0) - f A$. Starting an experiment at time $t=0$ in equilibrium and applying a field perturbation at time $t_w \ge 0$, the variable $\langle A(t) \rangle$ is measured at time $t \ge t_w$, where the brackets stand for an ensemble average. This setting provides a general form for the FDT:
\begin{equation}
  \chi_A(t,t_w)=\frac{\partial \langle A(t)\rangle}{\partial f(t_w)} =
  \frac{1}{T} [C_A(t,t)-C_A(t,t_w) ],
  \label{eq:FDT}
  \end{equation}
  with $C_A(t,t_w) = \langle A(t) A(t_w) \rangle - \langle A(t) \rangle \langle A(t_w) \rangle$. The proportionality constant, $1/T$, between left and right hand functions illustrates the fact that the FDT is indeed explicitely derived in the conditions of thermodynamic equilibrium. In addition, in equilibrium the functions depend on a single time variable, the time difference, $t'=t-t_w$. Taking $t' \rightarrow \infty$ we find the static limit, $\chi_A=\frac{\partial \langle A\rangle}{\partial f} =  \frac{1}{T} [ \langle A^2\rangle - \langle A\rangle^2 ]$.

Defining the normalised response and auto-correlation functions
\begin{eqnarray}
\tilde{\chi}_A(t,t_w)&=& \frac{\chi_A(t,t_w)}{C_A(t,t)} , \nonumber \\
\tilde{C}_A(t,t_w)&=& \frac{C_A(t,t_w)}{C_A(t,t)},
\label{Eq-FDT}
\end{eqnarray}
the FDT in Eq.~(\ref{eq:FDT}) takes the very simple form $T \tilde{\chi}_A = 1-\tilde{C}_A$ and can be tested graphically by making parametric plots of $T\tilde{\chi}_A$ as a function of $1-\tilde{C}_A$, with FDT satisfaction corresponding to a straight line of unit slope. 

Starting at $t=0$ from a non-equilibrium situation such as a thermal or field quench~\cite{Paulsen2014}, the FDT may in principle be violated, requiring the insertion into Eq.~(\ref{Eq-FDT}) of a further parameter, $X_A(t,t_w)$, called the fluctuation-dissipation ratio, such that 
\begin{equation}
  T\tilde{\chi}_A=X_A(t,t_w)(1-\tilde{C}_A),
  \label{eq:defX}
\end{equation}
with $X_A(t,t_w)=1$ at equilibrium. In principle, if one waits long enough, so that $t_w$ exceeds all relaxation times of the system, equilibrium behaviour and the FDT will be recovered with $X_A \rightarrow 1$. However, in glassy regimes the equilibrium relaxation times may exceed all possible observation times with the result that they appear permanently out of equilibrium, with corresponding violations of the FDT and $X_A \neq 1$. 

In following such protocols for systems where time-translation invariance is not satisfied, either $t$ or $t_w$ can be varied as the control parameter, although varying $t_w$ can be extremely time consuming, as each measurement requires an independent procedure for each value of $t_w$. This problem is particularly acute when the fluctuation-dissipation $X_A(t,t_w)$ has a non-trivial time-dependence, since its mathematical definition via Eq.~(\ref{eq:defX}) requires using $t_w$ as the appropriate time variable~\cite{sollich2002fluctuation}. This problem can be fully circumvented in numerical simulations by using the no-field techniques~\cite{chatelain2003far,ricci2003measuring,berthier2007efficient} described in Appendix A, in which the linear response to the applied field can be extracted directly from derivatives of the Boltzmann weights in zero field. With this innovation, numerical simulations varying $t$ or $t_w$ become equivalent in computational effort as both correlation and response functions are evaluated in the same set of unperturbed simulations.

In certain model systems showing a thermodynamic glass transition, $X_A(t,t_w)$ can be shown to be a constant~\cite{cugliandolo1993analytical}, different from unity, allowing for the definition of an effective temperature~\cite{cugliandolo1997energy},
\begin{equation}
  T_{\rm eff} = \frac{T}{X_A},
  \label{eq:Teff}
\end{equation}
associated with out of equilibrium fluctuations. No such definition can be made in more realistic models showing glassy behaviour, but $X_A(t,t_w)$ does appear approximately linear in many numerical simulations~\cite{barrat1999fluctuation,barrat2000fluctuation,berthier2002nonequilibrium,berthier2007efficient,sciortino2001extension} and in some experiments~\cite{grigera1999observation,bellon2001violation}.

This phenomenology has been an important development over the last three decades as it provides a test of possible universal behaviour of the non-equilibrium dynamics of systems with very long relaxation times. This has been largely triggered by the analytic study of a broad family of mean-field glassy models in their aging regime~\cite{cugliandolo1993analytical,bouchaud1998out,cugliandolo1999thermal}. A unique effective temperature shared by all physical observables is found in supercooled liquids~\cite{barrat1999fluctuation,berthier2002shearing,berthier2002nonequilibrium}, which are characterised by a clear separation of timescales with a unique slow relaxation mechanism. Kinetically constrained and non-mean-field models of glasses partly share this phenomenology~\cite{sollich2002fluctuation}, but reveal in addition the possibility of an absolute negative effective temperature when considering energy fluctuations and response functions~\cite{mayer2006activated,leonard2007non,jack2006fluctuation,garriga2009negative}. This striking result directly follows from the thermally activated nature of the microscopic aging motion in such systems, whereby increasing the temperature leads to an acceleration of the energy decrease, and thus to negative response functions~\cite{mayer2006activated}.

By contrast, spin glasses are characterised, at least at mean-field level, by a more complex hierarchy of relaxation timescales~\cite{cugliandolo1994out}, leading to a fluctuation-dissipation ratio which becomes a function rather than a simple number. This finding has direct, deep connections to the Parisi overlap distribution describing the spin glass order in equilibrium at low temperatures~\cite{bouchaud1998out,PhysRevLett.81.1758,PhysRevE.63.016105,kurchan2021time}. Physically, one can associate an effective temperature to each relaxation timescale, and the emergence of multiple timescales suggests the possibility that distinct effective temperatures can also emerge.  

Critical dynamics in pure ferromagnets also display remarkable universal violation of the FDT with the fluctuation-dissipation ratio taking a universal value uniquely dependent of the universality class~\cite{godreche2000response,henkel2001aging,berthier2001nonequilibrium,mayer2003fluctuation,calabrese2005ageing}. Coarsening and domain growth processes in ordered phases are characterised by an effective temnperature which asymptotically diverges, or equivalently, by a vanishing fluctuation-dissipation ratio~\cite{barrat1998monte,berthier1999response}.    

With few notable exceptions~\cite{herisson2002fluctuation,herisson2004off} most experiments dedicated to tests of FDT violations in aging glassy materials are not performed in the time domain but rather in Fourier space~\cite{refregier1987dynamic,grigera1999observation,bellon2001violation,bellon2002experimental,buisson2003intermittency,PhysRevLett.93.160603}. In this case, the FDT relates the fluctuation spectral density $S(\omega)$ of a given observable at frequency $\omega$ to the out of phase linear susceptibility $\chi''(\omega)$ at the same frequency:
\begin{equation}
  S(\omega) = \frac{2 k_B T}{\pi \omega} \chi''(\omega),
\label{eq:noise}
\end{equation}
and its generalisation introduces a frequency-dependent fluctuation-dissipation ratio as the ratio between the right and left hand sides of Eq.~(\ref{eq:noise}). In the case where the fluctuation-dissipation ratio is a simple number in the slow regime, then the time and frequency domain approaches yield similar results~\cite{cugliandolo1997energy}, with a linear relation between response and fluctuations with an effective temperature replacing the thermal bath temperature.  

\subsection{Correlation and response functions in spin ice}

We adapt the FDT formalism to three different observables in the dumbbell model of spin ice. The first is the local moment on each site, $S_{\kappa}$. This is an Ising-like variable whose conjugate field, $h_\kappa$, lies parallel or anti-parallel to the axis joining the two tetrahedron centres which the spin connects. The relevant correlation function 
\begin{equation}
  \tilde{C}_S (t,t_w) = \frac{\langle S_{\kappa}(t)S_{\kappa}(t_w) \rangle -\langle S_{\kappa}(t) \rangle \langle S_{\kappa}(t_w) \rangle}{\langle S_{\kappa}(t)S_{\kappa}(t) \rangle -\langle S_{\kappa}(t) \rangle \langle S_{\kappa}(t) \rangle},
  \label{C_S}
\end{equation}
is related to the local spin susceptibility, $\chi_S=\frac{\partial\langle S_{\kappa}\rangle}{\partial h_\kappa}$ and measurements are taken from a configurational average over the $N$ spins.

Secondly, we consider fluctuations of the configurational energy $E$ calculated from Eq.~(\ref{H-dumb}) through the response and correlation functions $\tilde{\chi}_E(t,t_w)$ and $\tilde{C}_E (t,t_w)$. In the language of a fluid in the grand ensemble, one might call this a number enthalpy as for each configuration $E = U_C - \mu \mathcal{N} $  contains both the Coulomb energy, $U_C$  and the energy cost of creating $\mathcal{N}=\sum_i |\hat{n}_i|$ magnetic monopoles. In principle one should also take into account the cost of creating double monopoles (tetrahedra with all spins pointing in or out) but for the temperatures used in this paper these can be safely neglected. The conjugate field for the energy is a dimensionless temperature change, $\alpha = \frac{\delta T}{T}$. The response function in the static limit is then closely related to the specific heat at constant chemical potential, $C_{\mu}$, which is equivalent to the magnetic specific heat in zero field, $\chi_E=\frac{\partial E}{\partial\alpha}= TC_{\mu}$.

Finally, we consider fluctuations in the monopole concentration, choosing as for the spins a local measure, $|\hat{n}_i|$, with conjugate field a local chemical potential, $\mu_i=\mu+\delta \mu_i$ and with the response and correlation functions $\tilde{\chi}_Q(t,t_w)$ and $\tilde{C}_Q (t,t_w)$.
The response function, $\chi_Q=\frac{\partial |\hat{n}_i|}{\partial\mu_i}$, is closely related to the monopole compressibility, $\frac{\partial \mathcal{N}}{\partial\mu}$.

\section{Dynamics at equilibrium}

\label{sec:eq}

\subsection{Low-temperature dynamics}

The evolution of magnetic time scales extracted from {\it ac} susceptibility measurements~\cite{Snyder2004} on Dy$_2$Ti$_2$O$_7$ is largely captured by stochastic monopole dynamics in the dumbbell model~\cite{Jaubert2009} and through spin dynamics in the DSI~\cite{Jaubert11,Revell2012,Takatsu2013,dusad2019magnetic}. On cooling, the measured time scale exhibits a plateau-like region between $10$~K and $4$~K before entering a regime of rapid change in which the characteristic time increases faster than an Arrhenius law, before leaving the experimental time window for temperatures below around $0.65$ K~\cite{Snyder2004}. The latter also corresponds to the ergodicity breaking temperature between field-cooled and zero-field cooled protocols for {\it dc} susceptibility measurements. The increasing time scale is directly related to the fall in monopole density, with the non-Arrhenius behaviour mostly due to the Coulomb interaction between monopoles and the increased screening length as temperature is reduced. As Coulomb screening falls to zero at low temperature, one might expect expect the time scale to approach an Arrhenius law, $\tau=\tau_0 \exp(-\mu/T )$, with $\mu=-4.35$ K \cite{Castelnovo2011,Jaubert11} and $\tau_0$ a microscopic timescale.

Experimental results differ in two details. Firstly, the characteristic time slows down beyond this limit \cite{Matsuhira2001,Snyder2004,Quilliam2011,Matsuhira2011,Yaraskavitch2012,Revell2012} and, secondly, rather than a single time scale, a spread of times appears at each temperature \cite{Bovo2013} and some measurements suggest stretched exponential behaviour~\cite{Revell2012,eyvazov2018common}. Possible corrections to the monopole picture to account for these differences include moving from the dumbbell model back to the spins of the DSI which reintroduces a finite band width for the Pauling states~\cite{Jaubert11,Revell2012,Blundell2012}, considering open rather than periodic boundaries which can introduce multiple time scales and stretched exponential decay, or the effects of defects or configuration-dependent microscopic tunnelling rates, both of which increase time scales at low temperature~\cite{Revell2012,Takatsu2013}. 

Here we show that finite-size effects, even in this simple model, also provide corrections to the above theoretical result and longer than expected time scales at low temperature. We calculate the spin-spin autocorrelation function, $\tilde{C}_S (t,0)=\tilde{C}_S (t)$, with a configurational average  made over the $N$ spins. As this is a local mesure, it contains the diagonal terms of a magnetic correlation function only. Previous studies have shown that inclusion of off-diagonal terms makes no significant difference to the measured decay of correlations with time~\cite{Jaubert2009,Revell2012}, at least down to the magnetic ergodicity temperature, while use of the local function allows for much better statistics allowing us to extract data at lower temperature.  

\begin{figure}
  \includegraphics[width=7cm]{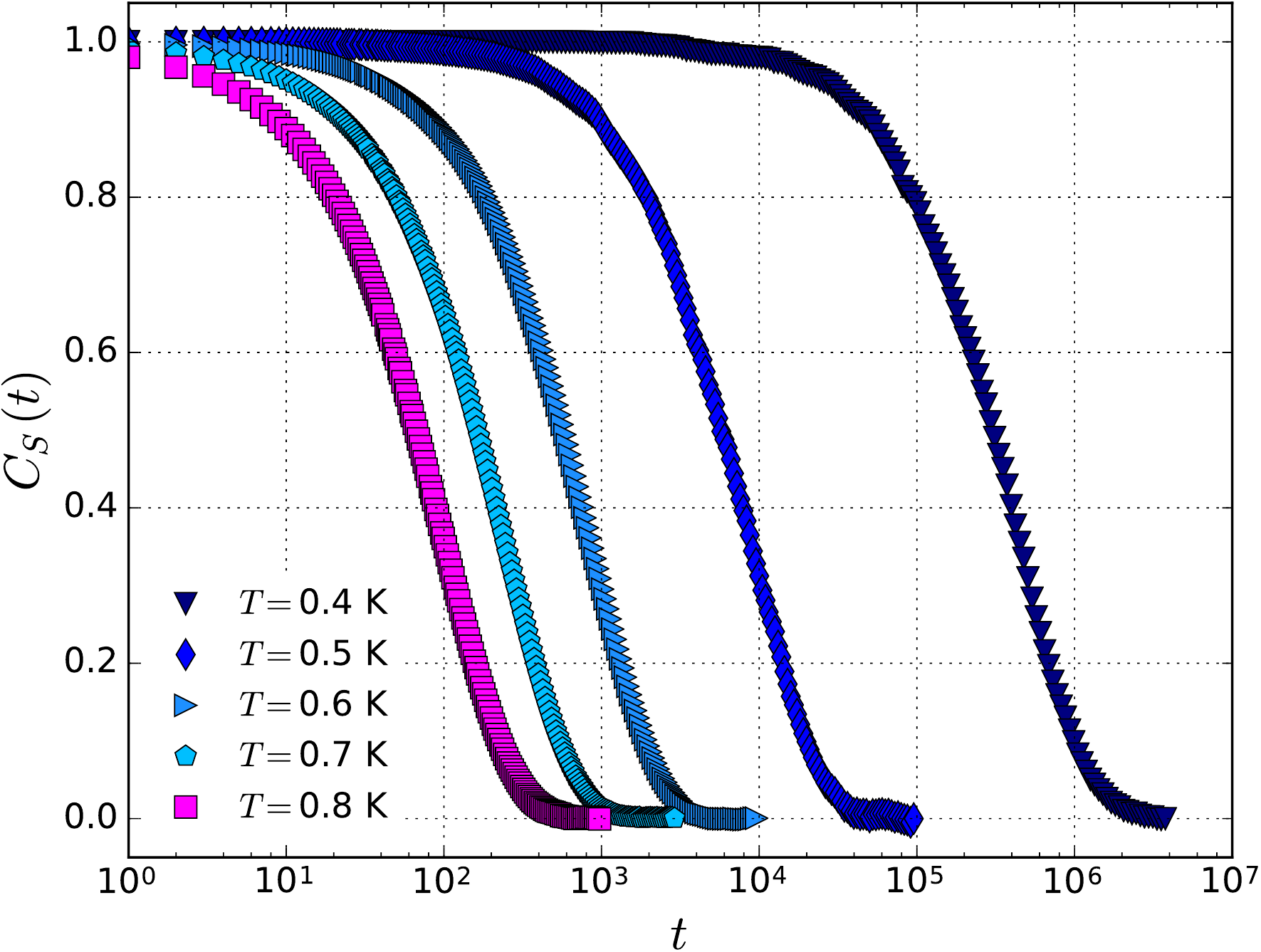}
  \includegraphics[width=7.1cm]{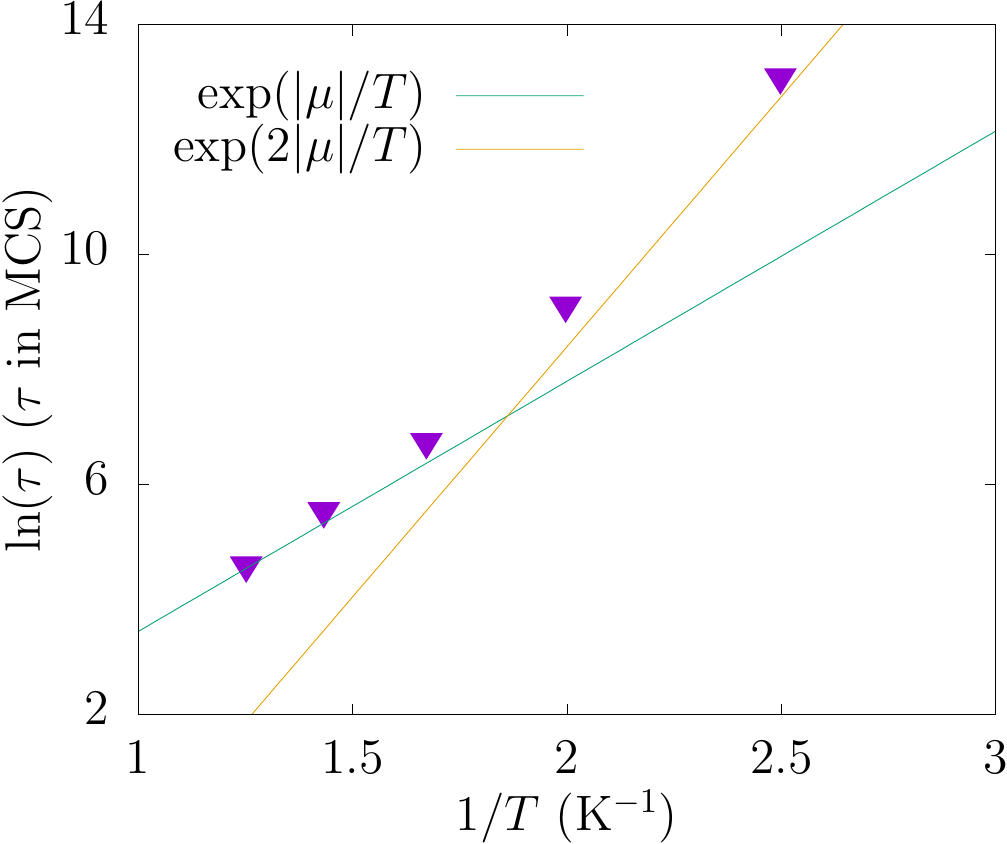}
  \caption{Top: Normalised time autocorrelation function of the spin degrees of freedom at equilibrium. Temperatures are shown in Kelvin and the energy scale is set by parameters from Dy$_2$Ti$_2$O$_7$.
  Bottom: Correlation time for the spin degrees of freedom versus the inverse of the temperature for temperature between 0.8 and 0.4 K (points). The lines show Arrhenius laws with energy scales $|\mu |$ and $2|\mu |$ with $\mu=-4.35$~K.}
\label{CorrSEq}
\end{figure}

Data are shown in Fig.~\ref{CorrSEq} for temperatures between $0.8$ and $0.4$ K, for a fixed system size, $L=6$, with correlations averaged over 2000 initial equilibrium configurations. As can be seen from the logarithmic time axis, at each temperature the correlation function falls to zero at a single, well defined time scale $\tau$ and the decay is well-represented by an exponential decay, $C_S (t) \approx \exp{-(t/\tau)}$. The value of $\tau$ increases rapidly with decreasing temperature and exceeds $10^7$ MCS for $T \leq 0.4$~K. Given that the microscopic tunnelling rate is estimated to be in the millisecond range~\cite{Jaubert2009}, this would correspond to experimental time scales of several hours. 

The evolution of $\tau$ with inverse temperature is also shown in Fig.~\ref{CorrSEq}. Linear behaviour would correspond to an Arrhenius law,  $\tau=\tau_0 \exp(\Delta E/T)$, with a slope giving the characteristic energy scale $\Delta E$. Shown also is Arrhenius behaviour for both $\Delta E=|\mu|$ and $\Delta E=2|\mu|$. It is clear that, in this temperature range, the behaviour remains non-Arrhenius and the energy scale at the lowest temperature considerably exceeds that set by the monopole chemical potential. The reason is that, for $L=6$ ($1728$ diamond lattice sites), on reducing the temperature, one rapidly enters a regime where the average number of monopoles in the system is less than two, so that for most configurations there are none. Putting $n=\frac{2}{8L^3}=\frac{4}{3}\exp{(\mu/T)}$, the density expected for the Coulomb fluid at low temperature~\cite{Kaiser2018}, gives a temperature $T=0.62$ K, which is also the threshold temperature of earlier studies. Below this temperature crossover, monopoles are confined by the periodic boundaries and the system must be excited through an energy scale of $2|\mu |=8.7$ K to create monopoles that are then free to decorrelate the system. This is the crossover we are observing in Fig.~\ref{CorrSEq}. This finite-size effect is rather stylised to be directly relevant for experiment but it is consistent with the effect of adding defects to the DSI, which appear to generate a longer time scale in the decay of correlation functions~\cite{Revell2012}. It also illustrates how mosaics and grain boundaries could lead to non-universal long-time behaviour, as seen in the ensemble of experimental results.

Despite this change of regime to finite-size, or mosaic-driven long time scales, we are able to equilibrate down to temperatures somewhat below the experimental ergodic threshold, consistently with the most recent low temperature neutron scattering and specific heat measurements on Dy$_2$Ti$_2$O$_7$~\cite{giblin2018specificheat}. This observation begs a rather subtil question concerning the emergence of a topological time scale characterising magnetic relaxation which is different from the local equilibrium time scale exposed here. At low temperature, a change in the magnetisation requires the flipping of an extended string of spins, giving ultimately a change in topological sector as the monopole concentration falls to zero~\cite{Jaubert2013}. Although a detailed analysis is beyond the scope of the present paper, these results suggest the possibility of an ergodic time scale for such sector fluctuations that exceeds that for local equilibrium, in which local measures find their zero-field equilibrium values. Calculation of the response function for the local spin variable, rather than the bulk magnetisation, allows us to bypass this delicate question, which we leave for future work.

\subsection{The FDT at equilibrium}

\begin{figure}
\centering{\includegraphics[width=7.cm]{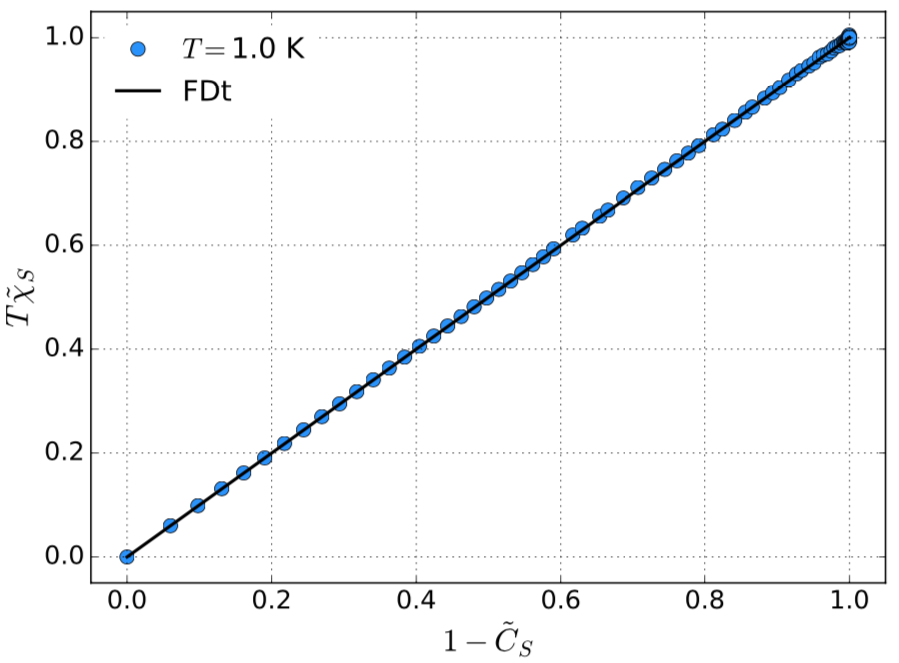}}
\centering{\includegraphics[width=7.cm]{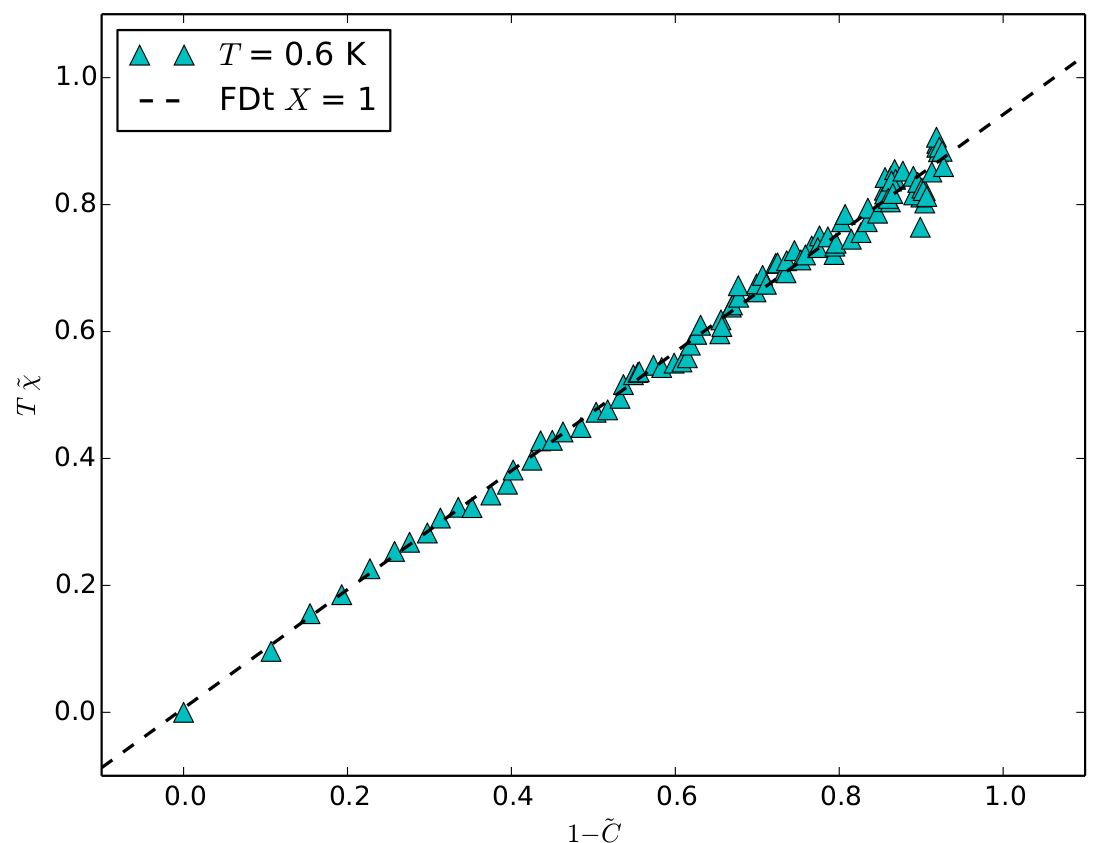}}
\caption{FDT for spin ice in equilibrium.
Top: Parametric plot of $T\tilde{\chi}_S$ {\it vs.} $1-\tilde{C}_S$, for the spin-spin autocorrelation function and spin susceptibility for the dumbbell model for $T=1$ K. The line shows the predicted FDT result.
Bottom:  Parametric plot of $T\tilde{\chi}_E$ {\it vs.} $1-\tilde{C}_E$, for the energy autocorrelation function and susceptibility for the dumbbell model for $T=0.6$~K. The line shows the predicted FDT result.
}
\label{FTD-SE}
\end{figure}

In Fig.~\ref{FTD-SE}, we show the parametric plot for the local spin correlation and response functions, $\tilde{\chi}_S$ and $\tilde{C}_S$ for $T=1$ K where equilibration is easily achieved. The data is taken by varying total time $t$ with $t_w$ held fixed, with $L=6$. The system is started at $t=0$ in an equilibrium configuration and results are averaged over $5000$ independent runs. The longest times correspond to $1-\tilde{C}_S\approx 1$, so that spins are fully decorrelated. As expected, the FDT is satisfied with a high degree of accuracy over the whole spectrum of correlation and response. Lowering the temperature, the correlations times increase but as shown above, local equilibrium can be achieved down to $T=0.4$~K, with corresponding satisfaction of the FDT. This is illustrated in Fig.~\ref{FTD-SE}, lower panel, where we show data for the energy $E$ at $T=0.6$~K, temperature close to the magnetic ergodicity breaking and for which the FDT is also clearly satisfied. 

However, the exponentially diverging time scale guarantees that equilibration, local or otherwise is excluded at lower temperatures. For example, our results suggest that moving to $T=0.2$ K would give a relaxation time scale, $\tau\sim\tau_0 \exp(40)\sim 10^7$ years for dysprosium titanate. Over the vast majority of this time scale a finite system would have a monopole population of zero, which ensures that a system near equilibrium would be completely frozen throughout any experimental time window. As a consequence, low temperature states with measurable dynamics must, by construction be far from equilibrium. 

\section{Non-equilibrium dynamics}

\label{sec:noneq}

\subsection{Non-contractable pairs}

\begin{figure}
\begin{center}
\includegraphics[width=5cm,clip]{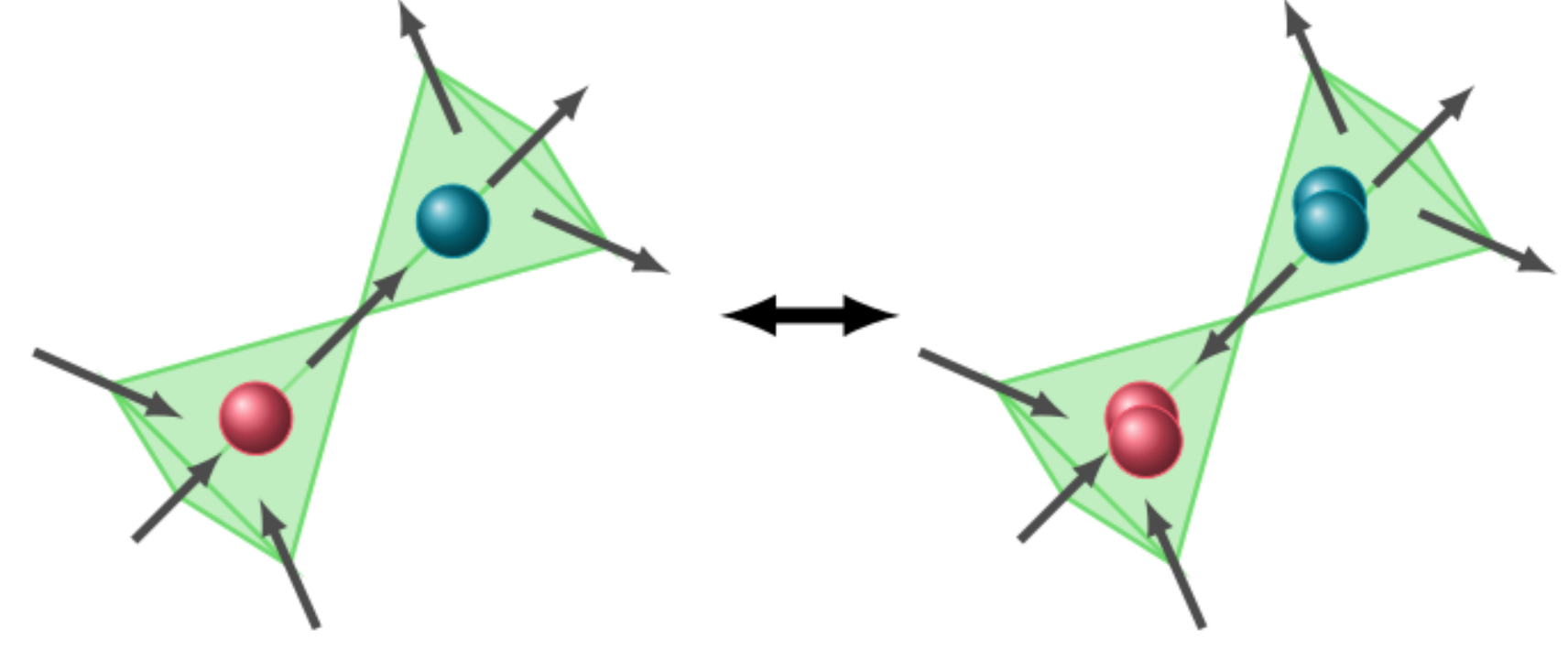}
\caption{A non-contractable pair of charges (left). Flipping the central spin does not annihilate the charges but instead creates two double charges (right).}
\label{pair}
\end{center}
\end{figure}

A class of non-equilibrium states showing measurable dynamics at ultra-low temperatures are those with a finite concentration of `non-contractable' monopole pairs~\cite{castelnovo2010thermal}, which are illustrated in Fig.~\ref{pair}. Here, annihilating the nearest neighbour monopole pair would lead to an energy gain, $|2\mu|-u=5.82$ K, so that far below this energy scale such defects would disappear at equilibrium. However, the constraints of spin ice are such that flipping the spin separating the monopole pair creates a pair of double monopoles rather than annihilating the charge. As a consequence, charge annihilation requires movement of the quasi-particles around an external path and passage over an energy barrier. \rev{As a result such a pair can be locked in a metastable state over a finite period of time after the system is quenched from a high temperature state (the opposite temperature protocol would not yield any interesting metastability).} A detailed discussion of this process was provided in Ref.~[\onlinecite{castelnovo2010thermal}].

Here we confirm the key role played by the Coulomb interaction through a study of the dumbbell model. A minimal path for monopole annihilation is a hexagon of six tetrahedra with a maximum separation of third nearest neighbour on the diamond lattice. Using the parameters quoted here for Dy$_2$Ti$_2$0$_7$ the energy barrier, $\Delta E$ for the annihilation of an isolated non-contractable pair is around $1.4$ K. Such a barrier would give an Arrhenius time scale, $\tau\sim\tau_0 \exp(14)\sim 10^6$ MCS at $T=0.1$ K which suggests the existence of slow, but accessible dynamics down to this temperature scale.

\begin{figure}
\includegraphics[width=7cm]{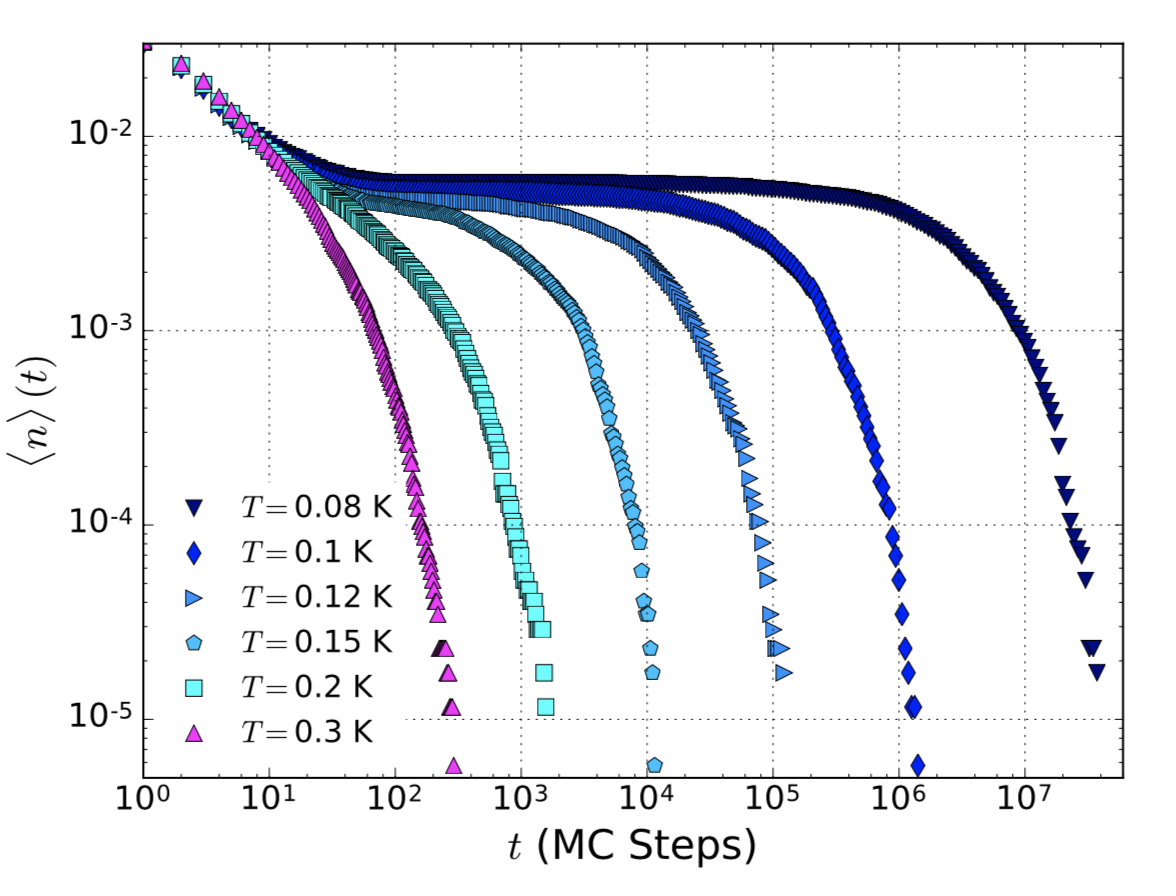}
\includegraphics[width=7cm]{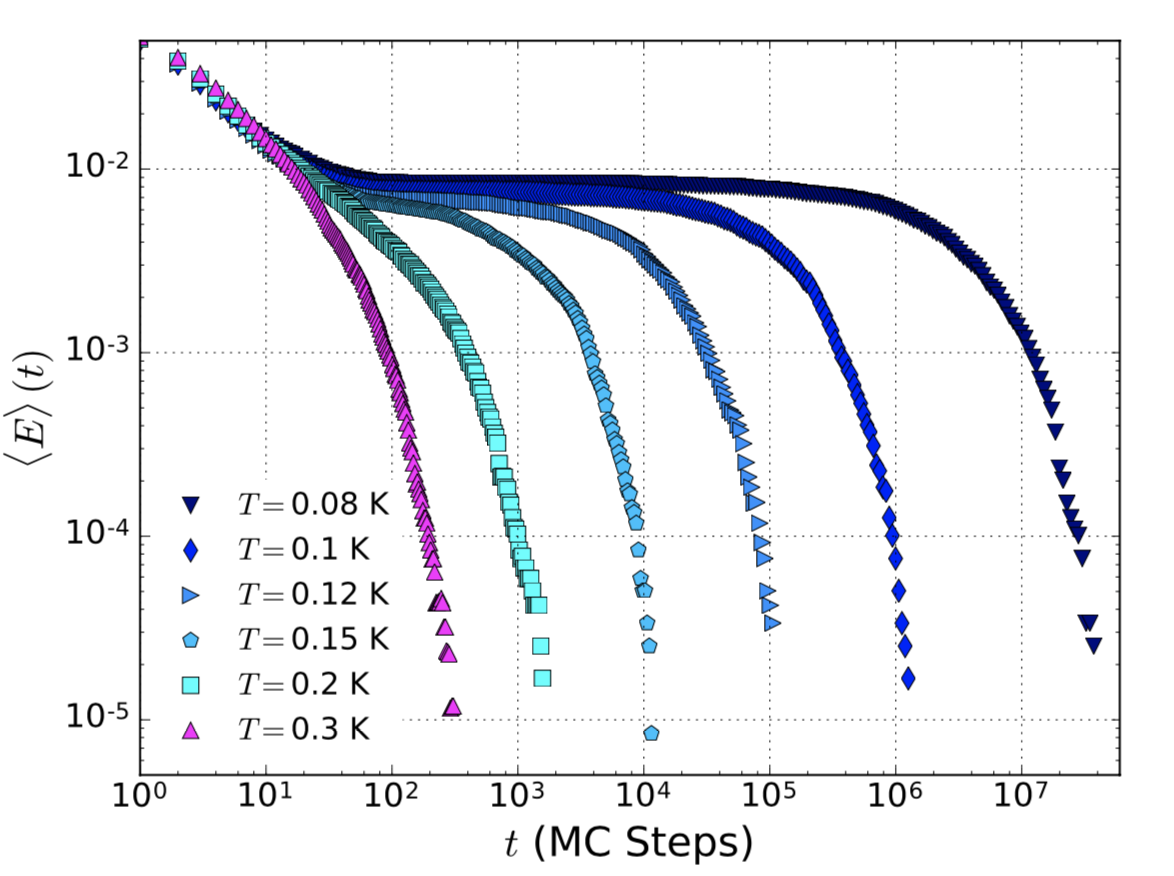}
\caption{Monopole concentration (top) and energy (bottom) following a quench at $t=0$ from $T_0= 1$~K to a final temperature $T$. For a quench below 0.3 K, a plateau regime appears for both quantities.}
\label{quenchNRJ}
\end{figure}

\begin{figure}
\includegraphics[width=7cm]{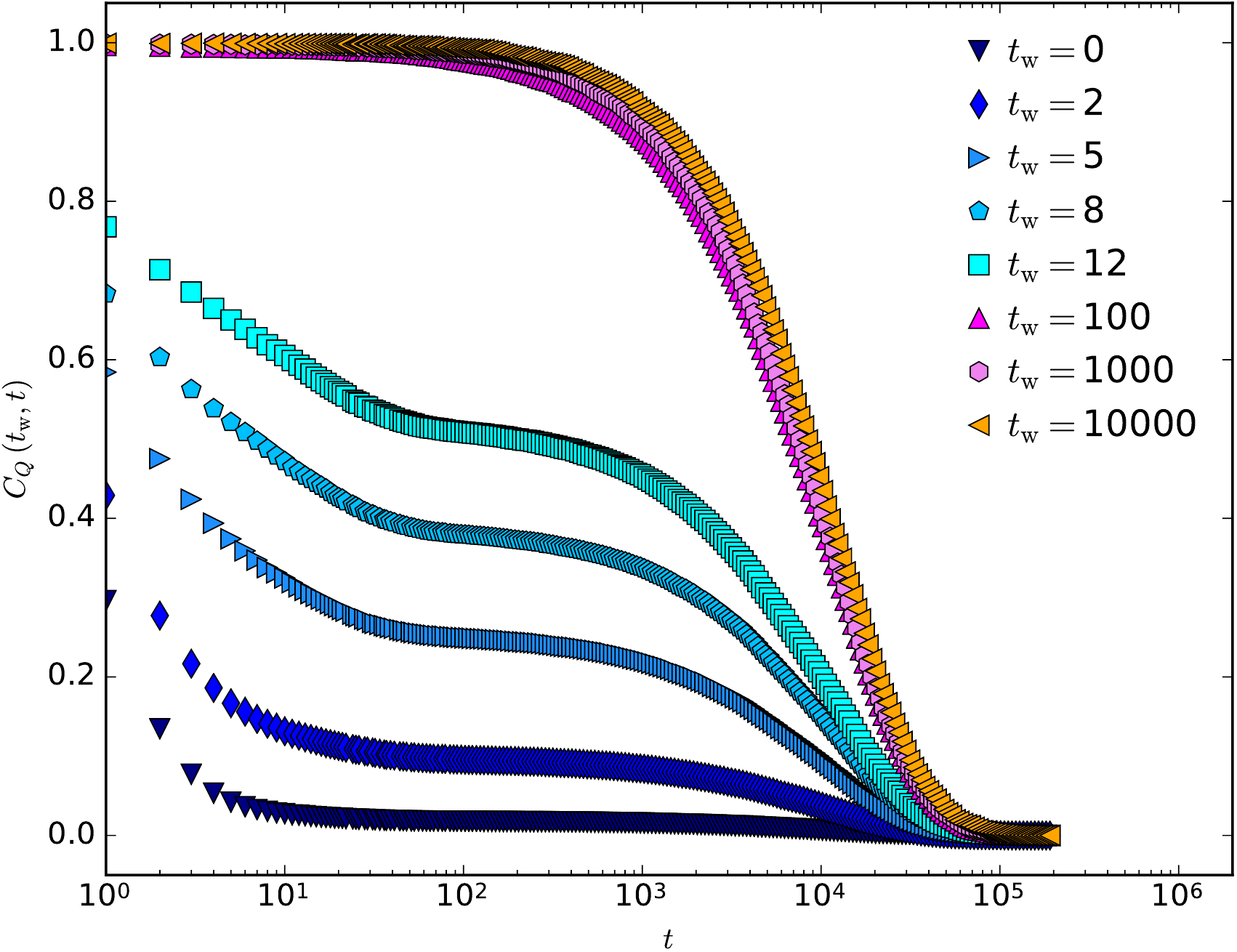}
\caption{Normalised time autocorrelation functions of the local charge as a function of total time $t$ for different waiting time $t_{w}$ after a quench from $T = 1$ K to $T= 0.12$ K. The aging of the correlation is trivial in the sense that the correlation reaches its equilibrium behaviour very quickly (about a hundred MCS) compared to the equilibration time.}
\label{AgingS}
\end{figure}

\rev{Monopole-monopole interactions are the key to  the pair formation. A nearest neighbour monopole interaction corresponds to second neighbour spin interactions, while the isotropic nature of the monopole picture comes explicitly from summing terms of the original DSI to infinite distance \cite{denHertog2000}. 
Hence, even though truncating the Coulomb potential could speed up the numerics the long range interactions would still be playing a crucial role.}

To access such states we performed thermal quenches~\cite{castelnovo2010thermal} from an initial temperature $T_0=1$ K where the equilibrium monopole concentration is relatively high, to a target, much lower temperature $T$. In Fig.~\ref{quenchNRJ} we show the time evolution of $\langle n\rangle=\langle{\mathcal{N}}/{N}\rangle$ and $\langle E\rangle$ with the time spent since the quench. Starting from the equilibrium monopole concentration at $T_0=1$ K, $n\sim 0.2$ per tetrahedron, $\langle n(t)\rangle$ falls rapidly, for all temperature. For $T=0.3$ K and above, the monopole concentration falls towards its (very low) equilibrium value in a few hundreds of Monte Carlo steps. Below this crossover temperature, the quench is sufficiently violent to trap a finite concentration of non-contractable pairs, forming a concentration plateau whose lifetime increases rapidly with decreasing the quench temperature $T$. This behaviour is accurately paralleled in the energy function as, once the free particles have been annihilated, the energy depends, almost entirely on the small concentration of isolated and locally-confined charge pairs, see Fig.~\ref{quenchNRJ}.

The apparition of two time scales following a quench means that spin ice undergoes a simple form of aging, whereby correlation functions depend on both times $t$ and $t_w$ following the quench and not just the time difference, $t-t_w$, as in equilibrium. In Fig.~\ref{AgingS} we show the normalised local charge correlation function, $\tilde{C}_Q (t,t_w)$, following a quench to $T=0.12$ K for different fixed values of $t_w$. In the initial regime in which free monopoles are annihilated, the correlation function depends strongly on $t_w$ but it becomes independent, for $t_w>100$ Monte Carlo Steps,  as one enters the regime of metastable pair confinement. In this regime the correlation function depends on $t-t_w$ only, as in equilibrium, although we stress that the system remains far from equilibrium. Similar behaviour is found for both energy and spin correlations, despite the fact that most spins spend the majority of their time completely frozen.

\begin{figure}
\includegraphics[width=7cm]{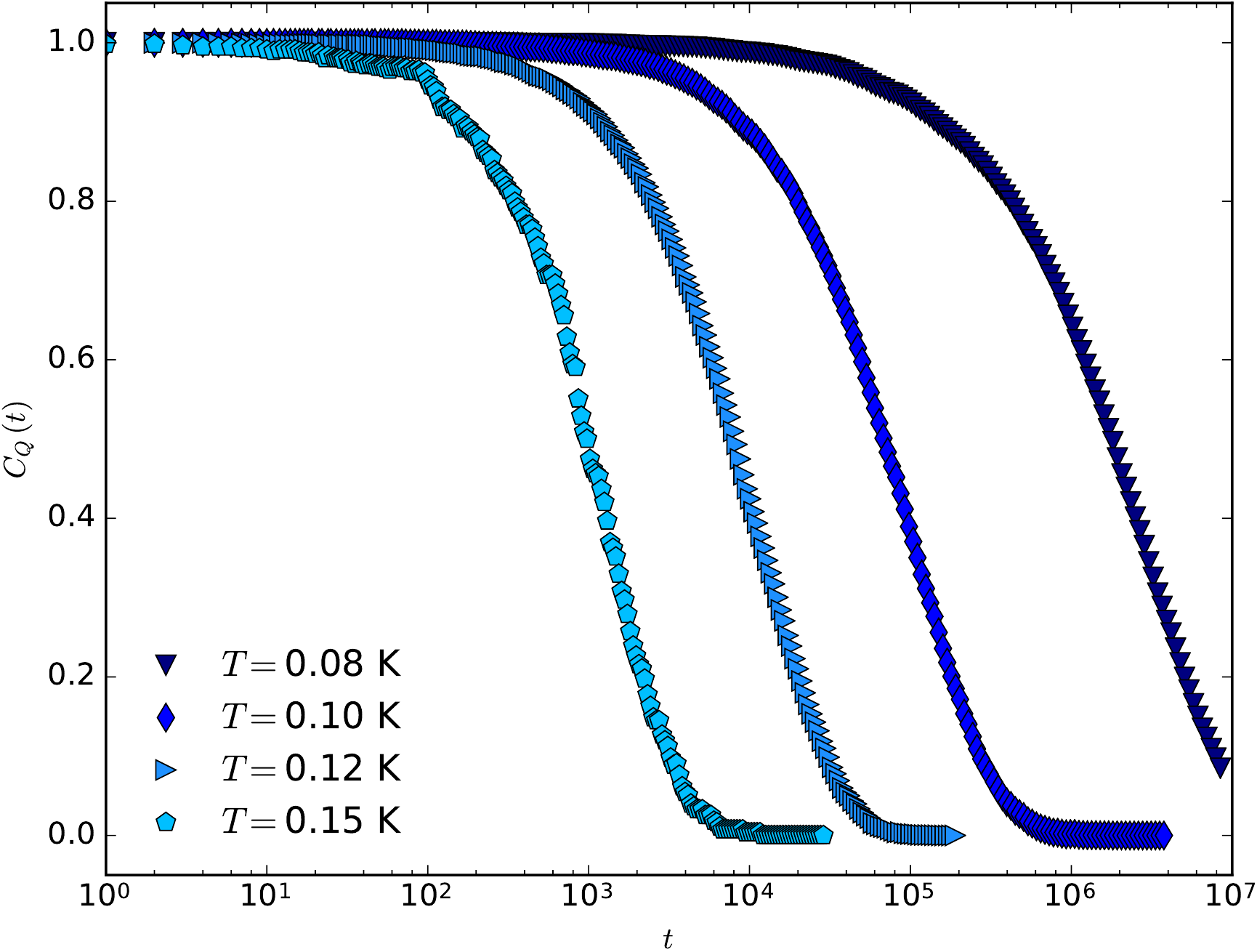}
\includegraphics[width=7cm]{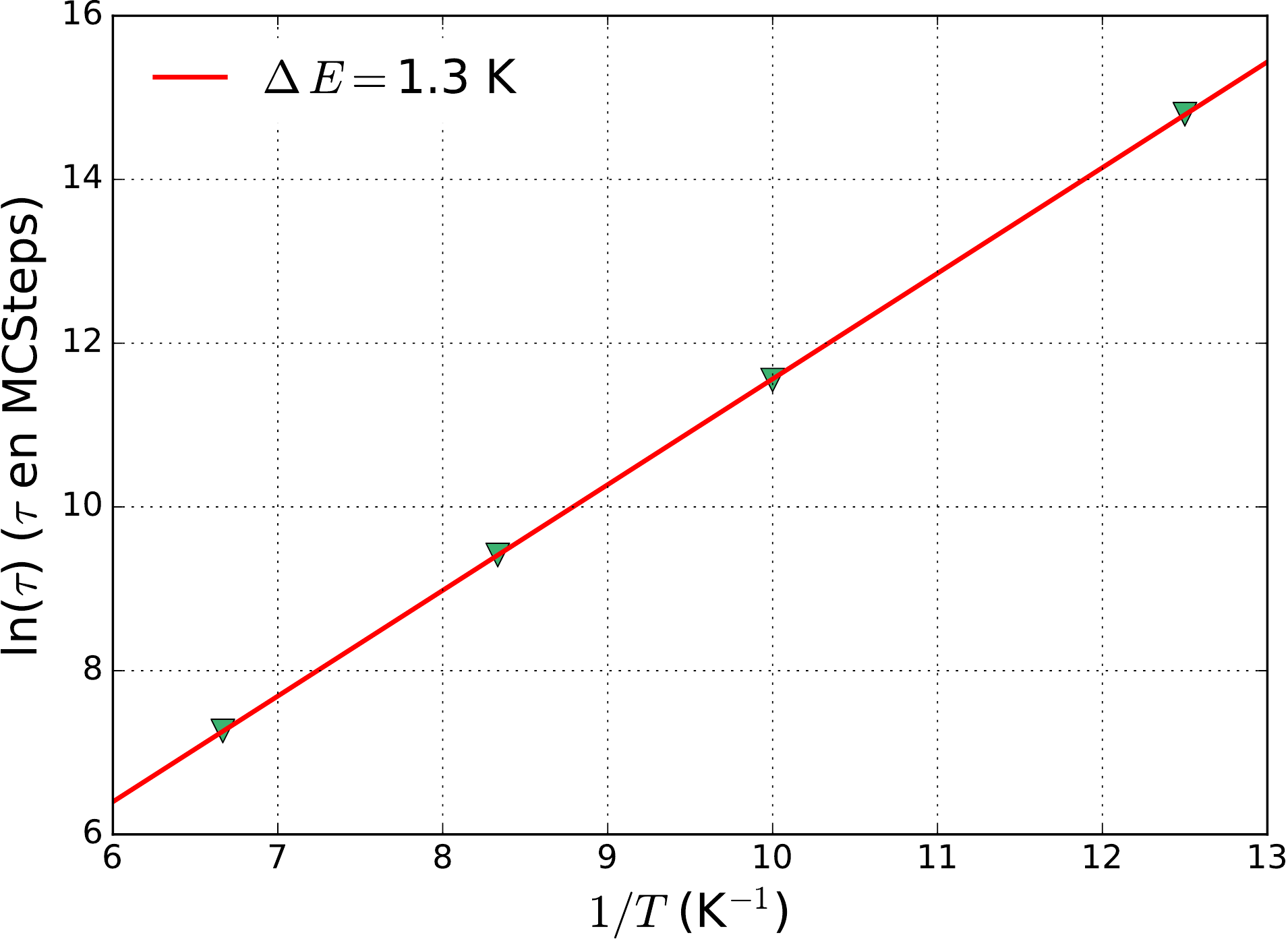}
\caption{Top: Normalised time autocorrelation functions of the local charge as a function of total time $t$ for fixed waiting time $t_{w}=200$ after a quench from $T = 1$ K.
  Bottom: Estimated relaxation times $\tau_Q$ from the data in the upper panel are well described by an Arrhenius law with energy scale $\Delta E \approx 1.3$ K.}
\label{CQ-quasi-eq}
\end{figure}

In Fig.~\ref{CQ-quasi-eq} we show $\tilde{C}_Q (t,t_w)$ for different quench temperatures, for fixed $t_w=2000$ MCS. As anticipated~\cite{castelnovo2010thermal}, the relaxations closely follow an exponential decay from which one can extract an associated time scale, $\tau_Q$. As shown in the lower panel, the temperature evolution of $\tau_Q(T)$ closely follows an Arrhenius law, with and energy barrier $\Delta E_Q \approx 1.3$~K, close to but slightly lower than the estimated Coulomb energy barrier ($\approx 1.4$~K) for activated monopole hopping around a hexagon ring. This small difference could be explained by taking into account the interactions between the bound monopole pairs. Taking this to be a random dipole interaction of order
\begin{equation}
\delta E \sim \frac{\mu_0 m^2}{\pi L^3} ,
\end{equation}
gives an energy scale  $\delta E\sim 0.1$ K. This scale, which is the correct order of magnitude could then be used as a parameter in a stochastic decay model~\cite{castelnovo2010thermal} that captures the small shift in time scales from the Coulomb estimate.

\subsection{FDT violations in the non-contractable pair regime}

\begin{figure}
\includegraphics[width=7cm]{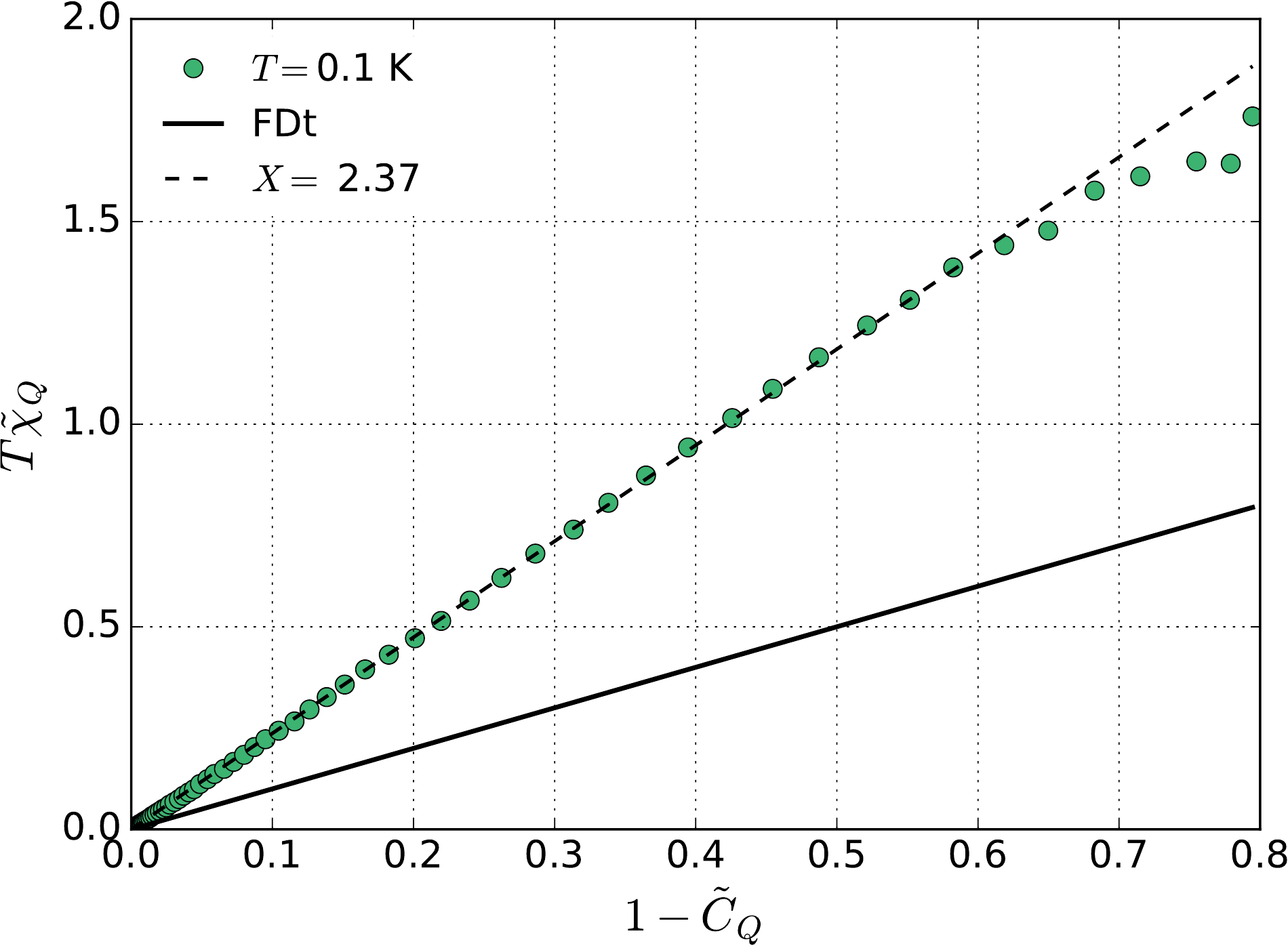}
\includegraphics[width=7cm]{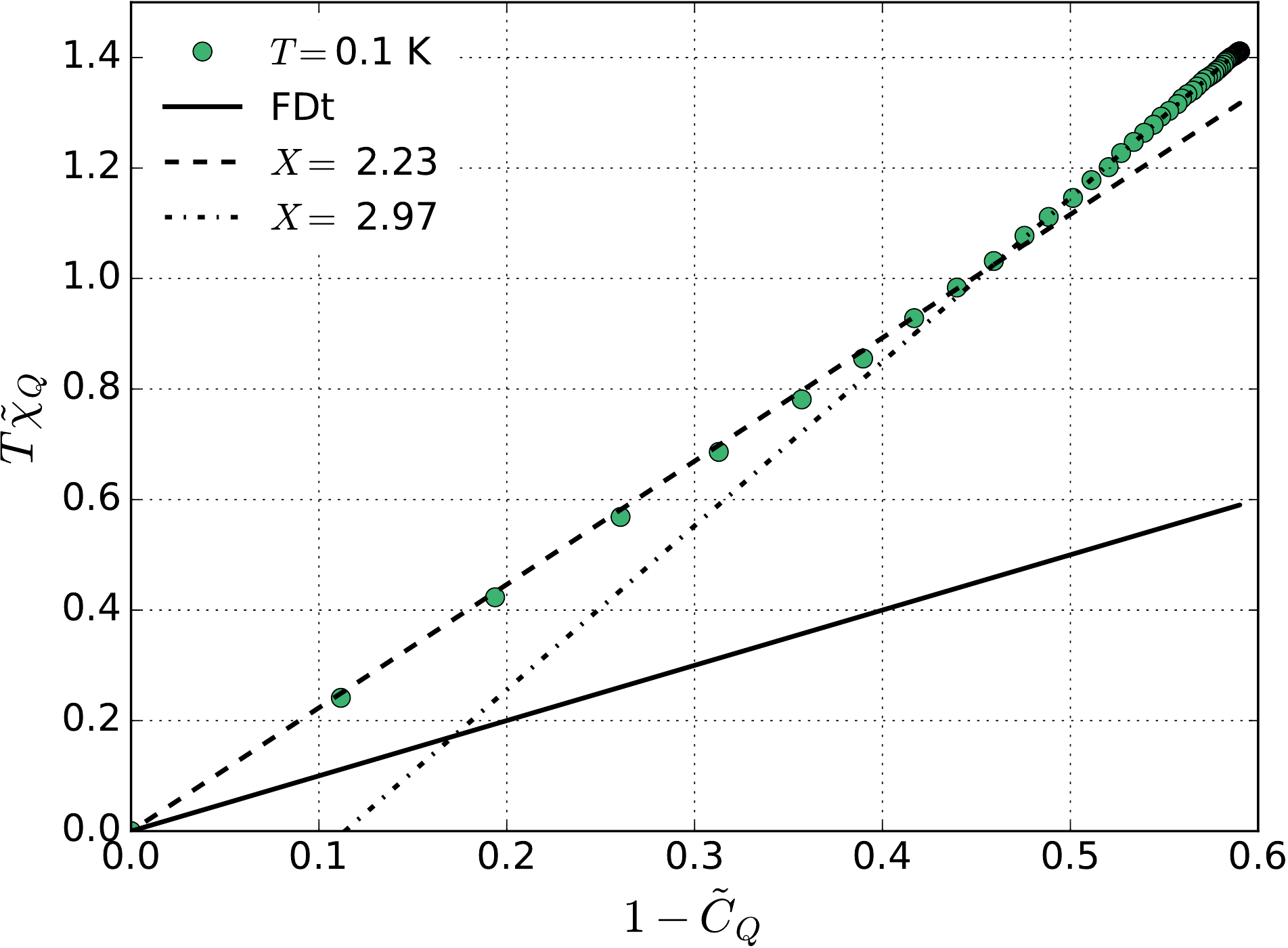}
\caption{Top: Parametric fluctuation-dissipation plot for the local charge density, $|\hat{n}_i|$ following a quench to $T=0.1$ K (see Sec.~\ref{sec:FDT}). The variable parameter is the waiting time $500< t_w<t$ at fixed $t=3.6 \times 10^5$ MCS . Results are for $L=6$ and data are averaged over $4 \times 10^5$ initial configurations. The solid black line corresponds to equilibrium FDT with dotted lines representing constant $X$ values as shown. Bottom: As for upper panel with fixed $t_w=500$ MCS and variable parameter $t_w<t< 8\times 10^5$ MCS.}
\label{FDT-Q}
\end{figure}

\begin{figure}
\includegraphics[width=7cm]{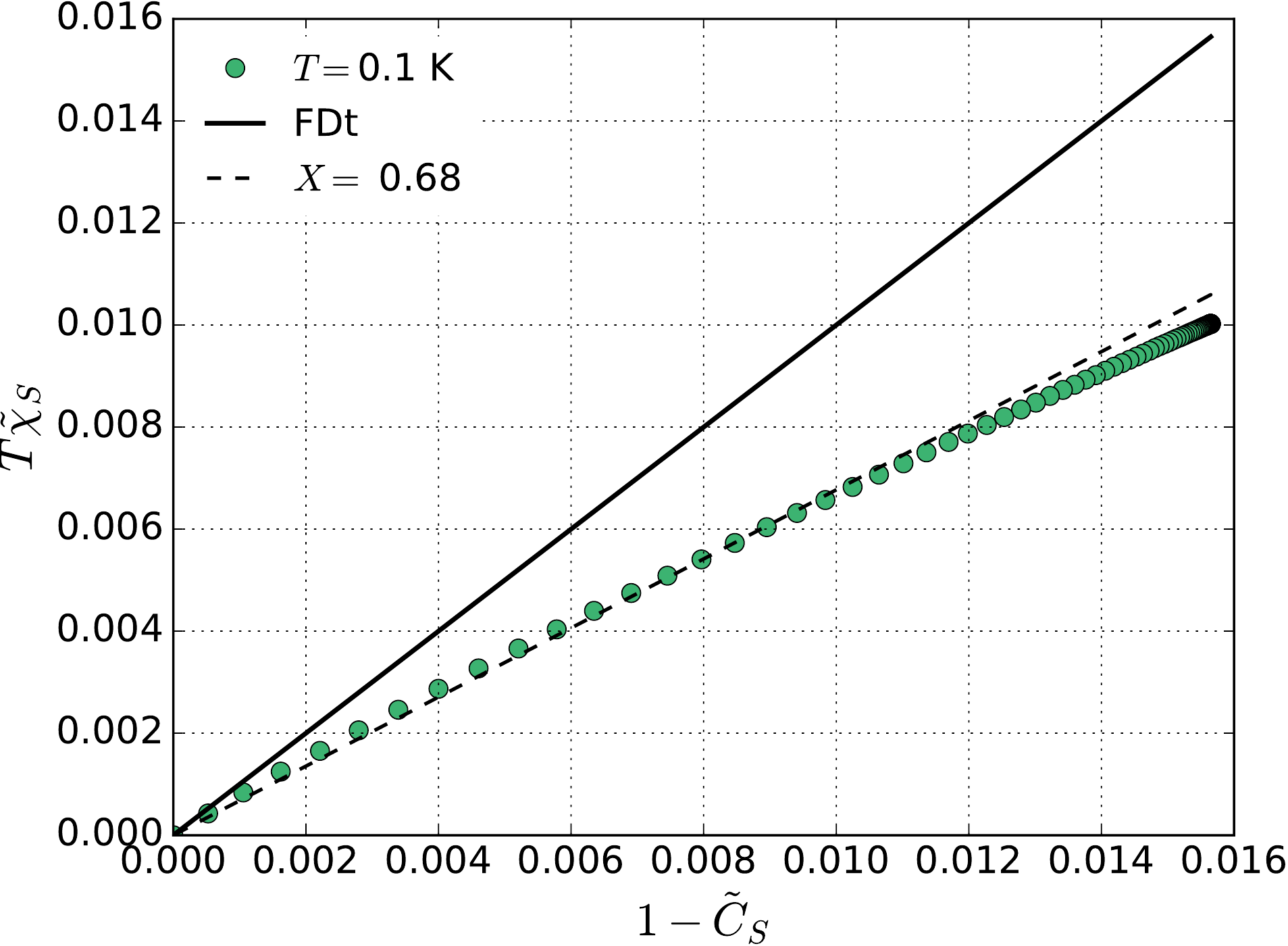}
\includegraphics[width=7cm]{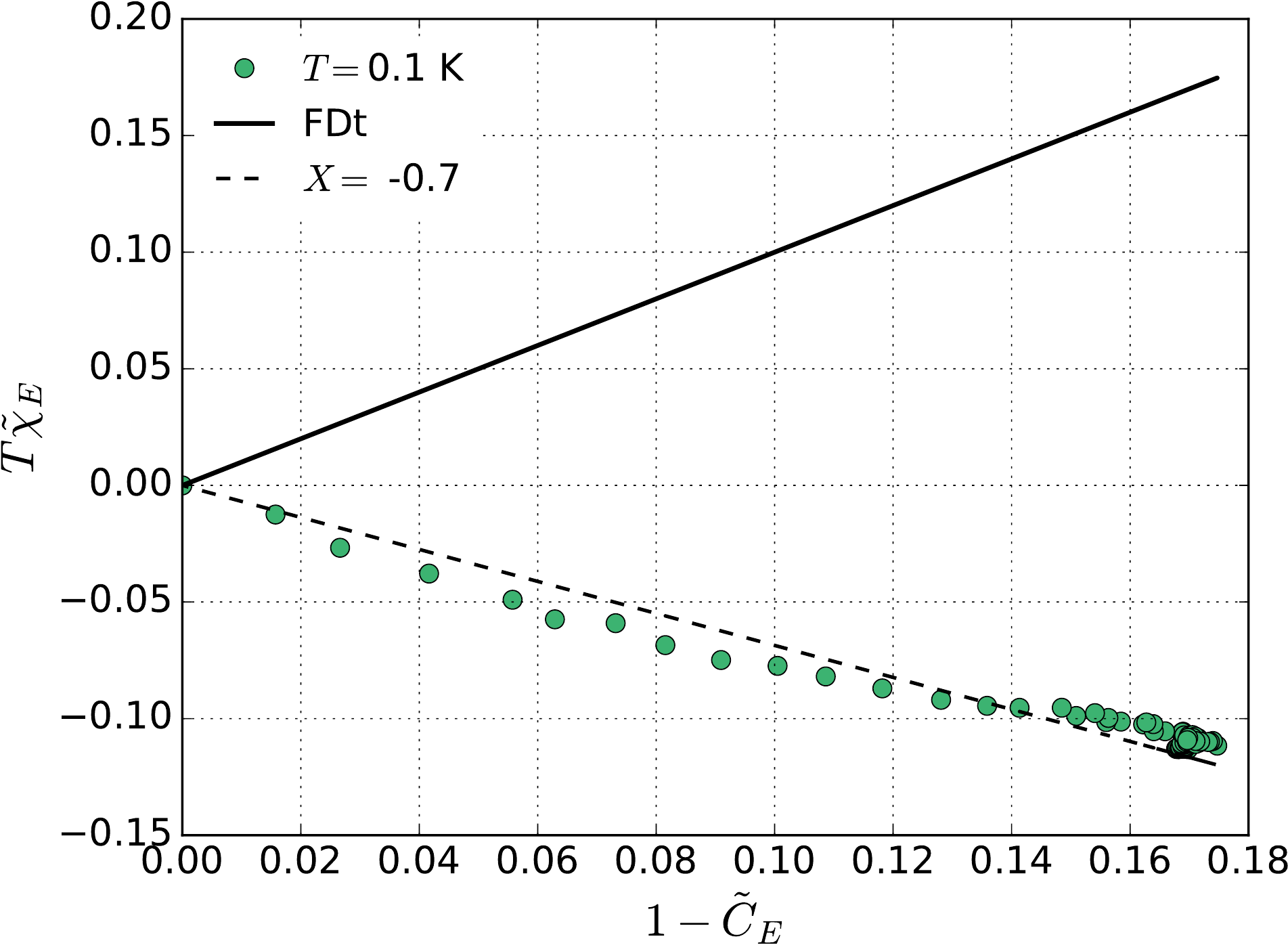}
\caption{Top: Parametric fluctuation-dissipation plot for the local spin variable, $S_{\kappa}$ following a quench to $T=0.1$ K (see Sec.~\ref{sec:FDT}). The variable parameter is the waiting time $500< t_w<t$ at fixed $t=3.6 \times 10^5$ MCS . Results are for $L=6$ and data are averaged over $4 \times 10^5$ initial configurations. The solid black line corresponds to equilibrium FDT with dotted line representing constant $X$ values as shown. Bottom: As for upper panel for the energy variable, $E$.}
\label{FDT-SE}
\end{figure}

The existence of a non-equilibrium regime with measurable fluctuations allows us to test the FDT in a non-equilibrium environment for the three designated variables, $|\hat{n_i}|$, $S_{\kappa}$ and $E$. In each case data was collected both for fixed $t$ varying $t_w\le t$ and for fixed $t_w$ varying $t\ge t_w$. 
The resulting parametric plots are shown in Figs.~\ref{FDT-Q} and \ref{FDT-SE}. To help with interpretation, we remind the reader that a short time difference $t-t_w$ corresponds to a small value of the horizontal axis, $1-\tilde{C}_A \sim 0$, and that the equilibrium FDT appears as a straight line of slope unity in this representation.

The FDT is clearly violated in all cases. Remarkably, in each case we observe that the fluctuation-dissipation ratio, $X(t,t_w)$, is approximately constant over the accessible range of correlations, allowing for the association of an effective non-equilibrium temperature, Eq.~(\ref{eq:Teff}). However, this temperature appears different for each variable, which shows that the non-equilibrium problem does not simply translate to an effective equilibrium problem, with phenomenology reproduced by a minimal variational principle. On the contrary, this is a more complex situation in which each quantity has its own fluctuation amplitude without the constraints set by the second law of thermodynamics. This type of finding is consistent with results for glassy systems characterised either by well-separated time scales~\cite{cugliandolo1997energy,cugliandolo1999thermal,kurchan2021time}, or systems with independent degrees of freedom~\cite{doi:10.1063/1.4901526,PhysRevLett.104.195701}. 

Figure~\ref{FDT-Q} shows data for charge fluctuations, calculated using both $t_w$ (upper) and $t$ (lower) as the variable parameter. The data are consistent for short times, giving approximately linear behaviour and a constant value for $X_Q$ around $X_Q \approx 2.3$. Hence the data in this region depends on the time difference $t-t_w$ only, as in equilibrium. For $1-\tilde{C}_Q$ greater than $0.5$ the two methods give different results, indicating a clear two time dependence. Mathematically the fluctuation-dissipation ratio is defined by using $t_w$ as the variable~\cite{sollich2002fluctuation}, and this representation should therefore be preferred. Indeed, the data indicate a much simpler structure of the FDT plot in this representation with a well-defined effective temperature characterising the charge degrees of freedom over the entire time regime. 

Fig.~\ref{FDT-SE} upper panel shows data for the local spin variables $S_{\kappa}$ with $t_w$ as the variable parameter. The small scale on both the horizontal and vertical axes reflects the inaccessible relaxation time scale for the spin degrees of freedom as most of the spins remain completely static over the simulation time scale. As a result only small corrections from $\tilde{C}_S=1$ are observable. The parametric plot of response and correlations can nevertheless be approximated reasonably well over this correlation range by a constant fluctuation dissipation ratio, $X \approx  0.7$, with some very weak departure from linearity. Data collected varying $t$ for fixed $t_w$ (not shown) are very similar. We have also tested magnetic fluctuations through the total moment $\vec M(t)$. The data is much harder to collect for the reasons discussed above, but our results are compatible with those for the local spins suggesting that local and global spin degrees of freedom are characterised by similar violations of the FDT with an equal fluctuation-dissipation ratio, as also seen in pure Ising models at criticality~\cite{mayer2003fluctuation}.

A fluctuation ratio, $X < 1$ ($X>1$) corresponds to an effective temperature that is higher (lower) than the equilibrium temperature. We find that the effective temperature for the charge variable $|\hat{n}_i|$ is lower than the equilibrium temperature, $T_{\rm eff}= T / X_Q  \approx 0.04$~K while that for the spin variable, $S_{\kappa}$ is higher with $T_{\rm eff} = T / X_S \approx 0.14$~K for quenches performed at the bath temperature $T=0.1$~K. 

Given that the monopoles are objects made up of spin textures it seems surprising that the non-equilibrium response to them and to the spin degrees of freedom are different. The difference finds its origin in the emergent Helmholtz decomposition of the moments into longitudinal and transverse fragments. The two fragments interact only weakly through the constraint fixed by the total spin density~\cite{lhotel2020fragmentation}, allowing an apparent separation of the magnetic charge made from the longitudinal component and the total spin. 

In equilibrium, the monopoles are deconfined objects, free to move independently, interacting only via a Coulomb force which falls to zero at large distance. However, in the non-contractable pair regime, the monopoles are locally confined and can only diffuse via double moves in which the first separates the particles to a second neighbour distance and the second restores the nearest neighbour pair. The double move requires the crossing of an energy barrier with a consequent drastic reduction in monopole mobility. It is possible that the reduced effective temperature reflects this reduced mobility. Effective temperatures that are lower than the bath temperature have been observed in a few other systems as well~\cite{PhysRevLett.110.035701,berthier2001nonequilibrium}. However, in parallel with the reduced monopole mobility there is an overall enhancement of spin fluctuations associated with the enhanced monopole concentration, leading to an increased effective temperature for the complete spin degrees of freedom. Such an increase in effective spin temperature is compatible with the case of glasses and spin glasses following a temperature quench~\cite{marinari1998violation,barrat1999fluctuation,sciortino2001extension,herisson2002fluctuation}.

The response to non-equilibrium energy fluctuations is at first sight even more surprising. The data shown in Fig.~\ref{FDT-SE} (lower panel) yield a negative slope, indicating a negative effective temperature. We observe $X_E \approx -0.7$, and so $T_{\rm eff}= T / X_E \approx - 0.14$~K, again using $t_w$ as the appropriate variable parameter. Varying $t$ at fixed $t_w$ yields qualitatively similar results. Negative temperatures have also been observed in kinetically constrained models of glasses showing slow dynamics~\cite{mayer2006activated,leonard2007non,jack2006fluctuation}. Here, the negative value for $X_E$ was associated with a dynamic evolution via thermally activated processes involving localised defects rather than by the evolution within a hierarchical free energy landscape provided by a mean-field analysis. This is indeed the case also here: non-contractable pairs are essentially isolated, decaying via the thermally activated process discussed above. Intuitively, increasing the temperature (which is the field conjugate to the energy) therefore increases the rate of relaxation of the number of pairs. Consequently, after the initial decay following  the quench, a linear increase of the applied temperature at $t_w$ lowers the observed energy after a fixed time $t-t_w$ as energy relaxes faster at higher temperature, which then naturally gives a negative response to a positive impulse field. As a result, the fluctuation-dissipation ratio between response and correlation becomes negative. It is a non-trivial finding that the negative fluctuation-dissipation ratio takes, to a good approximation, a unique value, thus mimicking an equilibrium system with a negative absolute effective temperature.  

\section{Discussion}

\label{sec:conclusion}

In this paper we have shown numerically that the dumbbell model of spin ice can be forced into a non-equilibrium regime in which the fluctuation-dissipation theorem is violated. This is the non-contractable pair regime~\cite{castelnovo2010thermal} in which neutral pairs of magnetic monopoles are locally confined by the constraints of spin ice. The pairs relax via the passage over an energy barrier of order one Kelvin, allowing for slow but measurable dynamics down to $100$ mK. We have measured FDT violations associated with three quantities: local spin variables, local monopole occupation number, and total energy. Despite the expected FDT violation, in each case the relation between fluctuations and response was found to be approximately linear, allowing for the definition of effective non-equilibrium temperature. This very simple structure of the FDT violation is not trivial at all and could not have been anticipated on simple grounds. 
It therefore seems remarkable that a unique quantity, the fluctuation-dissipation ratio, is able to characterise the time-dependent relation between response and correlation functions in the far from equilibrium aging regime of spin ice. The effective temperatures  were found to be different in the three cases. The spin variables appear to be at a higher temperature than that of the heat reservoir, the monopole concentration responds at an apparent lower temperature, while the energy has an effective temperature which is negative. 

The difference in response temperature for the spin and monopole variables appears as a direct consequence of the fractionalisation of dipolar elements into magnetic monopoles~\cite{castelnovo2008magnetic}. As the monopoles only eat up a small fraction of the total spin density~\cite{brooks2014magnetic}, the monopoles and total spin can behave quasi-independently, which is the case here. The spin activity is enhanced with respect to equilibrium at such low temperatures of approximately 100 mK which reflects the higher effective temperature but the monopoles are locally confined into pairs and have considerably reduced mobility. It is possible that this is reflected in the reduced effective temperature. In aging and sheared supercooled liquids characterised by a single glassy timescale, a unique value of the effective temperature is observed~\cite{barrat1999fluctuation,berthier2002nonequilibrium}, but in that case all observables are dynamically strongly coupled. Clearly, spin ice has a richer physical behaviour with two independent variables. 

In kinetic models of glass-forming materials, the negative energy response to an increase in the temperature is a direct result of the thermally activated relaxation over energy barriers. In these models~\cite{mayer2006activated,leonard2007non,jack2006fluctuation,garriga2009negative}, microscopic or mesoscopic objects indeed relax independently via local energy barriers. As a consequence, on heating, the energy is relaxed more easily towards its lower, equilibrium value via the faster motion of the defects, thus producing negative response functions and negative fluctuation-dissipation ratio. Clearly, the bound pair regime of spin ice is another example of this physics as the bound pairs are essentially independent, two-body metastable states. The similarity of the results for the energy degree of freedom shows that the analogy extends to the out of equilibrium regime. It would be interesting to see whether the measured value of the fluctuation-dissipation ratio, $X_E \approx -0.7$ can be theoretically rationalised in the framework of the diffusion-annihilation process proposed in Ref.~\cite{castelnovo2010thermal}, following the type of field-theoretical analysis done for kinetic glass models~\cite{mayer2006activated,leonard2007non,jack2006fluctuation}. 

\rev{It would also be interesting to study FDT relations in artificial spin ice~\cite{nisoli-artificial2013,skjaervo-artifical2020,levis2012out,levis2013defects} and colloidal ice systems \cite{Libal2020}. These mesoscopic systems have the advantage that monopoles are directly observable in real space and that the accuracy of the monopole picture can be tuned to some extent \cite{Perrin2016,Ostman2018}. In these experimental systems the initial paradigm is slightly different in that attaining equilibrium, real or effective \cite{Macauley-thermal2020,Sendetskyi2019} for any situation is a formidable challenge so that an experiment in which the FDT were satisfied would already be a very interesting result. From here, performing quenches into the non-contractable pair regime with observation of FDT violations would seem a feasible project. }

The above analogies with glass-forming materials and kinetic glass models therefore confirm that despite FDT violations, the non-contractable pair regime does not have the kind of hierarchical energy landscape structure that characterises spin glasses~\cite{castelnovo2010thermal}. The non-contractable pairs do themselves carry a magnetic dipole moment so that there is a collective dipolar interaction between them. It is therefore conceivable that they could show dipolar glass behaviour on a much lower temperature scale, but the separation of energy scales here is such that a regime of this kind is beyond the scope of our work. 

The bound pairs are formed following a quench from $1$~K where the monopole concentration is still high, to target temperatures below $0.3$ K. Above this quench temperature the pairs fail to form and the concentration falls rapidly to the equilibrium value as if the system had been annealed. For a finite size system in this temperature range most microstates of the equilibrium measure have zero monopoles so that spin fluctuations essentially stop here. On annealing we were able to equilibrate down to around $0.4$ K and the FDT was satisfied down to this limit, consistently with recent specific heat measurements~\cite{giblin2018specificheat} and with preliminary magnetic noise measurement~\cite{Cathelin2020}. Hence, for the chosen variables, following an annealing protocol, we find that the system progresses from an equilibrium to a frozen state without an intermediate regime showing non-equilibrium dynamics. 

There is of course the potential for the development of hierarchical dynamics through the existence of closed loops of spin flips moving the system from one constrained spin ice configuration to another through the band of Pauling states. However, in the dumbbell model, like in the nearest-neighbour spin ice model, such states are degenerate so that any hierarchy must be entropy driven. The time and length scales necessary to address such subtle questions are at present out of reach. Moving back to the dipolar spin ice model lifts this degeneracy, opening up the possibility of a loop glass and hierarchical behaviour on the energy scale of this band width; that is, around $150$ mK~\cite{melko2001long}. Re-opening the bandwidth takes us beyond the monopole picture~\cite{denHertog2000,isakov2005projective,kassner2015supercooled} and into a realm in which defects and realistic boundaries will also be essential ingredients~\cite{Revell2012}. That is, glassy behaviour could be possible on this small energy scale but our results suggest that it will be extrinsic, depending on corrections to the monopole picture and to the translational symmetry of the pyrochlore lattice. 

In preliminary experiments~\cite{Cathelin2020} measuring magnetic noise and response in the frequency domain, the FDT is confirmed over the full frequency spectrum down to $0.6$ K. There is partial confirmation down to $0.4$ K but at this lowest temperature, time scales are so long that measurements are limited to the relatively high frequency, short time scale regime. These preliminary measurements therefore confirm the results from the monopole picture that exclude FDT violations in this temperature range. Corrections to this picture, although of great interest will be subtle and present considerable experimental challenges. 

Far from equilibrium states in spin ice materials, with high monopole density are attainable through field assisted monopole avalanche quench protocols~\cite{Paulsen2014,paulsen2016wien,paulsen2019nuclear} with quenches going well below the required 300 mK.
Magnetic noise measurement below this temperature range is a challenging experimental problem but our results suggest that the effort could be worth it as it will offer a robust regime of non-equilibrium response with measurable FDT violations. The origin of these violations can be found in the simplest monopole description of spin ice and they should resist the many corrections to this picture offered here and in the wider literature. 

\section*{Acknowledgments}
It is a pleasure to thank Elsa Lhotel and Claudio Castelnovo for fruitful discussions and for collaborations on related topics. This work was supported by ANR, France, Grant  No.ANR-19-CE30-0040 and by a Grant from the Simons Foundation (Grant No. 454933, L.B.).

\appendix
 
\section{Computation of response functions without introducing a field}

\label{Response-F}

The usual way to determine the response function of a given observable $A$ is to run two different kinds of simulations: the first one with the Hamiltonian without perturbation, $\mathcal{H}_0$, and the second one with the perturbed hamiltonian, $\mathcal{H}_0 - f\, A$ which defines the field $f$ conjugate to the observable $A$. We wish to measure the time-integrated response function, and so the time evolution of the field $f$ is
\begin{equation} 
f = 
\left \lbrace
  \begin{array}{rr}
    0 \quad \text{if} \quad t < t_{\mathrm{w}} \\
    \varepsilon \quad \text{if} \quad t \geqslant t_{\mathrm{w}}\\
  \end{array}
\right. 
\end{equation}

In both kinds of simulations, the time evolution of $A(t)$ is computed, and then an ensemble average is done over a series of independent initial configurations. The response function is eventually obtained as the finite derivative
\begin{equation} 
\chi(t,t_{\mathrm{w}}) = \frac{\partial{\langle A (t) \rangle_{\varepsilon}}}{\partial f} \underset{\varepsilon \rightarrow 0}{\approx} \frac{1}{\varepsilon}\, \Big\lbrace \langle A (t) \rangle_{\varepsilon} - \langle A (t) \rangle_0 \Big\rbrace_{t \geqslant t_{\mathrm{w}}}
\end{equation}

Numerically, $\varepsilon$ has to be small so that the finite derivative is a good approximation to the true derivative (linear response regime), but it can not be too small, otherwise the signal would be smaller than the statistical noise.

Another point to mention is the fact that each $t_{\mathrm{w}}$ requires two sets of simulations (with and without the conjugated field), which ultimately results in a heavy computational load. This led us to implement instead a no-field method to obtained the directly response functions from unperturbed simulations~\cite{ricci2003measuring, chatelain2003far, berthier2007efficient}.

If $P\big(\lbrace \sigma \rbrace \rightarrow \lbrace \sigma' \rbrace\big)$ is the probability to accept a Monte Carlo move from the configuration $\lbrace \sigma \rbrace$ to the configuration $\lbrace \sigma' \rbrace$, then 
\begin{equation}
\langle A (t) \rangle = \underset{\rm ic}{\sum} A(t)\,  P\big(\lbrace \sigma(t_{\mathrm{w}}) \rbrace \rightarrow \lbrace \sigma(t) \rbrace\big)
\end{equation}
where the sum runs over independent initial configurations. The integrated linear response is thus given by
\begin{equation}
\chi(t,t_{\mathrm{w}}) = \underset{\rm ic}{\sum} A (t)\, \frac{\partial}{\partial f} P\big(\lbrace \sigma(t_{\mathrm{w}}) \rbrace \rightarrow \lbrace \sigma(t) \rbrace\big),
\label{eq:appchi}
\end{equation}
where the derivative with respect to the field can be evaluated exactly.  

Using a markov chain of single spin flip with Metropolis rates, one can express the total probablity as a product over each attempted Monte Carlo move, 
\begin{equation}
 P\big(\lbrace \sigma(t_{\mathrm{w}}) \rbrace \rightarrow \lbrace \sigma(t) \rbrace\big) = \prod_{t'=t_{\mathrm{w}}}^{t} \mathcal{W}(t'\rightarrow t'+1)
\end{equation}
with the probabilities $\mathcal{W}(t\rightarrow t+1)$ directly following from the Metropolis acceptance rate, namely:
\begin{equation*}
\left \lbrace
  \begin{array}{lll}
    0 & \text{if} & \sigma(t+1) \text{ can not been reached from } \sigma(t)\\
    1 &\text{if } & \Delta \mathcal{H} < 0\\
    \mathrm{e}^{-\beta\, \Delta \mathcal{H}} & \text{if} & \Delta \mathcal{H} \geqslant 0 \text{ and the spin flip is accepted}  \\
    1 - \mathrm{e}^{-\beta\, \Delta \mathcal{H}} & \text{if} & \Delta \mathcal{H} \geqslant 0 \text{ and the spin flip is rejected}  \\
  \end{array}
\right.
\end{equation*}
One can then express the derivative in Eq.~(\ref{eq:appchi}) to get  
\begin{align}
& \frac{\partial}{\partial f} P\big(\lbrace \sigma(t_{\mathrm{w}}) \rbrace \rightarrow \lbrace \sigma(t) \rbrace\big) = \nonumber \\ 
& P\big(\lbrace \sigma(t_{\mathrm{w}}) \rbrace \rightarrow \lbrace \sigma(t) \rbrace\big)\sum_{t'=t_{\mathrm{w}}}^{t}  \, \frac{\partial}{\partial f}\,  \mathrm{ln}\, \mathcal{W}(t'\rightarrow t'+1)
\end{align}
so that finally
\begin{align}
\chi(t,t_{\mathrm{w}}) &=  \underset{\rm ic}{\sum}\ A (t)\, \bigg(\sum_{t'=t_{\mathrm{w}}}^{t}  \, \frac{\partial}{\partial f}\,  \mathrm{ln}\, \mathcal{W}(t'\rightarrow t'+1) \bigg)\  \times \nonumber \\
&  \times P\big(\lbrace \sigma(t_{\mathrm{w}}) \rbrace \rightarrow \lbrace \sigma(t) \rbrace\big) \nonumber \\
                       & \equiv   \langle A(t)\, \mathcal{R}(t_{\mathrm{w}},t) \rangle
                         \label{toto}
\end{align}
which defines the quantity 
\begin{equation}
\mathcal{R}(t_{\mathrm{w}},t) = \sum_{t'=t_{\mathrm{w}}}^{t}  \, \frac{\partial}{\partial f}\,  \mathrm{ln}\, \mathcal{W}(t'\rightarrow t'+1).
\end{equation}

If the observable is $A$, and its conjugate field is $f$ then $-\beta \mathcal{H} \rightarrow -\beta \mathcal{H} + \beta f A$, then one can express the derivatives explicitely
\[\hspace{-3.5cm} \frac{\partial}{\partial f}\,  \mathrm{ln}\, \mathcal{W}(t'\rightarrow t'+1) = \]
\[ 
\left \lbrace
  \begin{array}{ll}
    + \beta\, \Delta A & \text{if }  \Delta \mathcal{H}\geqslant 0 \text{ and the spin flip is accepted}  \\[2mm]
    - \frac{\beta\, \Delta A}{\mathrm{e}^{\beta \Delta \mathcal{H}} - 1} & \text{if } \Delta \mathcal{H} \geqslant 0 \text{ and the spin flip is rejected}  \\[2mm]
  0 &\text{otherwise} . \\
  \end{array}
\right. \]
The key point of the method is the evaluation of integrated response functions  in Eq.~(\ref{toto}) which involves the product of two physical observables that can be evaluated explicitely during the course of the simulation followed by an ensemble average with no perturbing field. Thereore, linear response is valid by construction and a single set of simulations is needed as the perturbing field is nver applied. In addition, several physical observables can be analysed in the course of a single simulations. The only drawback is that the statistical noise affecting Eq.~(\ref{toto}) can become quite large, especially if long time scales are needed and this may imply using a large number of initical configurations.

\bibliography{article}

\begin{thebibliography}{104}%
\makeatletter
\providecommand \@ifxundefined [1]{%
 \@ifx{#1\undefined}
}%
\providecommand \@ifnum [1]{%
 \ifnum #1\expandafter \@firstoftwo
 \else \expandafter \@secondoftwo
 \fi
}%
\providecommand \@ifx [1]{%
 \ifx #1\expandafter \@firstoftwo
 \else \expandafter \@secondoftwo
 \fi
}%
\providecommand \natexlab [1]{#1}%
\providecommand \enquote  [1]{``#1''}%
\providecommand \bibnamefont  [1]{#1}%
\providecommand \bibfnamefont [1]{#1}%
\providecommand \citenamefont [1]{#1}%
\providecommand \href@noop [0]{\@secondoftwo}%
\providecommand \href [0]{\begingroup \@sanitize@url \@href}%
\providecommand \@href[1]{\@@startlink{#1}\@@href}%
\providecommand \@@href[1]{\endgroup#1\@@endlink}%
\providecommand \@sanitize@url [0]{\catcode `\\12\catcode `\$12\catcode
  `\&12\catcode `\#12\catcode `\^12\catcode `\_12\catcode `\%12\relax}%
\providecommand \@@startlink[1]{}%
\providecommand \@@endlink[0]{}%
\providecommand \url  [0]{\begingroup\@sanitize@url \@url }%
\providecommand \@url [1]{\endgroup\@href {#1}{\urlprefix }}%
\providecommand \urlprefix  [0]{URL }%
\providecommand \Eprint [0]{\href }%
\providecommand \doibase [0]{http://dx.doi.org/}%
\providecommand \selectlanguage [0]{\@gobble}%
\providecommand \bibinfo  [0]{\@secondoftwo}%
\providecommand \bibfield  [0]{\@secondoftwo}%
\providecommand \translation [1]{[#1]}%
\providecommand \BibitemOpen [0]{}%
\providecommand \bibitemStop [0]{}%
\providecommand \bibitemNoStop [0]{.\EOS\space}%
\providecommand \EOS [0]{\spacefactor3000\relax}%
\providecommand \BibitemShut  [1]{\csname bibitem#1\endcsname}%
\let\auto@bib@innerbib\@empty
\bibitem [{\citenamefont {Jaubert}\ and\ \citenamefont
  {Udagawa}(2021)}]{spinicebook2021}%
  \BibitemOpen
  \bibfield  {author} {\bibinfo {author} {\bibfnamefont {L.~D.~C.}\
  \bibnamefont {Jaubert}}\ and\ \bibinfo {author} {\bibfnamefont
  {M.}~\bibnamefont {Udagawa}},\ }\href@noop {} {\emph {\bibinfo {title} {Spin
  Ice}}}\ (\bibinfo  {publisher} {Springer},\ \bibinfo {year}
  {2021})\BibitemShut {NoStop}%
\bibitem [{\citenamefont {M{\'e}zard}\ \emph {et~al.}(1987)\citenamefont
  {M{\'e}zard}, \citenamefont {Parisi},\ and\ \citenamefont
  {Virasoro}}]{mezard1987spin}%
  \BibitemOpen
  \bibfield  {author} {\bibinfo {author} {\bibfnamefont {Marc}\ \bibnamefont
  {M{\'e}zard}}, \bibinfo {author} {\bibfnamefont {Giorgio}\ \bibnamefont
  {Parisi}}, \ and\ \bibinfo {author} {\bibfnamefont {Miguel~Angel}\
  \bibnamefont {Virasoro}},\ }\href@noop {} {\emph {\bibinfo {title} {Spin
  glass theory and beyond: An Introduction to the Replica Method and Its
  Applications}}},\ Vol.~\bibinfo {volume} {9}\ (\bibinfo  {publisher} {World
  Scientific Publishing Company},\ \bibinfo {year} {1987})\BibitemShut
  {NoStop}%
\bibitem [{\citenamefont {Berthier}\ and\ \citenamefont
  {Biroli}(2011)}]{berthier2011theoretical}%
  \BibitemOpen
  \bibfield  {author} {\bibinfo {author} {\bibfnamefont {Ludovic}\ \bibnamefont
  {Berthier}}\ and\ \bibinfo {author} {\bibfnamefont {Giulio}\ \bibnamefont
  {Biroli}},\ }\bibfield  {title} {\enquote {\bibinfo {title} {Theoretical
  perspective on the glass transition and amorphous materials},}\ }\href
  {\doibase 10.1103/RevModPhys.83.587} {\bibfield  {journal} {\bibinfo
  {journal} {Rev. Mod. Phys.}\ }\textbf {\bibinfo {volume} {83}},\ \bibinfo
  {pages} {587--645} (\bibinfo {year} {2011})}\BibitemShut {NoStop}%
\bibitem [{\citenamefont {Matsuhira}\ \emph {et~al.}(2001)\citenamefont
  {Matsuhira}, \citenamefont {Hinatsu},\ and\ \citenamefont
  {Sakakibara}}]{Matsuhira2001}%
  \BibitemOpen
  \bibfield  {author} {\bibinfo {author} {\bibfnamefont {K}~\bibnamefont
  {Matsuhira}}, \bibinfo {author} {\bibfnamefont {Y}~\bibnamefont {Hinatsu}}, \
  and\ \bibinfo {author} {\bibfnamefont {T}~\bibnamefont {Sakakibara}},\
  }\bibfield  {title} {\enquote {\bibinfo {title} {Novel dynamical magnetic
  properties in the spin ice compound dy2ti2o7},}\ }\href {\doibase
  10.1088/0953-8984/13/31/101} {\bibfield  {journal} {\bibinfo  {journal}
  {Journal of Physics: Condensed Matter}\ }\textbf {\bibinfo {volume} {13}},\
  \bibinfo {pages} {L737--L746} (\bibinfo {year} {2001})}\BibitemShut {NoStop}%
\bibitem [{\citenamefont {Snyder}\ \emph {et~al.}(2004)\citenamefont {Snyder},
  \citenamefont {Ueland}, \citenamefont {Slusky}, \citenamefont {Karunadasa},
  \citenamefont {Cava},\ and\ \citenamefont {Schiffer}}]{Snyder2004}%
  \BibitemOpen
  \bibfield  {author} {\bibinfo {author} {\bibfnamefont {J.}~\bibnamefont
  {Snyder}}, \bibinfo {author} {\bibfnamefont {B.G.}\ \bibnamefont {Ueland}},
  \bibinfo {author} {\bibfnamefont {J.S.}\ \bibnamefont {Slusky}}, \bibinfo
  {author} {\bibfnamefont {H.}~\bibnamefont {Karunadasa}}, \bibinfo {author}
  {\bibfnamefont {R.J.}\ \bibnamefont {Cava}}, \ and\ \bibinfo {author}
  {\bibfnamefont {P.}~\bibnamefont {Schiffer}},\ }\bibfield  {title} {\enquote
  {\bibinfo {title} {Low-temperature spin freezing in the {Dy2Ti2O7} spin
  ice},}\ }\href {\doibase 10.1103/PhysRevB.69.064414} {\bibfield  {journal}
  {\bibinfo  {journal} {Physical Review B}\ }\textbf {\bibinfo {volume} {69}},\
  \bibinfo {pages} {064414} (\bibinfo {year} {2004})}\BibitemShut {NoStop}%
\bibitem [{\citenamefont {Jaubert}\ and\ \citenamefont
  {Holdsworth}(2009)}]{Jaubert2009}%
  \BibitemOpen
  \bibfield  {author} {\bibinfo {author} {\bibfnamefont {L.~D.~C.}\
  \bibnamefont {Jaubert}}\ and\ \bibinfo {author} {\bibfnamefont {P.~C.~W.}\
  \bibnamefont {Holdsworth}},\ }\bibfield  {title} {\enquote {\bibinfo {title}
  {Signature of magnetic monopole and dirac string dynamics in spin ice},}\
  }\href {\doibase 10.1038/nphys1227} {\bibfield  {journal} {\bibinfo
  {journal} {Nature Physics}\ }\textbf {\bibinfo {volume} {5}},\ \bibinfo
  {pages} {258--261} (\bibinfo {year} {2009})}\BibitemShut {NoStop}%
\bibitem [{\citenamefont {Quilliam}\ \emph {et~al.}(2011)\citenamefont
  {Quilliam}, \citenamefont {Yaraskavitch}, \citenamefont {Dabkowska},
  \citenamefont {Gaulin},\ and\ \citenamefont {Kycia}}]{Quilliam2011}%
  \BibitemOpen
  \bibfield  {author} {\bibinfo {author} {\bibfnamefont {J.~A.}\ \bibnamefont
  {Quilliam}}, \bibinfo {author} {\bibfnamefont {L.~R.}\ \bibnamefont
  {Yaraskavitch}}, \bibinfo {author} {\bibfnamefont {H.~A.}\ \bibnamefont
  {Dabkowska}}, \bibinfo {author} {\bibfnamefont {B.~D.}\ \bibnamefont
  {Gaulin}}, \ and\ \bibinfo {author} {\bibfnamefont {J.~B.}\ \bibnamefont
  {Kycia}},\ }\bibfield  {title} {\enquote {\bibinfo {title} {Dynamics of the
  magnetic susceptibility deep in the coulomb phase of the dipolar spin ice
  material ho${}_{2}$ti${}_{2}$o${}_{7}$},}\ }\href {\doibase
  10.1103/PhysRevB.83.094424} {\bibfield  {journal} {\bibinfo  {journal} {Phys.
  Rev. B}\ }\textbf {\bibinfo {volume} {83}},\ \bibinfo {pages} {094424}
  (\bibinfo {year} {2011})}\BibitemShut {NoStop}%
\bibitem [{\citenamefont {Matsuhira}\ \emph {et~al.}(2011)\citenamefont
  {Matsuhira}, \citenamefont {Paulsen}, \citenamefont {Lhotel}, \citenamefont
  {Sekine}, \citenamefont {Hiroi},\ and\ \citenamefont
  {Takagi}}]{Matsuhira2011}%
  \BibitemOpen
  \bibfield  {author} {\bibinfo {author} {\bibfnamefont {Kazuyuki}\
  \bibnamefont {Matsuhira}}, \bibinfo {author} {\bibfnamefont {Carley}\
  \bibnamefont {Paulsen}}, \bibinfo {author} {\bibfnamefont {Elsa}\
  \bibnamefont {Lhotel}}, \bibinfo {author} {\bibfnamefont {Chihiro}\
  \bibnamefont {Sekine}}, \bibinfo {author} {\bibfnamefont {Zenji}\
  \bibnamefont {Hiroi}}, \ and\ \bibinfo {author} {\bibfnamefont {Seishi}\
  \bibnamefont {Takagi}},\ }\bibfield  {title} {\enquote {\bibinfo {title}
  {{Spin Dynamics at Very Low Temperature in Spin Ice Dy$_2$Ti$_2$O$_7$}},}\
  }\href {\doibase DOI 10.1143/JPSJ.80.123711} {\bibfield  {journal} {\bibinfo
  {journal} {Journal of the Physical Society of Japan}\ }\textbf {\bibinfo
  {volume} {80}},\ \bibinfo {pages} {123711} (\bibinfo {year}
  {2011})}\BibitemShut {NoStop}%
\bibitem [{\citenamefont {Yaraskavitch}\ \emph {et~al.}(2012)\citenamefont
  {Yaraskavitch}, \citenamefont {Revell}, \citenamefont {Meng}, \citenamefont
  {Ross}, \citenamefont {Noad}, \citenamefont {Dabkowska}, \citenamefont
  {Gaulin},\ and\ \citenamefont {Kycia}}]{Yaraskavitch2012}%
  \BibitemOpen
  \bibfield  {author} {\bibinfo {author} {\bibfnamefont {L~R}\ \bibnamefont
  {Yaraskavitch}}, \bibinfo {author} {\bibfnamefont {H~M}\ \bibnamefont
  {Revell}}, \bibinfo {author} {\bibfnamefont {S}~\bibnamefont {Meng}},
  \bibinfo {author} {\bibfnamefont {K~A}\ \bibnamefont {Ross}}, \bibinfo
  {author} {\bibfnamefont {H~M~L}\ \bibnamefont {Noad}}, \bibinfo {author}
  {\bibfnamefont {H~A}\ \bibnamefont {Dabkowska}}, \bibinfo {author}
  {\bibfnamefont {B~D}\ \bibnamefont {Gaulin}}, \ and\ \bibinfo {author}
  {\bibfnamefont {J~B}\ \bibnamefont {Kycia}},\ }\bibfield  {title} {\enquote
  {\bibinfo {title} {{Spin dynamics in the frozen state of the dipolar spin ice
  material Dy$_2$Ti$_2$O$_7$}},}\ }\href {\doibase DOI
  10.1103/PhysRevB.85.020410} {\bibfield  {journal} {\bibinfo  {journal} {Phys.
  Rev. B}\ }\textbf {\bibinfo {volume} {85}},\ \bibinfo {pages} {20410}
  (\bibinfo {year} {2012})}\BibitemShut {NoStop}%
\bibitem [{\citenamefont {Revell}\ \emph {et~al.}(2012)\citenamefont {Revell},
  \citenamefont {Yaraskavitch}, \citenamefont {Mason}, \citenamefont {Ross},
  \citenamefont {Dabkowska}, \citenamefont {Gaulin}, \citenamefont {Henelius},\
  and\ \citenamefont {Kycia}}]{Revell2012}%
  \BibitemOpen
  \bibfield  {author} {\bibinfo {author} {\bibfnamefont {H.~M.}\ \bibnamefont
  {Revell}}, \bibinfo {author} {\bibfnamefont {L.~R.}\ \bibnamefont
  {Yaraskavitch}}, \bibinfo {author} {\bibfnamefont {J.~D.}\ \bibnamefont
  {Mason}}, \bibinfo {author} {\bibfnamefont {H.~M.~L.}\ \bibnamefont {Ross},
  \bibfnamefont {K.~A. AU~Noad}}, \bibinfo {author} {\bibfnamefont {H.~A.}\
  \bibnamefont {Dabkowska}}, \bibinfo {author} {\bibfnamefont {B.~D.}\
  \bibnamefont {Gaulin}}, \bibinfo {author} {\bibfnamefont {P.}~\bibnamefont
  {Henelius}}, \ and\ \bibinfo {author} {\bibfnamefont {J.~B.}\ \bibnamefont
  {Kycia}},\ }\bibfield  {title} {\enquote {\bibinfo {title} {Evidence of
  impurity and boundary effects on magnetic monopole dynamics in spin ice},}\
  }\href {\doibase 10.1038/nphys2466} {\bibfield  {journal} {\bibinfo
  {journal} {Nat. Phys.}\ }\textbf {\bibinfo {volume} {9}},\ \bibinfo {pages}
  {34--37} (\bibinfo {year} {2012})}\BibitemShut {NoStop}%
\bibitem [{\citenamefont {Takatsu}\ \emph {et~al.}(2013)\citenamefont
  {Takatsu}, \citenamefont {Goto}, \citenamefont {Otsuka}, \citenamefont
  {Higashinaka}, \citenamefont {Matsubayashi}, \citenamefont {Uwatoko},\ and\
  \citenamefont {Kadowaki}}]{Takatsu2013}%
  \BibitemOpen
  \bibfield  {author} {\bibinfo {author} {\bibfnamefont {Hiroshi}\ \bibnamefont
  {Takatsu}}, \bibinfo {author} {\bibfnamefont {Kazuki}\ \bibnamefont {Goto}},
  \bibinfo {author} {\bibfnamefont {Hiromi}\ \bibnamefont {Otsuka}}, \bibinfo
  {author} {\bibfnamefont {Ryuji}\ \bibnamefont {Higashinaka}}, \bibinfo
  {author} {\bibfnamefont {Kazuyuki}\ \bibnamefont {Matsubayashi}}, \bibinfo
  {author} {\bibfnamefont {Yoshiya}\ \bibnamefont {Uwatoko}}, \ and\ \bibinfo
  {author} {\bibfnamefont {Hiroaki}\ \bibnamefont {Kadowaki}},\ }\bibfield
  {title} {\enquote {\bibinfo {title} {Ac susceptibility of the dipolar spin
  ice dy2ti2o7: Experiments and monte carlo simulations},}\ }\href {\doibase
  10.7566/JPSJ.82.104710} {\bibfield  {journal} {\bibinfo  {journal} {Journal
  of the Physical Society of Japan}\ }\textbf {\bibinfo {volume} {82}},\
  \bibinfo {pages} {104710} (\bibinfo {year} {2013})}\BibitemShut {NoStop}%
\bibitem [{\citenamefont {Eyvazov}\ \emph {et~al.}(2018)\citenamefont
  {Eyvazov}, \citenamefont {Dusad}, \citenamefont {Munsie}, \citenamefont
  {Dabkowska}, \citenamefont {Luke}, \citenamefont {Kassner}, \citenamefont
  {Davis},\ and\ \citenamefont {Eyal}}]{eyvazov2018common}%
  \BibitemOpen
  \bibfield  {author} {\bibinfo {author} {\bibfnamefont {Azar~B.}\ \bibnamefont
  {Eyvazov}}, \bibinfo {author} {\bibfnamefont {Ritika}\ \bibnamefont {Dusad}},
  \bibinfo {author} {\bibfnamefont {Timothy J.~S.}\ \bibnamefont {Munsie}},
  \bibinfo {author} {\bibfnamefont {Hanna~A.}\ \bibnamefont {Dabkowska}},
  \bibinfo {author} {\bibfnamefont {Graeme~M.}\ \bibnamefont {Luke}}, \bibinfo
  {author} {\bibfnamefont {Ethan~R.}\ \bibnamefont {Kassner}}, \bibinfo
  {author} {\bibfnamefont {J.~C.~Seamus}\ \bibnamefont {Davis}}, \ and\
  \bibinfo {author} {\bibfnamefont {Anna}\ \bibnamefont {Eyal}},\ }\bibfield
  {title} {\enquote {\bibinfo {title} {Common glass-forming spin-liquid state
  in the pyrochlore magnets
  ${\mathrm{dy}}_{2}{\mathrm{ti}}_{2}{\mathrm{o}}_{7}$ and
  ${\mathrm{ho}}_{2}{\mathrm{ti}}_{2}{\mathrm{o}}_{7}$},}\ }\href {\doibase
  10.1103/PhysRevB.98.214430} {\bibfield  {journal} {\bibinfo  {journal} {Phys.
  Rev. B}\ }\textbf {\bibinfo {volume} {98}},\ \bibinfo {pages} {214430}
  (\bibinfo {year} {2018})}\BibitemShut {NoStop}%
\bibitem [{\citenamefont {Kassner}\ \emph {et~al.}(2015)\citenamefont
  {Kassner}, \citenamefont {Eyvazov}, \citenamefont {Pichler}, \citenamefont
  {Munsie}, \citenamefont {Dabkowska}, \citenamefont {Luke},\ and\
  \citenamefont {Davis}}]{kassner2015supercooled}%
  \BibitemOpen
  \bibfield  {author} {\bibinfo {author} {\bibfnamefont {Ethan~R}\ \bibnamefont
  {Kassner}}, \bibinfo {author} {\bibfnamefont {Azar~B}\ \bibnamefont
  {Eyvazov}}, \bibinfo {author} {\bibfnamefont {Benjamin}\ \bibnamefont
  {Pichler}}, \bibinfo {author} {\bibfnamefont {Timothy~JS}\ \bibnamefont
  {Munsie}}, \bibinfo {author} {\bibfnamefont {Hanna~A}\ \bibnamefont
  {Dabkowska}}, \bibinfo {author} {\bibfnamefont {Graeme~M}\ \bibnamefont
  {Luke}}, \ and\ \bibinfo {author} {\bibfnamefont {JC~S{\'e}amus}\
  \bibnamefont {Davis}},\ }\bibfield  {title} {\enquote {\bibinfo {title}
  {Supercooled spin liquid state in the frustrated pyrochlore dy2ti2o7},}\
  }\href@noop {} {\bibfield  {journal} {\bibinfo  {journal} {Proceedings of the
  National Academy of Sciences}\ }\textbf {\bibinfo {volume} {112}},\ \bibinfo
  {pages} {8549--8554} (\bibinfo {year} {2015})}\BibitemShut {NoStop}%
\bibitem [{\citenamefont {Chandler}(1987)}]{chandler1987introduction}%
  \BibitemOpen
  \bibfield  {author} {\bibinfo {author} {\bibfnamefont {David}\ \bibnamefont
  {Chandler}},\ }\bibfield  {title} {\enquote {\bibinfo {title} {Introduction
  to modern statistical},}\ }\href@noop {} {\bibfield  {journal} {\bibinfo
  {journal} {Mechanics. Oxford University Press, Oxford, UK}\ }\textbf
  {\bibinfo {volume} {5}} (\bibinfo {year} {1987})}\BibitemShut {NoStop}%
\bibitem [{\citenamefont {Bouchaud}\ \emph {et~al.}(1998)\citenamefont
  {Bouchaud}, \citenamefont {Cugliandolo}, \citenamefont {Kurchan},\ and\
  \citenamefont {Mezard}}]{bouchaud1998out}%
  \BibitemOpen
  \bibfield  {author} {\bibinfo {author} {\bibfnamefont {Jean-Philippe}\
  \bibnamefont {Bouchaud}}, \bibinfo {author} {\bibfnamefont {Leticia~F}\
  \bibnamefont {Cugliandolo}}, \bibinfo {author} {\bibfnamefont {Jorge}\
  \bibnamefont {Kurchan}}, \ and\ \bibinfo {author} {\bibfnamefont {Marc}\
  \bibnamefont {Mezard}},\ }\bibfield  {title} {\enquote {\bibinfo {title} {Out
  of equilibrium dynamics in spin-glasses and other glassy systems},}\
  }\href@noop {} {\bibfield  {journal} {\bibinfo  {journal} {Spin glasses and
  random fields}\ ,\ \bibinfo {pages} {161--223}} (\bibinfo {year}
  {1998})}\BibitemShut {NoStop}%
\bibitem [{\citenamefont {Cugliandolo}\ and\ \citenamefont
  {Kurchan}(1993)}]{cugliandolo1993analytical}%
  \BibitemOpen
  \bibfield  {author} {\bibinfo {author} {\bibfnamefont {L.~F.}\ \bibnamefont
  {Cugliandolo}}\ and\ \bibinfo {author} {\bibfnamefont {J.}~\bibnamefont
  {Kurchan}},\ }\bibfield  {title} {\enquote {\bibinfo {title} {Analytical
  solution of the off-equilibrium dynamics of a long-range spin-glass model},}\
  }\href {\doibase 10.1103/PhysRevLett.71.173} {\bibfield  {journal} {\bibinfo
  {journal} {Phys. Rev. Lett.}\ }\textbf {\bibinfo {volume} {71}},\ \bibinfo
  {pages} {173--176} (\bibinfo {year} {1993})}\BibitemShut {NoStop}%
\bibitem [{\citenamefont {Cugliandolo}(2011)}]{cugliandolo2011effective}%
  \BibitemOpen
  \bibfield  {author} {\bibinfo {author} {\bibfnamefont {Leticia~F}\
  \bibnamefont {Cugliandolo}},\ }\bibfield  {title} {\enquote {\bibinfo {title}
  {The effective temperature},}\ }\href@noop {} {\bibfield  {journal} {\bibinfo
   {journal} {Journal of Physics A: Mathematical and Theoretical}\ }\textbf
  {\bibinfo {volume} {44}},\ \bibinfo {pages} {483001} (\bibinfo {year}
  {2011})}\BibitemShut {NoStop}%
\bibitem [{\citenamefont {Cugliandolo}\ \emph {et~al.}(1997)\citenamefont
  {Cugliandolo}, \citenamefont {Kurchan},\ and\ \citenamefont
  {Peliti}}]{cugliandolo1997energy}%
  \BibitemOpen
  \bibfield  {author} {\bibinfo {author} {\bibfnamefont {Leticia~F.}\
  \bibnamefont {Cugliandolo}}, \bibinfo {author} {\bibfnamefont {Jorge}\
  \bibnamefont {Kurchan}}, \ and\ \bibinfo {author} {\bibfnamefont {Luca}\
  \bibnamefont {Peliti}},\ }\bibfield  {title} {\enquote {\bibinfo {title}
  {Energy flow, partial equilibration, and effective temperatures in systems
  with slow dynamics},}\ }\href {\doibase 10.1103/PhysRevE.55.3898} {\bibfield
  {journal} {\bibinfo  {journal} {Phys. Rev. E}\ }\textbf {\bibinfo {volume}
  {55}},\ \bibinfo {pages} {3898--3914} (\bibinfo {year} {1997})}\BibitemShut
  {NoStop}%
\bibitem [{\citenamefont {Crisanti}\ and\ \citenamefont
  {Ritort}(2003)}]{crisanti2003violation}%
  \BibitemOpen
  \bibfield  {author} {\bibinfo {author} {\bibfnamefont {Andrea}\ \bibnamefont
  {Crisanti}}\ and\ \bibinfo {author} {\bibfnamefont {Felix}\ \bibnamefont
  {Ritort}},\ }\bibfield  {title} {\enquote {\bibinfo {title} {Violation of the
  fluctuation--dissipation theorem in glassy systems: basic notions and the
  numerical evidence},}\ }\href@noop {} {\bibfield  {journal} {\bibinfo
  {journal} {Journal of Physics A: Mathematical and General}\ }\textbf
  {\bibinfo {volume} {36}},\ \bibinfo {pages} {R181} (\bibinfo {year}
  {2003})}\BibitemShut {NoStop}%
\bibitem [{\citenamefont {Kurchan}(2005)}]{kurchan2005and}%
  \BibitemOpen
  \bibfield  {author} {\bibinfo {author} {\bibfnamefont {Jorge}\ \bibnamefont
  {Kurchan}},\ }\bibfield  {title} {\enquote {\bibinfo {title} {In and out of
  equilibrium},}\ }\href@noop {} {\bibfield  {journal} {\bibinfo  {journal}
  {Nature}\ }\textbf {\bibinfo {volume} {433}},\ \bibinfo {pages} {222--225}
  (\bibinfo {year} {2005})}\BibitemShut {NoStop}%
\bibitem [{\citenamefont {Struik}(1977)}]{struik1977physical}%
  \BibitemOpen
  \bibfield  {author} {\bibinfo {author} {\bibfnamefont {Leendert
  Cornelis~Elisa}\ \bibnamefont {Struik}},\ }\bibfield  {title} {\enquote
  {\bibinfo {title} {Physical aging in plastics and other glassy materials},}\
  }\href@noop {} {\bibfield  {journal} {\bibinfo  {journal} {Polymer
  Engineering \& Science}\ }\textbf {\bibinfo {volume} {17}},\ \bibinfo {pages}
  {165--173} (\bibinfo {year} {1977})}\BibitemShut {NoStop}%
\bibitem [{\citenamefont {Marinari}\ \emph {et~al.}(1998)\citenamefont
  {Marinari}, \citenamefont {Parisi}, \citenamefont {Ricci-Tersenghi},\ and\
  \citenamefont {Ruiz-Lorenzo}}]{marinari1998violation}%
  \BibitemOpen
  \bibfield  {author} {\bibinfo {author} {\bibfnamefont {Enzo}\ \bibnamefont
  {Marinari}}, \bibinfo {author} {\bibfnamefont {Giorgio}\ \bibnamefont
  {Parisi}}, \bibinfo {author} {\bibfnamefont {Federico}\ \bibnamefont
  {Ricci-Tersenghi}}, \ and\ \bibinfo {author} {\bibfnamefont {Juan~J}\
  \bibnamefont {Ruiz-Lorenzo}},\ }\bibfield  {title} {\enquote {\bibinfo
  {title} {Violation of the fluctuation-dissipation theorem in
  finite-dimensional spin glasses},}\ }\href@noop {} {\bibfield  {journal}
  {\bibinfo  {journal} {Journal of Physics A: Mathematical and General}\
  }\textbf {\bibinfo {volume} {31}},\ \bibinfo {pages} {2611} (\bibinfo {year}
  {1998})}\BibitemShut {NoStop}%
\bibitem [{\citenamefont {Barrat}\ and\ \citenamefont
  {Berthier}(2001)}]{barrat2001real}%
  \BibitemOpen
  \bibfield  {author} {\bibinfo {author} {\bibfnamefont {Alain}\ \bibnamefont
  {Barrat}}\ and\ \bibinfo {author} {\bibfnamefont {Ludovic}\ \bibnamefont
  {Berthier}},\ }\bibfield  {title} {\enquote {\bibinfo {title} {Real-space
  application of the mean-field description of spin-glass dynamics},}\ }\href
  {\doibase 10.1103/PhysRevLett.87.087204} {\bibfield  {journal} {\bibinfo
  {journal} {Phys. Rev. Lett.}\ }\textbf {\bibinfo {volume} {87}},\ \bibinfo
  {pages} {087204} (\bibinfo {year} {2001})}\BibitemShut {NoStop}%
\bibitem [{\citenamefont {Ricci-Tersenghi}(2003)}]{ricci2003measuring}%
  \BibitemOpen
  \bibfield  {author} {\bibinfo {author} {\bibfnamefont {F.}~\bibnamefont
  {Ricci-Tersenghi}},\ }\bibfield  {title} {\enquote {\bibinfo {title}
  {Measuring the fluctuation-dissipation ratio in glassy systems with no
  perturbing field},}\ }\href {\doibase 10.1103/PhysRevE.68.065104} {\bibfield
  {journal} {\bibinfo  {journal} {Phys. Rev. E}\ }\textbf {\bibinfo {volume}
  {68}},\ \bibinfo {pages} {065104} (\bibinfo {year} {2003})}\BibitemShut
  {NoStop}%
\bibitem [{\citenamefont {H\'erisson}\ and\ \citenamefont
  {Ocio}(2002)}]{herisson2002fluctuation}%
  \BibitemOpen
  \bibfield  {author} {\bibinfo {author} {\bibfnamefont {D.}~\bibnamefont
  {H\'erisson}}\ and\ \bibinfo {author} {\bibfnamefont {M.}~\bibnamefont
  {Ocio}},\ }\bibfield  {title} {\enquote {\bibinfo {title}
  {Fluctuation-dissipation ratio of a spin glass in the aging regime},}\ }\href
  {\doibase 10.1103/PhysRevLett.88.257202} {\bibfield  {journal} {\bibinfo
  {journal} {Phys. Rev. Lett.}\ }\textbf {\bibinfo {volume} {88}},\ \bibinfo
  {pages} {257202} (\bibinfo {year} {2002})}\BibitemShut {NoStop}%
\bibitem [{\citenamefont {Parisi}(1997)}]{PhysRevLett.79.3660}%
  \BibitemOpen
  \bibfield  {author} {\bibinfo {author} {\bibfnamefont {Giorgio}\ \bibnamefont
  {Parisi}},\ }\bibfield  {title} {\enquote {\bibinfo {title} {Off-equilibrium
  fluctuation-dissipation relation in fragile glasses},}\ }\href {\doibase
  10.1103/PhysRevLett.79.3660} {\bibfield  {journal} {\bibinfo  {journal}
  {Phys. Rev. Lett.}\ }\textbf {\bibinfo {volume} {79}},\ \bibinfo {pages}
  {3660--3663} (\bibinfo {year} {1997})}\BibitemShut {NoStop}%
\bibitem [{\citenamefont {Barrat}\ and\ \citenamefont
  {Kob}(1999)}]{barrat1999fluctuation}%
  \BibitemOpen
  \bibfield  {author} {\bibinfo {author} {\bibfnamefont {J-L}\ \bibnamefont
  {Barrat}}\ and\ \bibinfo {author} {\bibfnamefont {Walter}\ \bibnamefont
  {Kob}},\ }\bibfield  {title} {\enquote {\bibinfo {title}
  {Fluctuation-dissipation ratio in an aging lennard-jones glass},}\
  }\href@noop {} {\bibfield  {journal} {\bibinfo  {journal} {EPL (Europhysics
  Letters)}\ }\textbf {\bibinfo {volume} {46}},\ \bibinfo {pages} {637}
  (\bibinfo {year} {1999})}\BibitemShut {NoStop}%
\bibitem [{\citenamefont {Grigera}\ and\ \citenamefont
  {Israeloff}(1999)}]{grigera1999observation}%
  \BibitemOpen
  \bibfield  {author} {\bibinfo {author} {\bibfnamefont {Tom\'as~S.}\
  \bibnamefont {Grigera}}\ and\ \bibinfo {author} {\bibfnamefont {N.~E.}\
  \bibnamefont {Israeloff}},\ }\bibfield  {title} {\enquote {\bibinfo {title}
  {Observation of fluctuation-dissipation-theorem violations in a structural
  glass},}\ }\href {\doibase 10.1103/PhysRevLett.83.5038} {\bibfield  {journal}
  {\bibinfo  {journal} {Phys. Rev. Lett.}\ }\textbf {\bibinfo {volume} {83}},\
  \bibinfo {pages} {5038--5041} (\bibinfo {year} {1999})}\BibitemShut {NoStop}%
\bibitem [{\citenamefont {Bellon}\ \emph {et~al.}(2001)\citenamefont {Bellon},
  \citenamefont {Ciliberto},\ and\ \citenamefont
  {Laroche}}]{bellon2001violation}%
  \BibitemOpen
  \bibfield  {author} {\bibinfo {author} {\bibfnamefont {Ludovic}\ \bibnamefont
  {Bellon}}, \bibinfo {author} {\bibfnamefont {Sergio}\ \bibnamefont
  {Ciliberto}}, \ and\ \bibinfo {author} {\bibfnamefont {Claude}\ \bibnamefont
  {Laroche}},\ }\bibfield  {title} {\enquote {\bibinfo {title} {Violation of
  the fluctuation-dissipation relation during the formation of a colloidal
  glass},}\ }\href@noop {} {\bibfield  {journal} {\bibinfo  {journal} {EPL
  (Europhysics Letters)}\ }\textbf {\bibinfo {volume} {53}},\ \bibinfo {pages}
  {511} (\bibinfo {year} {2001})}\BibitemShut {NoStop}%
\bibitem [{\citenamefont {Berthier}(2007)}]{berthier2007efficient}%
  \BibitemOpen
  \bibfield  {author} {\bibinfo {author} {\bibfnamefont {Ludovic}\ \bibnamefont
  {Berthier}},\ }\bibfield  {title} {\enquote {\bibinfo {title} {Efficient
  measurement of linear susceptibilities in molecular simulations: Application
  to aging supercooled liquids},}\ }\href {\doibase
  10.1103/PhysRevLett.98.220601} {\bibfield  {journal} {\bibinfo  {journal}
  {Phys. Rev. Lett.}\ }\textbf {\bibinfo {volume} {98}},\ \bibinfo {pages}
  {220601} (\bibinfo {year} {2007})}\BibitemShut {NoStop}%
\bibitem [{\citenamefont {Barrat}(1998)}]{barrat1998monte}%
  \BibitemOpen
  \bibfield  {author} {\bibinfo {author} {\bibfnamefont {A.}~\bibnamefont
  {Barrat}},\ }\bibfield  {title} {\enquote {\bibinfo {title} {Monte carlo
  simulations of the violation of the fluctuation-dissipation theorem in domain
  growth processes},}\ }\href {\doibase 10.1103/PhysRevE.57.3629} {\bibfield
  {journal} {\bibinfo  {journal} {Phys. Rev. E}\ }\textbf {\bibinfo {volume}
  {57}},\ \bibinfo {pages} {3629--3632} (\bibinfo {year} {1998})}\BibitemShut
  {NoStop}%
\bibitem [{\citenamefont {Berthier}\ \emph {et~al.}(1999)\citenamefont
  {Berthier}, \citenamefont {Barrat},\ and\ \citenamefont
  {Kurchan}}]{berthier1999response}%
  \BibitemOpen
  \bibfield  {author} {\bibinfo {author} {\bibfnamefont {Ludovic}\ \bibnamefont
  {Berthier}}, \bibinfo {author} {\bibfnamefont {Jean-Louis}\ \bibnamefont
  {Barrat}}, \ and\ \bibinfo {author} {\bibfnamefont {Jorge}\ \bibnamefont
  {Kurchan}},\ }\bibfield  {title} {\enquote {\bibinfo {title} {Response
  function of coarsening systems},}\ }\href
  {https://link.springer.com/article/10.1007/s100510051192} {\bibfield
  {journal} {\bibinfo  {journal} {The European Physical Journal B-Condensed
  Matter and Complex Systems}\ }\textbf {\bibinfo {volume} {11}},\ \bibinfo
  {pages} {635--641} (\bibinfo {year} {1999})}\BibitemShut {NoStop}%
\bibitem [{\citenamefont {Godreche}\ and\ \citenamefont
  {Luck}(2000)}]{godreche2000response}%
  \BibitemOpen
  \bibfield  {author} {\bibinfo {author} {\bibfnamefont {C}~\bibnamefont
  {Godreche}}\ and\ \bibinfo {author} {\bibfnamefont {JM}~\bibnamefont
  {Luck}},\ }\bibfield  {title} {\enquote {\bibinfo {title} {Response of
  non-equilibrium systems at criticality: exact results for the glauber-ising
  chain},}\ }\href@noop {} {\bibfield  {journal} {\bibinfo  {journal} {Journal
  of Physics A: Mathematical and General}\ }\textbf {\bibinfo {volume} {33}},\
  \bibinfo {pages} {1151} (\bibinfo {year} {2000})}\BibitemShut {NoStop}%
\bibitem [{\citenamefont {Henkel}\ \emph {et~al.}(2001)\citenamefont {Henkel},
  \citenamefont {Pleimling}, \citenamefont {Godr\`eche},\ and\ \citenamefont
  {Luck}}]{henkel2001aging}%
  \BibitemOpen
  \bibfield  {author} {\bibinfo {author} {\bibfnamefont {Malte}\ \bibnamefont
  {Henkel}}, \bibinfo {author} {\bibfnamefont {Michel}\ \bibnamefont
  {Pleimling}}, \bibinfo {author} {\bibfnamefont {Claude}\ \bibnamefont
  {Godr\`eche}}, \ and\ \bibinfo {author} {\bibfnamefont {Jean-Marc}\
  \bibnamefont {Luck}},\ }\bibfield  {title} {\enquote {\bibinfo {title}
  {Aging, phase ordering, and conformal invariance},}\ }\href {\doibase
  10.1103/PhysRevLett.87.265701} {\bibfield  {journal} {\bibinfo  {journal}
  {Phys. Rev. Lett.}\ }\textbf {\bibinfo {volume} {87}},\ \bibinfo {pages}
  {265701} (\bibinfo {year} {2001})}\BibitemShut {NoStop}%
\bibitem [{\citenamefont {Berthier}\ \emph {et~al.}(2001)\citenamefont
  {Berthier}, \citenamefont {Holdsworth},\ and\ \citenamefont
  {Sellitto}}]{berthier2001nonequilibrium}%
  \BibitemOpen
  \bibfield  {author} {\bibinfo {author} {\bibfnamefont {Ludovic}\ \bibnamefont
  {Berthier}}, \bibinfo {author} {\bibfnamefont {Peter~CW}\ \bibnamefont
  {Holdsworth}}, \ and\ \bibinfo {author} {\bibfnamefont {Mauro}\ \bibnamefont
  {Sellitto}},\ }\bibfield  {title} {\enquote {\bibinfo {title} {Nonequilibrium
  critical dynamics of the two-dimensional xy model},}\ }\href
  {https://iopscience.iop.org/article/10.1088/0305-4470/34/9/301/meta}
  {\bibfield  {journal} {\bibinfo  {journal} {Journal of Physics A:
  Mathematical and General}\ }\textbf {\bibinfo {volume} {34}},\ \bibinfo
  {pages} {1805} (\bibinfo {year} {2001})}\BibitemShut {NoStop}%
\bibitem [{\citenamefont {Mayer}\ \emph {et~al.}(2003)\citenamefont {Mayer},
  \citenamefont {Berthier}, \citenamefont {Garrahan},\ and\ \citenamefont
  {Sollich}}]{mayer2003fluctuation}%
  \BibitemOpen
  \bibfield  {author} {\bibinfo {author} {\bibfnamefont {Peter}\ \bibnamefont
  {Mayer}}, \bibinfo {author} {\bibfnamefont {Ludovic}\ \bibnamefont
  {Berthier}}, \bibinfo {author} {\bibfnamefont {Juan~P.}\ \bibnamefont
  {Garrahan}}, \ and\ \bibinfo {author} {\bibfnamefont {Peter}\ \bibnamefont
  {Sollich}},\ }\bibfield  {title} {\enquote {\bibinfo {title}
  {Fluctuation-dissipation relations in the nonequilibrium critical dynamics of
  ising models},}\ }\href {\doibase 10.1103/PhysRevE.68.016116} {\bibfield
  {journal} {\bibinfo  {journal} {Phys. Rev. E}\ }\textbf {\bibinfo {volume}
  {68}},\ \bibinfo {pages} {016116} (\bibinfo {year} {2003})}\BibitemShut
  {NoStop}%
\bibitem [{\citenamefont {Calabrese}\ and\ \citenamefont
  {Gambassi}(2005)}]{calabrese2005ageing}%
  \BibitemOpen
  \bibfield  {author} {\bibinfo {author} {\bibfnamefont {Pasquale}\
  \bibnamefont {Calabrese}}\ and\ \bibinfo {author} {\bibfnamefont {Andrea}\
  \bibnamefont {Gambassi}},\ }\bibfield  {title} {\enquote {\bibinfo {title}
  {Ageing properties of critical systems},}\ }\href@noop {} {\bibfield
  {journal} {\bibinfo  {journal} {Journal of Physics A: Mathematical and
  General}\ }\textbf {\bibinfo {volume} {38}},\ \bibinfo {pages} {R133}
  (\bibinfo {year} {2005})}\BibitemShut {NoStop}%
\bibitem [{\citenamefont {Barrat}\ and\ \citenamefont
  {Berthier}(2000)}]{barrat2000fluctuation}%
  \BibitemOpen
  \bibfield  {author} {\bibinfo {author} {\bibfnamefont {Jean-Louis}\
  \bibnamefont {Barrat}}\ and\ \bibinfo {author} {\bibfnamefont {Ludovic}\
  \bibnamefont {Berthier}},\ }\bibfield  {title} {\enquote {\bibinfo {title}
  {Fluctuation-dissipation relation in a sheared fluid},}\ }\href {\doibase
  10.1103/PhysRevE.63.012503} {\bibfield  {journal} {\bibinfo  {journal} {Phys.
  Rev. E}\ }\textbf {\bibinfo {volume} {63}},\ \bibinfo {pages} {012503}
  (\bibinfo {year} {2000})}\BibitemShut {NoStop}%
\bibitem [{\citenamefont {Berthier}\ and\ \citenamefont
  {Barrat}(2002{\natexlab{a}})}]{berthier2002nonequilibrium}%
  \BibitemOpen
  \bibfield  {author} {\bibinfo {author} {\bibfnamefont {Ludovic}\ \bibnamefont
  {Berthier}}\ and\ \bibinfo {author} {\bibfnamefont {Jean-Louis}\ \bibnamefont
  {Barrat}},\ }\bibfield  {title} {\enquote {\bibinfo {title} {Nonequilibrium
  dynamics and fluctuation-dissipation relation in a sheared fluid},}\ }\href
  {https://aip.scitation.org/doi/abs/10.1063/1.1460862} {\bibfield  {journal}
  {\bibinfo  {journal} {The Journal of Chemical Physics}\ }\textbf {\bibinfo
  {volume} {116}},\ \bibinfo {pages} {6228--6242} (\bibinfo {year}
  {2002}{\natexlab{a}})}\BibitemShut {NoStop}%
\bibitem [{\citenamefont {Berthier}\ and\ \citenamefont
  {Barrat}(2002{\natexlab{b}})}]{berthier2002shearing}%
  \BibitemOpen
  \bibfield  {author} {\bibinfo {author} {\bibfnamefont {Ludovic}\ \bibnamefont
  {Berthier}}\ and\ \bibinfo {author} {\bibfnamefont {Jean-Louis}\ \bibnamefont
  {Barrat}},\ }\bibfield  {title} {\enquote {\bibinfo {title} {Shearing a
  glassy material: Numerical tests of nonequilibrium mode-coupling approaches
  and experimental proposals},}\ }\href {\doibase
  10.1103/PhysRevLett.89.095702} {\bibfield  {journal} {\bibinfo  {journal}
  {Phys. Rev. Lett.}\ }\textbf {\bibinfo {volume} {89}},\ \bibinfo {pages}
  {095702} (\bibinfo {year} {2002}{\natexlab{b}})}\BibitemShut {NoStop}%
\bibitem [{\citenamefont {Loi}\ \emph {et~al.}(2008)\citenamefont {Loi},
  \citenamefont {Mossa},\ and\ \citenamefont {Cugliandolo}}]{loi2008effective}%
  \BibitemOpen
  \bibfield  {author} {\bibinfo {author} {\bibfnamefont {Davide}\ \bibnamefont
  {Loi}}, \bibinfo {author} {\bibfnamefont {Stefano}\ \bibnamefont {Mossa}}, \
  and\ \bibinfo {author} {\bibfnamefont {Leticia~F.}\ \bibnamefont
  {Cugliandolo}},\ }\bibfield  {title} {\enquote {\bibinfo {title} {Effective
  temperature of active matter},}\ }\href {\doibase 10.1103/PhysRevE.77.051111}
  {\bibfield  {journal} {\bibinfo  {journal} {Phys. Rev. E}\ }\textbf {\bibinfo
  {volume} {77}},\ \bibinfo {pages} {051111} (\bibinfo {year}
  {2008})}\BibitemShut {NoStop}%
\bibitem [{\citenamefont {Levis}\ and\ \citenamefont
  {Berthier}(2015)}]{levis2015single}%
  \BibitemOpen
  \bibfield  {author} {\bibinfo {author} {\bibfnamefont {Demian}\ \bibnamefont
  {Levis}}\ and\ \bibinfo {author} {\bibfnamefont {Ludovic}\ \bibnamefont
  {Berthier}},\ }\bibfield  {title} {\enquote {\bibinfo {title} {From
  single-particle to collective effective temperatures in an active fluid of
  self-propelled particles},}\ }\href
  {https://iopscience.iop.org/article/10.1209/0295-5075/111/60006/meta}
  {\bibfield  {journal} {\bibinfo  {journal} {EPL (Europhysics Letters)}\
  }\textbf {\bibinfo {volume} {111}},\ \bibinfo {pages} {60006} (\bibinfo
  {year} {2015})}\BibitemShut {NoStop}%
\bibitem [{\citenamefont {Castelnovo}\ \emph {et~al.}(2010)\citenamefont
  {Castelnovo}, \citenamefont {Moessner},\ and\ \citenamefont
  {Sondhi}}]{castelnovo2010thermal}%
  \BibitemOpen
  \bibfield  {author} {\bibinfo {author} {\bibfnamefont {C.}~\bibnamefont
  {Castelnovo}}, \bibinfo {author} {\bibfnamefont {R.}~\bibnamefont
  {Moessner}}, \ and\ \bibinfo {author} {\bibfnamefont {S.~L.}\ \bibnamefont
  {Sondhi}},\ }\bibfield  {title} {\enquote {\bibinfo {title} {Thermal quenches
  in spin ice},}\ }\href {\doibase 10.1103/PhysRevLett.104.107201} {\bibfield
  {journal} {\bibinfo  {journal} {Phys. Rev. Lett.}\ }\textbf {\bibinfo
  {volume} {104}},\ \bibinfo {pages} {107201} (\bibinfo {year}
  {2010})}\BibitemShut {NoStop}%
\bibitem [{\citenamefont {Mayer}\ \emph {et~al.}(2006)\citenamefont {Mayer},
  \citenamefont {L\'eonard}, \citenamefont {Berthier}, \citenamefont
  {Garrahan},\ and\ \citenamefont {Sollich}}]{mayer2006activated}%
  \BibitemOpen
  \bibfield  {author} {\bibinfo {author} {\bibfnamefont {Peter}\ \bibnamefont
  {Mayer}}, \bibinfo {author} {\bibfnamefont {S\'ebastien}\ \bibnamefont
  {L\'eonard}}, \bibinfo {author} {\bibfnamefont {Ludovic}\ \bibnamefont
  {Berthier}}, \bibinfo {author} {\bibfnamefont {Juan~P.}\ \bibnamefont
  {Garrahan}}, \ and\ \bibinfo {author} {\bibfnamefont {Peter}\ \bibnamefont
  {Sollich}},\ }\bibfield  {title} {\enquote {\bibinfo {title} {Activated aging
  dynamics and negative fluctuation-dissipation ratios},}\ }\href {\doibase
  10.1103/PhysRevLett.96.030602} {\bibfield  {journal} {\bibinfo  {journal}
  {Phys. Rev. Lett.}\ }\textbf {\bibinfo {volume} {96}},\ \bibinfo {pages}
  {030602} (\bibinfo {year} {2006})}\BibitemShut {NoStop}%
\bibitem [{\citenamefont {L{\'e}onard}\ \emph {et~al.}(2007)\citenamefont
  {L{\'e}onard}, \citenamefont {Mayer}, \citenamefont {Sollich}, \citenamefont
  {Berthier},\ and\ \citenamefont {Garrahan}}]{leonard2007non}%
  \BibitemOpen
  \bibfield  {author} {\bibinfo {author} {\bibfnamefont {S{\'e}bastien}\
  \bibnamefont {L{\'e}onard}}, \bibinfo {author} {\bibfnamefont {Peter}\
  \bibnamefont {Mayer}}, \bibinfo {author} {\bibfnamefont {Peter}\ \bibnamefont
  {Sollich}}, \bibinfo {author} {\bibfnamefont {Ludovic}\ \bibnamefont
  {Berthier}}, \ and\ \bibinfo {author} {\bibfnamefont {Juan~P}\ \bibnamefont
  {Garrahan}},\ }\bibfield  {title} {\enquote {\bibinfo {title}
  {Non-equilibrium dynamics of spin facilitated glass models},}\ }\href
  {https://iopscience.iop.org/article/10.1088/1742-5468/2007/07/P07017/meta}
  {\bibfield  {journal} {\bibinfo  {journal} {Journal of statistical mechanics:
  theory and experiment}\ }\textbf {\bibinfo {volume} {2007}},\ \bibinfo
  {pages} {P07017} (\bibinfo {year} {2007})}\BibitemShut {NoStop}%
\bibitem [{\citenamefont {Jack}\ \emph {et~al.}(2006)\citenamefont {Jack},
  \citenamefont {Berthier},\ and\ \citenamefont
  {Garrahan}}]{jack2006fluctuation}%
  \BibitemOpen
  \bibfield  {author} {\bibinfo {author} {\bibfnamefont {Robert~L}\
  \bibnamefont {Jack}}, \bibinfo {author} {\bibfnamefont {Ludovic}\
  \bibnamefont {Berthier}}, \ and\ \bibinfo {author} {\bibfnamefont {Juan~P}\
  \bibnamefont {Garrahan}},\ }\bibfield  {title} {\enquote {\bibinfo {title}
  {Fluctuation-dissipation relations in plaquette spin systems with multi-stage
  relaxation},}\ }\href
  {https://iopscience.iop.org/article/10.1088/1742-5468/2006/12/P12005/meta}
  {\bibfield  {journal} {\bibinfo  {journal} {Journal of Statistical Mechanics:
  Theory and Experiment}\ }\textbf {\bibinfo {volume} {2006}},\ \bibinfo
  {pages} {P12005} (\bibinfo {year} {2006})}\BibitemShut {NoStop}%
\bibitem [{\citenamefont {Garriga}\ \emph {et~al.}(2009)\citenamefont
  {Garriga}, \citenamefont {Pagonabarraga},\ and\ \citenamefont
  {Ritort}}]{garriga2009negative}%
  \BibitemOpen
  \bibfield  {author} {\bibinfo {author} {\bibfnamefont {A.}~\bibnamefont
  {Garriga}}, \bibinfo {author} {\bibfnamefont {I.}~\bibnamefont
  {Pagonabarraga}}, \ and\ \bibinfo {author} {\bibfnamefont {F.}~\bibnamefont
  {Ritort}},\ }\bibfield  {title} {\enquote {\bibinfo {title} {Negative
  fluctuation-dissipation ratios in the backgammon model},}\ }\href {\doibase
  10.1103/PhysRevE.79.041122} {\bibfield  {journal} {\bibinfo  {journal} {Phys.
  Rev. E}\ }\textbf {\bibinfo {volume} {79}},\ \bibinfo {pages} {041122}
  (\bibinfo {year} {2009})}\BibitemShut {NoStop}%
\bibitem [{\citenamefont {Ryzhkin}(2005)}]{Ryzhkin2005}%
  \BibitemOpen
  \bibfield  {author} {\bibinfo {author} {\bibfnamefont {I.~A.}\ \bibnamefont
  {Ryzhkin}},\ }\bibfield  {title} {\enquote {\bibinfo {title} {Magnetic
  relaxation in rare-earth oxide pyrochlores},}\ }\href
  {http://link.springer.com/article/10.1134/1.2103216} {\bibfield  {journal}
  {\bibinfo  {journal} {Journal of Experimental and Theoretical Physics}\
  }\textbf {\bibinfo {volume} {101}},\ \bibinfo {pages} {481{\textendash}486}
  (\bibinfo {year} {2005})}\BibitemShut {NoStop}%
\bibitem [{\citenamefont {Castelnovo}\ \emph {et~al.}(2008)\citenamefont
  {Castelnovo}, \citenamefont {Moessner},\ and\ \citenamefont
  {Sondhi}}]{castelnovo2008magnetic}%
  \BibitemOpen
  \bibfield  {author} {\bibinfo {author} {\bibfnamefont {C}~\bibnamefont
  {Castelnovo}}, \bibinfo {author} {\bibfnamefont {R}~\bibnamefont {Moessner}},
  \ and\ \bibinfo {author} {\bibfnamefont {SL}~\bibnamefont {Sondhi}},\
  }\bibfield  {title} {\enquote {\bibinfo {title} {Magnetic monopoles in spin
  ice.}}\ }\href@noop {} {\bibfield  {journal} {\bibinfo  {journal} {Nature}\
  }\textbf {\bibinfo {volume} {451}},\ \bibinfo {pages} {42--45} (\bibinfo
  {year} {2008})}\BibitemShut {NoStop}%
\bibitem [{\citenamefont {M\"oller}\ and\ \citenamefont
  {Moessner}(2006)}]{molller2006needle}%
  \BibitemOpen
  \bibfield  {author} {\bibinfo {author} {\bibfnamefont {G.}~\bibnamefont
  {M\"oller}}\ and\ \bibinfo {author} {\bibfnamefont {R.}~\bibnamefont
  {Moessner}},\ }\bibfield  {title} {\enquote {\bibinfo {title} {Artificial
  square ice and related dipolar nanoarrays},}\ }\href {\doibase
  10.1103/PhysRevLett.96.237202} {\bibfield  {journal} {\bibinfo  {journal}
  {Phys. Rev. Lett.}\ }\textbf {\bibinfo {volume} {96}},\ \bibinfo {pages}
  {237202} (\bibinfo {year} {2006})}\BibitemShut {NoStop}%
\bibitem [{\citenamefont {Castelnovo}\ and\ \citenamefont
  {Holdsworth}(2021)}]{castelnovo2021modelling}%
  \BibitemOpen
  \bibfield  {author} {\bibinfo {author} {\bibfnamefont {C}~\bibnamefont
  {Castelnovo}}\ and\ \bibinfo {author} {\bibfnamefont {PCW}\ \bibnamefont
  {Holdsworth}},\ }\bibfield  {title} {\enquote {\bibinfo {title} {Modelling of
  classical spin ice: Coulomb gas description of thermodynamic and dynamic
  properties},}\ }in\ \href@noop {} {\emph {\bibinfo {booktitle} {Spin Ice}}}\
  (\bibinfo  {publisher} {Springer},\ \bibinfo {year} {2021})\ pp.\ \bibinfo
  {pages} {143--188}\BibitemShut {NoStop}%
\bibitem [{\citenamefont {Jaubert}\ and\ \citenamefont
  {Holdsworth}(2011)}]{Jaubert11}%
  \BibitemOpen
  \bibfield  {author} {\bibinfo {author} {\bibfnamefont {L~D~C}\ \bibnamefont
  {Jaubert}}\ and\ \bibinfo {author} {\bibfnamefont {P~C~W}\ \bibnamefont
  {Holdsworth}},\ }\bibfield  {title} {\enquote {\bibinfo {title} {Magnetic
  monopole dynamics in spin ice},}\ }\href
  {http://stacks.iop.org/0953-8984/23/i=16/a=164222} {\bibfield  {journal}
  {\bibinfo  {journal} {Journal of Physics: Condensed Matter}\ }\textbf
  {\bibinfo {volume} {23}},\ \bibinfo {pages} {164222} (\bibinfo {year}
  {2011})}\BibitemShut {NoStop}%
\bibitem [{\citenamefont {Tomasello}\ \emph {et~al.}(2019)\citenamefont
  {Tomasello}, \citenamefont {Castelnovo}, \citenamefont {Moessner},\ and\
  \citenamefont {Quintanilla}}]{Tomasello2019}%
  \BibitemOpen
  \bibfield  {author} {\bibinfo {author} {\bibfnamefont {Bruno}\ \bibnamefont
  {Tomasello}}, \bibinfo {author} {\bibfnamefont {Claudio}\ \bibnamefont
  {Castelnovo}}, \bibinfo {author} {\bibfnamefont {Roderich}\ \bibnamefont
  {Moessner}}, \ and\ \bibinfo {author} {\bibfnamefont {Jorge}\ \bibnamefont
  {Quintanilla}},\ }\bibfield  {title} {\enquote {\bibinfo {title} {Correlated
  quantum tunneling of monopoles in spin ice},}\ }\href {\doibase
  10.1103/PhysRevLett.123.067204} {\bibfield  {journal} {\bibinfo  {journal}
  {Phys. Rev. Lett.}\ }\textbf {\bibinfo {volume} {123}},\ \bibinfo {pages}
  {067204} (\bibinfo {year} {2019})}\BibitemShut {NoStop}%
\bibitem [{\citenamefont {Harris}\ \emph {et~al.}(1997)\citenamefont {Harris},
  \citenamefont {Bramwell}, \citenamefont {McMorrow}, \citenamefont {Zeiske},\
  and\ \citenamefont {Godfrey}}]{harris1997geometrical}%
  \BibitemOpen
  \bibfield  {author} {\bibinfo {author} {\bibfnamefont {MJ}~\bibnamefont
  {Harris}}, \bibinfo {author} {\bibfnamefont {ST}~\bibnamefont {Bramwell}},
  \bibinfo {author} {\bibfnamefont {DF}~\bibnamefont {McMorrow}}, \bibinfo
  {author} {\bibfnamefont {Th}~\bibnamefont {Zeiske}}, \ and\ \bibinfo {author}
  {\bibfnamefont {KW}~\bibnamefont {Godfrey}},\ }\bibfield  {title} {\enquote
  {\bibinfo {title} {Geometrical frustration in the ferromagnetic pyrochlore ho
  2 ti 2 o 7},}\ }\href@noop {} {\bibfield  {journal} {\bibinfo  {journal}
  {Physical Review Letters}\ }\textbf {\bibinfo {volume} {79}},\ \bibinfo
  {pages} {2554} (\bibinfo {year} {1997})}\BibitemShut {NoStop}%
\bibitem [{\citenamefont {Bramwell}\ and\ \citenamefont
  {Gingras}(2001)}]{bramwell2001spin}%
  \BibitemOpen
  \bibfield  {author} {\bibinfo {author} {\bibfnamefont {Steven~T}\
  \bibnamefont {Bramwell}}\ and\ \bibinfo {author} {\bibfnamefont {Michel~JP}\
  \bibnamefont {Gingras}},\ }\bibfield  {title} {\enquote {\bibinfo {title}
  {Spin ice state in frustrated magnetic pyrochlore materials},}\ }\href@noop
  {} {\bibfield  {journal} {\bibinfo  {journal} {Science}\ }\textbf {\bibinfo
  {volume} {294}},\ \bibinfo {pages} {1495--1501} (\bibinfo {year}
  {2001})}\BibitemShut {NoStop}%
\bibitem [{\citenamefont {den Hertog}\ and\ \citenamefont
  {Gingras}(2000)}]{denHertog2000}%
  \BibitemOpen
  \bibfield  {author} {\bibinfo {author} {\bibfnamefont {Byron~C.}\
  \bibnamefont {den Hertog}}\ and\ \bibinfo {author} {\bibfnamefont {Michel
  J.~P.}\ \bibnamefont {Gingras}},\ }\bibfield  {title} {\enquote {\bibinfo
  {title} {Dipolar interactions and origin of spin ice in ising pyrochlore
  magnets},}\ }\href {\doibase 10.1103/PhysRevLett.84.3430} {\bibfield
  {journal} {\bibinfo  {journal} {Phys. Rev. Lett.}\ }\textbf {\bibinfo
  {volume} {84}},\ \bibinfo {pages} {3430--3433} (\bibinfo {year}
  {2000})}\BibitemShut {NoStop}%
\bibitem [{\citenamefont {Melko}\ and\ \citenamefont
  {Gingras}(2004)}]{melko2004monte}%
  \BibitemOpen
  \bibfield  {author} {\bibinfo {author} {\bibfnamefont {Roger~G}\ \bibnamefont
  {Melko}}\ and\ \bibinfo {author} {\bibfnamefont {Michel~JP}\ \bibnamefont
  {Gingras}},\ }\bibfield  {title} {\enquote {\bibinfo {title} {Monte carlo
  studies of the dipolar spin ice model},}\ }\href@noop {} {\bibfield
  {journal} {\bibinfo  {journal} {Journal of Physics: Condensed Matter}\
  }\textbf {\bibinfo {volume} {16}},\ \bibinfo {pages} {R1277} (\bibinfo {year}
  {2004})}\BibitemShut {NoStop}%
\bibitem [{\citenamefont {Bernal}\ and\ \citenamefont
  {Fowler}(1933)}]{doi:10.1063/1.1749327}%
  \BibitemOpen
  \bibfield  {author} {\bibinfo {author} {\bibfnamefont {J.~D.}\ \bibnamefont
  {Bernal}}\ and\ \bibinfo {author} {\bibfnamefont {R.~H.}\ \bibnamefont
  {Fowler}},\ }\bibfield  {title} {\enquote {\bibinfo {title} {A theory of
  water and ionic solution, with particular reference to hydrogen and hydroxyl
  ions},}\ }\href {\doibase 10.1063/1.1749327} {\bibfield  {journal} {\bibinfo
  {journal} {The Journal of Chemical Physics}\ }\textbf {\bibinfo {volume}
  {1}},\ \bibinfo {pages} {515--548} (\bibinfo {year} {1933})}\BibitemShut
  {NoStop}%
\bibitem [{\citenamefont {Isakov}\ \emph {et~al.}(2005)\citenamefont {Isakov},
  \citenamefont {Moessner},\ and\ \citenamefont
  {Sondhi}}]{isakov2005projective}%
  \BibitemOpen
  \bibfield  {author} {\bibinfo {author} {\bibfnamefont {S.~V.}\ \bibnamefont
  {Isakov}}, \bibinfo {author} {\bibfnamefont {R.}~\bibnamefont {Moessner}}, \
  and\ \bibinfo {author} {\bibfnamefont {S.~L.}\ \bibnamefont {Sondhi}},\
  }\bibfield  {title} {\enquote {\bibinfo {title} {Why spin ice obeys the ice
  rules},}\ }\href {\doibase 10.1103/PhysRevLett.95.217201} {\bibfield
  {journal} {\bibinfo  {journal} {Phys. Rev. Lett.}\ }\textbf {\bibinfo
  {volume} {95}},\ \bibinfo {pages} {217201} (\bibinfo {year}
  {2005})}\BibitemShut {NoStop}%
\bibitem [{\citenamefont {Melko}\ \emph {et~al.}(2001)\citenamefont {Melko},
  \citenamefont {den Hertog},\ and\ \citenamefont {Gingras}}]{melko2001long}%
  \BibitemOpen
  \bibfield  {author} {\bibinfo {author} {\bibfnamefont {Roger~G.}\
  \bibnamefont {Melko}}, \bibinfo {author} {\bibfnamefont {Byron~C.}\
  \bibnamefont {den Hertog}}, \ and\ \bibinfo {author} {\bibfnamefont {Michel
  J.~P.}\ \bibnamefont {Gingras}},\ }\bibfield  {title} {\enquote {\bibinfo
  {title} {Long-range order at low temperatures in dipolar spin ice},}\ }\href
  {\doibase 10.1103/PhysRevLett.87.067203} {\bibfield  {journal} {\bibinfo
  {journal} {Phys. Rev. Lett.}\ }\textbf {\bibinfo {volume} {87}},\ \bibinfo
  {pages} {067203} (\bibinfo {year} {2001})}\BibitemShut {NoStop}%
\bibitem [{\citenamefont {Brooks-Bartlett}\ \emph {et~al.}(2014)\citenamefont
  {Brooks-Bartlett}, \citenamefont {Banks}, \citenamefont {Jaubert},
  \citenamefont {Harman-Clarke},\ and\ \citenamefont
  {Holdsworth}}]{brooks2014magnetic}%
  \BibitemOpen
  \bibfield  {author} {\bibinfo {author} {\bibfnamefont {M.~E.}\ \bibnamefont
  {Brooks-Bartlett}}, \bibinfo {author} {\bibfnamefont {S.~T.}\ \bibnamefont
  {Banks}}, \bibinfo {author} {\bibfnamefont {L.~D.~C.}\ \bibnamefont
  {Jaubert}}, \bibinfo {author} {\bibfnamefont {A.}~\bibnamefont
  {Harman-Clarke}}, \ and\ \bibinfo {author} {\bibfnamefont {P.~C.~W.}\
  \bibnamefont {Holdsworth}},\ }\bibfield  {title} {\enquote {\bibinfo {title}
  {Magnetic-moment fragmentation and monopole crystallization},}\ }\href
  {\doibase 10.1103/PhysRevX.4.011007} {\bibfield  {journal} {\bibinfo
  {journal} {Phys. Rev. X}\ }\textbf {\bibinfo {volume} {4}},\ \bibinfo {pages}
  {011007} (\bibinfo {year} {2014})}\BibitemShut {NoStop}%
\bibitem [{\citenamefont {Lhotel}\ \emph {et~al.}(2020)\citenamefont {Lhotel},
  \citenamefont {Jaubert},\ and\ \citenamefont
  {Holdsworth}}]{lhotel2020fragmentation}%
  \BibitemOpen
  \bibfield  {author} {\bibinfo {author} {\bibfnamefont {Elsa}\ \bibnamefont
  {Lhotel}}, \bibinfo {author} {\bibfnamefont {Ludovic~DC}\ \bibnamefont
  {Jaubert}}, \ and\ \bibinfo {author} {\bibfnamefont {Peter~CW}\ \bibnamefont
  {Holdsworth}},\ }\bibfield  {title} {\enquote {\bibinfo {title}
  {Fragmentation in frustrated magnets: A review},}\ }\href@noop {} {\bibfield
  {journal} {\bibinfo  {journal} {Journal of Low Temperature Physics}\ ,\
  \bibinfo {pages} {1--28}} (\bibinfo {year} {2020})}\BibitemShut {NoStop}%
\bibitem [{\citenamefont {Kaiser}\ \emph {et~al.}(2018)\citenamefont {Kaiser},
  \citenamefont {Bloxsom}, \citenamefont {Bovo}, \citenamefont {Bramwell},
  \citenamefont {Holdsworth},\ and\ \citenamefont {Moessner}}]{Kaiser2018}%
  \BibitemOpen
  \bibfield  {author} {\bibinfo {author} {\bibfnamefont {V.}~\bibnamefont
  {Kaiser}}, \bibinfo {author} {\bibfnamefont {J.}~\bibnamefont {Bloxsom}},
  \bibinfo {author} {\bibfnamefont {L.}~\bibnamefont {Bovo}}, \bibinfo {author}
  {\bibfnamefont {S.~T.}\ \bibnamefont {Bramwell}}, \bibinfo {author}
  {\bibfnamefont {P.~C.~W.}\ \bibnamefont {Holdsworth}}, \ and\ \bibinfo
  {author} {\bibfnamefont {R.}~\bibnamefont {Moessner}},\ }\bibfield  {title}
  {\enquote {\bibinfo {title} {Emergent electrochemistry in spin ice:
  Debye-h\"uckel theory and beyond},}\ }\href {\doibase
  10.1103/PhysRevB.98.144413} {\bibfield  {journal} {\bibinfo  {journal} {Phys.
  Rev. B}\ }\textbf {\bibinfo {volume} {98}},\ \bibinfo {pages} {144413}
  (\bibinfo {year} {2018})}\BibitemShut {NoStop}%
\bibitem [{\citenamefont {Castelnovo}\ \emph {et~al.}(2011)\citenamefont
  {Castelnovo}, \citenamefont {Moessner},\ and\ \citenamefont
  {Sondhi}}]{Castelnovo2011}%
  \BibitemOpen
  \bibfield  {author} {\bibinfo {author} {\bibfnamefont {C.}~\bibnamefont
  {Castelnovo}}, \bibinfo {author} {\bibfnamefont {R.}~\bibnamefont
  {Moessner}}, \ and\ \bibinfo {author} {\bibfnamefont {S.~L.}\ \bibnamefont
  {Sondhi}},\ }\bibfield  {title} {\enquote {\bibinfo {title}
  {{Debye-H\"{u}ckel} theory for spin ice at low temperature},}\ }\href
  {\doibase 10.1103/PhysRevB.84.144435} {\bibfield  {journal} {\bibinfo
  {journal} {Physical Review B}\ }\textbf {\bibinfo {volume} {84}},\ \bibinfo
  {pages} {144435} (\bibinfo {year} {2011})}\BibitemShut {NoStop}%
\bibitem [{\citenamefont {Raban}\ \emph {et~al.}(2019)\citenamefont {Raban},
  \citenamefont {Suen}, \citenamefont {Berthier},\ and\ \citenamefont
  {Holdsworth}}]{raban2019multiple}%
  \BibitemOpen
  \bibfield  {author} {\bibinfo {author} {\bibfnamefont {V.}~\bibnamefont
  {Raban}}, \bibinfo {author} {\bibfnamefont {C.~T.}\ \bibnamefont {Suen}},
  \bibinfo {author} {\bibfnamefont {L.}~\bibnamefont {Berthier}}, \ and\
  \bibinfo {author} {\bibfnamefont {P.~C.~W.}\ \bibnamefont {Holdsworth}},\
  }\bibfield  {title} {\enquote {\bibinfo {title} {Multiple symmetry sustaining
  phase transitions in spin ice},}\ }\href {\doibase
  10.1103/PhysRevB.99.224425} {\bibfield  {journal} {\bibinfo  {journal} {Phys.
  Rev. B}\ }\textbf {\bibinfo {volume} {99}},\ \bibinfo {pages} {224425}
  (\bibinfo {year} {2019})}\BibitemShut {NoStop}%
\bibitem [{\citenamefont {Zhou}\ \emph {et~al.}(2012)\citenamefont {Zhou},
  \citenamefont {Cheng}, \citenamefont {Hallas}, \citenamefont {Wiebe},
  \citenamefont {Li}, \citenamefont {Balicas}, \citenamefont {Zhou},
  \citenamefont {Goodenough}, \citenamefont {Gardner},\ and\ \citenamefont
  {Choi}}]{zhou2012parameters}%
  \BibitemOpen
  \bibfield  {author} {\bibinfo {author} {\bibfnamefont {H.~D.}\ \bibnamefont
  {Zhou}}, \bibinfo {author} {\bibfnamefont {J.~G.}\ \bibnamefont {Cheng}},
  \bibinfo {author} {\bibfnamefont {A.~M.}\ \bibnamefont {Hallas}}, \bibinfo
  {author} {\bibfnamefont {C.~R.}\ \bibnamefont {Wiebe}}, \bibinfo {author}
  {\bibfnamefont {G.}~\bibnamefont {Li}}, \bibinfo {author} {\bibfnamefont
  {L.}~\bibnamefont {Balicas}}, \bibinfo {author} {\bibfnamefont {J.~S.}\
  \bibnamefont {Zhou}}, \bibinfo {author} {\bibfnamefont {J.~B.}\ \bibnamefont
  {Goodenough}}, \bibinfo {author} {\bibfnamefont {J.~S.}\ \bibnamefont
  {Gardner}}, \ and\ \bibinfo {author} {\bibfnamefont {E.~S.}\ \bibnamefont
  {Choi}},\ }\bibfield  {title} {\enquote {\bibinfo {title} {Chemical pressure
  effects on pyrochlore spin ice},}\ }\href {\doibase
  10.1103/PhysRevLett.108.207206} {\bibfield  {journal} {\bibinfo  {journal}
  {Phys. Rev. Lett.}\ }\textbf {\bibinfo {volume} {108}},\ \bibinfo {pages}
  {207206} (\bibinfo {year} {2012})}\BibitemShut {NoStop}%
\bibitem [{\citenamefont {Yavors'kii}\ \emph {et~al.}(2008)\citenamefont
  {Yavors'kii}, \citenamefont {Fennell}, \citenamefont {Gingras},\ and\
  \citenamefont {Bramwell}}]{Yavorskii2008}%
  \BibitemOpen
  \bibfield  {author} {\bibinfo {author} {\bibfnamefont {Taras}\ \bibnamefont
  {Yavors'kii}}, \bibinfo {author} {\bibfnamefont {Tom}\ \bibnamefont
  {Fennell}}, \bibinfo {author} {\bibfnamefont {Michel J.~P.}\ \bibnamefont
  {Gingras}}, \ and\ \bibinfo {author} {\bibfnamefont {Steven~T.}\ \bibnamefont
  {Bramwell}},\ }\bibfield  {title} {\enquote {\bibinfo {title}
  {${\mathrm{dy}}_{2}{\mathrm{ti}}_{2}{\mathrm{o}}_{7}$},}\ }\href {\doibase
  10.1103/PhysRevLett.101.037204} {\bibfield  {journal} {\bibinfo  {journal}
  {Phys. Rev. Lett.}\ }\textbf {\bibinfo {volume} {101}},\ \bibinfo {pages}
  {037204} (\bibinfo {year} {2008})}\BibitemShut {NoStop}%
\bibitem [{\citenamefont {Bramwell}(2017)}]{bramwell2017harmonic}%
  \BibitemOpen
  \bibfield  {author} {\bibinfo {author} {\bibfnamefont {Steven~T}\
  \bibnamefont {Bramwell}},\ }\bibfield  {title} {\enquote {\bibinfo {title}
  {Harmonic phase in polar liquids and spin ice},}\ }\href@noop {} {\bibfield
  {journal} {\bibinfo  {journal} {Nature communications}\ }\textbf {\bibinfo
  {volume} {8}},\ \bibinfo {pages} {1--9} (\bibinfo {year} {2017})}\BibitemShut
  {NoStop}%
\bibitem [{\citenamefont {Huse}\ \emph {et~al.}(2003)\citenamefont {Huse},
  \citenamefont {Krauth}, \citenamefont {Moessner},\ and\ \citenamefont
  {Sondhi}}]{PhysRevLett.91.167004}%
  \BibitemOpen
  \bibfield  {author} {\bibinfo {author} {\bibfnamefont {David~A.}\
  \bibnamefont {Huse}}, \bibinfo {author} {\bibfnamefont {Werner}\ \bibnamefont
  {Krauth}}, \bibinfo {author} {\bibfnamefont {R.}~\bibnamefont {Moessner}}, \
  and\ \bibinfo {author} {\bibfnamefont {S.~L.}\ \bibnamefont {Sondhi}},\
  }\bibfield  {title} {\enquote {\bibinfo {title} {Coulomb and liquid dimer
  models in three dimensions},}\ }\href {\doibase
  10.1103/PhysRevLett.91.167004} {\bibfield  {journal} {\bibinfo  {journal}
  {Phys. Rev. Lett.}\ }\textbf {\bibinfo {volume} {91}},\ \bibinfo {pages}
  {167004} (\bibinfo {year} {2003})}\BibitemShut {NoStop}%
\bibitem [{\citenamefont {Isakov}\ \emph {et~al.}(2004)\citenamefont {Isakov},
  \citenamefont {Gregor}, \citenamefont {Moessner},\ and\ \citenamefont
  {Sondhi}}]{Isakov2004}%
  \BibitemOpen
  \bibfield  {author} {\bibinfo {author} {\bibfnamefont {S~V}\ \bibnamefont
  {Isakov}}, \bibinfo {author} {\bibfnamefont {K}~\bibnamefont {Gregor}},
  \bibinfo {author} {\bibfnamefont {R}~\bibnamefont {Moessner}}, \ and\
  \bibinfo {author} {\bibfnamefont {S~L}\ \bibnamefont {Sondhi}},\ }\bibfield
  {title} {\enquote {\bibinfo {title} {{Dipolar spin correlations in classical
  pyrochlore magnets}},}\ }\href {\doibase ARTN 167204} {\bibfield  {journal}
  {\bibinfo  {journal} {Physical Review Letters}\ }\textbf {\bibinfo {volume}
  {93}},\ \bibinfo {pages} {167204} (\bibinfo {year} {2004})}\BibitemShut
  {NoStop}%
\bibitem [{\citenamefont {Paulsen}\ \emph {et~al.}(2014)\citenamefont
  {Paulsen}, \citenamefont {Jackson}, \citenamefont {Lhotel}, \citenamefont
  {Canals}, \citenamefont {Prabhakaran}, \citenamefont {Matsuhira},
  \citenamefont {Giblin},\ and\ \citenamefont {Bramwell}}]{Paulsen2014}%
  \BibitemOpen
  \bibfield  {author} {\bibinfo {author} {\bibfnamefont {C.}~\bibnamefont
  {Paulsen}}, \bibinfo {author} {\bibfnamefont {M.~J.}\ \bibnamefont
  {Jackson}}, \bibinfo {author} {\bibfnamefont {E.}~\bibnamefont {Lhotel}},
  \bibinfo {author} {\bibfnamefont {B.}~\bibnamefont {Canals}}, \bibinfo
  {author} {\bibfnamefont {D.}~\bibnamefont {Prabhakaran}}, \bibinfo {author}
  {\bibfnamefont {K.}~\bibnamefont {Matsuhira}}, \bibinfo {author}
  {\bibfnamefont {S.~R.}\ \bibnamefont {Giblin}}, \ and\ \bibinfo {author}
  {\bibfnamefont {S.~T.}\ \bibnamefont {Bramwell}},\ }\bibfield  {title}
  {\enquote {\bibinfo {title} {Far-from-equilibrium monopole dynamics in spin
  ice},}\ }\href {\doibase 10.1038/nphys2847} {\bibfield  {journal} {\bibinfo
  {journal} {Nature Physics}\ }\textbf {\bibinfo {volume} {10}},\ \bibinfo
  {pages} {135--139} (\bibinfo {year} {2014})}\BibitemShut {NoStop}%
\bibitem [{\citenamefont {Sollich}\ \emph {et~al.}(2002)\citenamefont
  {Sollich}, \citenamefont {Fielding},\ and\ \citenamefont
  {Mayer}}]{sollich2002fluctuation}%
  \BibitemOpen
  \bibfield  {author} {\bibinfo {author} {\bibfnamefont {Peter}\ \bibnamefont
  {Sollich}}, \bibinfo {author} {\bibfnamefont {Suzanne}\ \bibnamefont
  {Fielding}}, \ and\ \bibinfo {author} {\bibfnamefont {Peter}\ \bibnamefont
  {Mayer}},\ }\bibfield  {title} {\enquote {\bibinfo {title}
  {Fluctuation-dissipation relations and effective temperatures in simple
  non-mean field systems},}\ }\href@noop {} {\bibfield  {journal} {\bibinfo
  {journal} {Journal of Physics: Condensed Matter}\ }\textbf {\bibinfo {volume}
  {14}},\ \bibinfo {pages} {1683} (\bibinfo {year} {2002})}\BibitemShut
  {NoStop}%
\bibitem [{\citenamefont {Chatelain}(2003)}]{chatelain2003far}%
  \BibitemOpen
  \bibfield  {author} {\bibinfo {author} {\bibfnamefont {Christophe}\
  \bibnamefont {Chatelain}},\ }\bibfield  {title} {\enquote {\bibinfo {title}
  {A far-from-equilibrium fluctuation--dissipation relation for an
  ising--glauber-like model},}\ }\href@noop {} {\bibfield  {journal} {\bibinfo
  {journal} {Journal of Physics A: Mathematical and General}\ }\textbf
  {\bibinfo {volume} {36}},\ \bibinfo {pages} {10739} (\bibinfo {year}
  {2003})}\BibitemShut {NoStop}%
\bibitem [{\citenamefont {Sciortino}\ and\ \citenamefont
  {Tartaglia}(2001)}]{sciortino2001extension}%
  \BibitemOpen
  \bibfield  {author} {\bibinfo {author} {\bibfnamefont {Francesco}\
  \bibnamefont {Sciortino}}\ and\ \bibinfo {author} {\bibfnamefont {Piero}\
  \bibnamefont {Tartaglia}},\ }\bibfield  {title} {\enquote {\bibinfo {title}
  {Extension of the fluctuation-dissipation theorem to the physical aging of a
  model glass-forming liquid},}\ }\href {\doibase 10.1103/PhysRevLett.86.107}
  {\bibfield  {journal} {\bibinfo  {journal} {Phys. Rev. Lett.}\ }\textbf
  {\bibinfo {volume} {86}},\ \bibinfo {pages} {107--110} (\bibinfo {year}
  {2001})}\BibitemShut {NoStop}%
\bibitem [{\citenamefont {Cugliandolo}\ and\ \citenamefont
  {Kurchan}(1999)}]{cugliandolo1999thermal}%
  \BibitemOpen
  \bibfield  {author} {\bibinfo {author} {\bibfnamefont {Leticia}\ \bibnamefont
  {Cugliandolo}}\ and\ \bibinfo {author} {\bibfnamefont {Jorge}\ \bibnamefont
  {Kurchan}},\ }\bibfield  {title} {\enquote {\bibinfo {title} {Thermal
  properties of slow dynamics},}\ }\href@noop {} {\bibfield  {journal}
  {\bibinfo  {journal} {Physica A: Statistical Mechanics and its Applications}\
  }\textbf {\bibinfo {volume} {263}},\ \bibinfo {pages} {242--251} (\bibinfo
  {year} {1999})}\BibitemShut {NoStop}%
\bibitem [{\citenamefont {Cugliandolo}\ and\ \citenamefont
  {Kurchan}(1994)}]{cugliandolo1994out}%
  \BibitemOpen
  \bibfield  {author} {\bibinfo {author} {\bibfnamefont {Leticia~F}\
  \bibnamefont {Cugliandolo}}\ and\ \bibinfo {author} {\bibfnamefont {Jorge}\
  \bibnamefont {Kurchan}},\ }\bibfield  {title} {\enquote {\bibinfo {title} {On
  the out-of-equilibrium relaxation of the sherrington-kirkpatrick model},}\
  }\href@noop {} {\bibfield  {journal} {\bibinfo  {journal} {Journal of Physics
  A: Mathematical and General}\ }\textbf {\bibinfo {volume} {27}},\ \bibinfo
  {pages} {5749} (\bibinfo {year} {1994})}\BibitemShut {NoStop}%
\bibitem [{\citenamefont {Franz}\ \emph {et~al.}(1998)\citenamefont {Franz},
  \citenamefont {M\'ezard}, \citenamefont {Parisi},\ and\ \citenamefont
  {Peliti}}]{PhysRevLett.81.1758}%
  \BibitemOpen
  \bibfield  {author} {\bibinfo {author} {\bibfnamefont {Silvio}\ \bibnamefont
  {Franz}}, \bibinfo {author} {\bibfnamefont {Marc}\ \bibnamefont {M\'ezard}},
  \bibinfo {author} {\bibfnamefont {Giorgio}\ \bibnamefont {Parisi}}, \ and\
  \bibinfo {author} {\bibfnamefont {Luca}\ \bibnamefont {Peliti}},\ }\bibfield
  {title} {\enquote {\bibinfo {title} {Measuring equilibrium properties in
  aging systems},}\ }\href {\doibase 10.1103/PhysRevLett.81.1758} {\bibfield
  {journal} {\bibinfo  {journal} {Phys. Rev. Lett.}\ }\textbf {\bibinfo
  {volume} {81}},\ \bibinfo {pages} {1758--1761} (\bibinfo {year}
  {1998})}\BibitemShut {NoStop}%
\bibitem [{\citenamefont {Berthier}\ \emph {et~al.}(2000)\citenamefont
  {Berthier}, \citenamefont {Barrat},\ and\ \citenamefont
  {Kurchan}}]{PhysRevE.63.016105}%
  \BibitemOpen
  \bibfield  {author} {\bibinfo {author} {\bibfnamefont {Ludovic}\ \bibnamefont
  {Berthier}}, \bibinfo {author} {\bibfnamefont {Jean-Louis}\ \bibnamefont
  {Barrat}}, \ and\ \bibinfo {author} {\bibfnamefont {Jorge}\ \bibnamefont
  {Kurchan}},\ }\bibfield  {title} {\enquote {\bibinfo {title} {Dynamic
  ultrametricity in spin glasses},}\ }\href {\doibase
  10.1103/PhysRevE.63.016105} {\bibfield  {journal} {\bibinfo  {journal} {Phys.
  Rev. E}\ }\textbf {\bibinfo {volume} {63}},\ \bibinfo {pages} {016105}
  (\bibinfo {year} {2000})}\BibitemShut {NoStop}%
\bibitem [{\citenamefont {Kurchan}(2021)}]{kurchan2021time}%
  \BibitemOpen
  \bibfield  {author} {\bibinfo {author} {\bibfnamefont {Jorge}\ \bibnamefont
  {Kurchan}},\ }\bibfield  {title} {\enquote {\bibinfo {title}
  {Time-reparametrization invariances, multithermalization and the parisi
  scheme},}\ }\href@noop {} {\bibfield  {journal} {\bibinfo  {journal} {arXiv
  preprint arXiv:2101.12702}\ } (\bibinfo {year} {2021})}\BibitemShut {NoStop}%
\bibitem [{\citenamefont {H{\'e}risson}\ and\ \citenamefont
  {Ocio}(2004)}]{herisson2004off}%
  \BibitemOpen
  \bibfield  {author} {\bibinfo {author} {\bibfnamefont {Didier}\ \bibnamefont
  {H{\'e}risson}}\ and\ \bibinfo {author} {\bibfnamefont {Miguel}\ \bibnamefont
  {Ocio}},\ }\bibfield  {title} {\enquote {\bibinfo {title} {Off-equilibrium
  fluctuation-dissipation relation in a spin glass},}\ }\href@noop {}
  {\bibfield  {journal} {\bibinfo  {journal} {The European Physical Journal
  B-Condensed Matter and Complex Systems}\ }\textbf {\bibinfo {volume} {40}},\
  \bibinfo {pages} {283--294} (\bibinfo {year} {2004})}\BibitemShut {NoStop}%
\bibitem [{\citenamefont {Refregier}\ \emph {et~al.}(1987)\citenamefont
  {Refregier}, \citenamefont {Alba}, \citenamefont {Hammann},\ and\
  \citenamefont {Ocio}}]{refregier1987dynamic}%
  \BibitemOpen
  \bibfield  {author} {\bibinfo {author} {\bibfnamefont {Ph}~\bibnamefont
  {Refregier}}, \bibinfo {author} {\bibfnamefont {M}~\bibnamefont {Alba}},
  \bibinfo {author} {\bibfnamefont {J}~\bibnamefont {Hammann}}, \ and\ \bibinfo
  {author} {\bibfnamefont {M}~\bibnamefont {Ocio}},\ }\bibfield  {title}
  {\enquote {\bibinfo {title} {Dynamic behaviour of the insulating spin glass
  csnifef6},}\ }\href@noop {} {\bibfield  {journal} {\bibinfo  {journal}
  {Journal of Physics C: Solid State Physics}\ }\textbf {\bibinfo {volume}
  {20}},\ \bibinfo {pages} {5545} (\bibinfo {year} {1987})}\BibitemShut
  {NoStop}%
\bibitem [{\citenamefont {Bellon}\ and\ \citenamefont
  {Ciliberto}(2002)}]{bellon2002experimental}%
  \BibitemOpen
  \bibfield  {author} {\bibinfo {author} {\bibfnamefont {Ludovic}\ \bibnamefont
  {Bellon}}\ and\ \bibinfo {author} {\bibfnamefont {Sergio}\ \bibnamefont
  {Ciliberto}},\ }\bibfield  {title} {\enquote {\bibinfo {title} {Experimental
  study of the fluctuation dissipation relation during an aging process},}\
  }\href@noop {} {\bibfield  {journal} {\bibinfo  {journal} {Physica D:
  Nonlinear Phenomena}\ }\textbf {\bibinfo {volume} {168}},\ \bibinfo {pages}
  {325--335} (\bibinfo {year} {2002})}\BibitemShut {NoStop}%
\bibitem [{\citenamefont {Buisson}\ \emph {et~al.}(2003)\citenamefont
  {Buisson}, \citenamefont {Bellon},\ and\ \citenamefont
  {Ciliberto}}]{buisson2003intermittency}%
  \BibitemOpen
  \bibfield  {author} {\bibinfo {author} {\bibfnamefont {Lionel}\ \bibnamefont
  {Buisson}}, \bibinfo {author} {\bibfnamefont {Ludovic}\ \bibnamefont
  {Bellon}}, \ and\ \bibinfo {author} {\bibfnamefont {Sergio}\ \bibnamefont
  {Ciliberto}},\ }\bibfield  {title} {\enquote {\bibinfo {title} {Intermittency
  in ageing},}\ }\href@noop {} {\bibfield  {journal} {\bibinfo  {journal}
  {Journal of Physics: Condensed Matter}\ }\textbf {\bibinfo {volume} {15}},\
  \bibinfo {pages} {S1163} (\bibinfo {year} {2003})}\BibitemShut {NoStop}%
\bibitem [{\citenamefont {Abou}\ and\ \citenamefont
  {Gallet}(2004)}]{PhysRevLett.93.160603}%
  \BibitemOpen
  \bibfield  {author} {\bibinfo {author} {\bibfnamefont {B\'ereng\`ere}\
  \bibnamefont {Abou}}\ and\ \bibinfo {author} {\bibfnamefont {Francois}\
  \bibnamefont {Gallet}},\ }\bibfield  {title} {\enquote {\bibinfo {title}
  {Probing a nonequilibrium einstein relation in an aging colloidal glass},}\
  }\href {\doibase 10.1103/PhysRevLett.93.160603} {\bibfield  {journal}
  {\bibinfo  {journal} {Phys. Rev. Lett.}\ }\textbf {\bibinfo {volume} {93}},\
  \bibinfo {pages} {160603} (\bibinfo {year} {2004})}\BibitemShut {NoStop}%
\bibitem [{\citenamefont {Dusad}\ \emph {et~al.}(2019)\citenamefont {Dusad},
  \citenamefont {Kirschner}, \citenamefont {Hoke}, \citenamefont {Roberts},
  \citenamefont {Eyal}, \citenamefont {Flicker}, \citenamefont {Luke},
  \citenamefont {Blundell},\ and\ \citenamefont {Davis}}]{dusad2019magnetic}%
  \BibitemOpen
  \bibfield  {author} {\bibinfo {author} {\bibfnamefont {Ritika}\ \bibnamefont
  {Dusad}}, \bibinfo {author} {\bibfnamefont {Franziska~KK}\ \bibnamefont
  {Kirschner}}, \bibinfo {author} {\bibfnamefont {Jesse~C}\ \bibnamefont
  {Hoke}}, \bibinfo {author} {\bibfnamefont {Benjamin~R}\ \bibnamefont
  {Roberts}}, \bibinfo {author} {\bibfnamefont {Anna}\ \bibnamefont {Eyal}},
  \bibinfo {author} {\bibfnamefont {Felix}\ \bibnamefont {Flicker}}, \bibinfo
  {author} {\bibfnamefont {Graeme~M}\ \bibnamefont {Luke}}, \bibinfo {author}
  {\bibfnamefont {Stephen~J}\ \bibnamefont {Blundell}}, \ and\ \bibinfo
  {author} {\bibfnamefont {JC~S{\'e}amus}\ \bibnamefont {Davis}},\ }\bibfield
  {title} {\enquote {\bibinfo {title} {Magnetic monopole noise},}\ }\href@noop
  {} {\bibfield  {journal} {\bibinfo  {journal} {Nature}\ }\textbf {\bibinfo
  {volume} {571}},\ \bibinfo {pages} {234--239} (\bibinfo {year}
  {2019})}\BibitemShut {NoStop}%
\bibitem [{\citenamefont {Bovo}\ \emph {et~al.}(2013)\citenamefont {Bovo},
  \citenamefont {Bloxsom}, \citenamefont {Prabhakaran}, \citenamefont
  {Aeppli},\ and\ \citenamefont {Bramwell}}]{Bovo2013}%
  \BibitemOpen
  \bibfield  {author} {\bibinfo {author} {\bibfnamefont {L.}~\bibnamefont
  {Bovo}}, \bibinfo {author} {\bibfnamefont {{J.A.}}\ \bibnamefont {Bloxsom}},
  \bibinfo {author} {\bibfnamefont {D.}~\bibnamefont {Prabhakaran}}, \bibinfo
  {author} {\bibfnamefont {G.}~\bibnamefont {Aeppli}}, \ and\ \bibinfo {author}
  {\bibfnamefont {{S.T.}}\ \bibnamefont {Bramwell}},\ }\bibfield  {title}
  {\enquote {\bibinfo {title} {Brownian motion and quantum dynamics of magnetic
  monopoles in spin ice},}\ }\href {\doibase 10.1038/ncomms2551} {\bibfield
  {journal} {\bibinfo  {journal} {Nature Communications}\ }\textbf {\bibinfo
  {volume} {4}},\ \bibinfo {pages} {1535} (\bibinfo {year} {2013})}\BibitemShut
  {NoStop}%
\bibitem [{\citenamefont {Blundell}(2012)}]{Blundell2012}%
  \BibitemOpen
  \bibfield  {author} {\bibinfo {author} {\bibfnamefont {Stephen~J.}\
  \bibnamefont {Blundell}},\ }\bibfield  {title} {\enquote {\bibinfo {title}
  {Monopoles, magnetricity, and the stray field from spin ice},}\ }\href
  {\doibase 10.1103/PhysRevLett.108.147601} {\bibfield  {journal} {\bibinfo
  {journal} {Physical Review Letters}\ }\textbf {\bibinfo {volume} {108}},\
  \bibinfo {pages} {147601} (\bibinfo {year} {2012})}\BibitemShut {NoStop}%
\bibitem [{\citenamefont {Giblin}\ \emph {et~al.}(2018)\citenamefont {Giblin},
  \citenamefont {Twengstr\"om}, \citenamefont {Bovo}, \citenamefont {Ruminy},
  \citenamefont {Bartkowiak}, \citenamefont {Manuel}, \citenamefont {Andresen},
  \citenamefont {Prabhakaran}, \citenamefont {Balakrishnan}, \citenamefont
  {Pomjakushina}, \citenamefont {Paulsen}, \citenamefont {Lhotel},
  \citenamefont {Keller}, \citenamefont {Frontzek}, \citenamefont {Capelli},
  \citenamefont {Zaharko}, \citenamefont {McClarty}, \citenamefont {Bramwell},
  \citenamefont {Henelius},\ and\ \citenamefont
  {Fennell}}]{giblin2018specificheat}%
  \BibitemOpen
  \bibfield  {author} {\bibinfo {author} {\bibfnamefont {S.~R.}\ \bibnamefont
  {Giblin}}, \bibinfo {author} {\bibfnamefont {M.}~\bibnamefont
  {Twengstr\"om}}, \bibinfo {author} {\bibfnamefont {L.}~\bibnamefont {Bovo}},
  \bibinfo {author} {\bibfnamefont {M.}~\bibnamefont {Ruminy}}, \bibinfo
  {author} {\bibfnamefont {M.}~\bibnamefont {Bartkowiak}}, \bibinfo {author}
  {\bibfnamefont {P.}~\bibnamefont {Manuel}}, \bibinfo {author} {\bibfnamefont
  {J.~C.}\ \bibnamefont {Andresen}}, \bibinfo {author} {\bibfnamefont
  {D.}~\bibnamefont {Prabhakaran}}, \bibinfo {author} {\bibfnamefont
  {G.}~\bibnamefont {Balakrishnan}}, \bibinfo {author} {\bibfnamefont
  {E.}~\bibnamefont {Pomjakushina}}, \bibinfo {author} {\bibfnamefont
  {C.}~\bibnamefont {Paulsen}}, \bibinfo {author} {\bibfnamefont
  {E.}~\bibnamefont {Lhotel}}, \bibinfo {author} {\bibfnamefont
  {L.}~\bibnamefont {Keller}}, \bibinfo {author} {\bibfnamefont
  {M.}~\bibnamefont {Frontzek}}, \bibinfo {author} {\bibfnamefont {S.~C.}\
  \bibnamefont {Capelli}}, \bibinfo {author} {\bibfnamefont {O.}~\bibnamefont
  {Zaharko}}, \bibinfo {author} {\bibfnamefont {P.~A.}\ \bibnamefont
  {McClarty}}, \bibinfo {author} {\bibfnamefont {S.~T.}\ \bibnamefont
  {Bramwell}}, \bibinfo {author} {\bibfnamefont {P.}~\bibnamefont {Henelius}},
  \ and\ \bibinfo {author} {\bibfnamefont {T.}~\bibnamefont {Fennell}},\
  }\bibfield  {title} {\enquote {\bibinfo {title} {Pauling entropy,
  metastability, and equilibrium in
  ${\mathrm{dy}}_{2}{\mathrm{ti}}_{2}{\mathrm{o}}_{7}$ spin ice},}\ }\href
  {\doibase 10.1103/PhysRevLett.121.067202} {\bibfield  {journal} {\bibinfo
  {journal} {Phys. Rev. Lett.}\ }\textbf {\bibinfo {volume} {121}},\ \bibinfo
  {pages} {067202} (\bibinfo {year} {2018})}\BibitemShut {NoStop}%
\bibitem [{\citenamefont {Jaubert}\ \emph {et~al.}(2013)\citenamefont
  {Jaubert}, \citenamefont {Harris}, \citenamefont {Fennell}, \citenamefont
  {Melko}, \citenamefont {Bramwell},\ and\ \citenamefont
  {Holdsworth}}]{Jaubert2013}%
  \BibitemOpen
  \bibfield  {author} {\bibinfo {author} {\bibfnamefont {L.D.C.}\ \bibnamefont
  {Jaubert}}, \bibinfo {author} {\bibfnamefont {M.J.}\ \bibnamefont {Harris}},
  \bibinfo {author} {\bibfnamefont {T.}~\bibnamefont {Fennell}}, \bibinfo
  {author} {\bibfnamefont {R.G.}\ \bibnamefont {Melko}}, \bibinfo {author}
  {\bibfnamefont {S.T.}\ \bibnamefont {Bramwell}}, \ and\ \bibinfo {author}
  {\bibfnamefont {P.C.W.}\ \bibnamefont {Holdsworth}},\ }\bibfield  {title}
  {\enquote {\bibinfo {title} {{Topological-Sector} fluctuations and
  {Curie-Law} crossover in spin ice},}\ }\href {\doibase
  10.1103/PhysRevX.3.011014} {\bibfield  {journal} {\bibinfo  {journal}
  {Physical Review X}\ }\textbf {\bibinfo {volume} {3}},\ \bibinfo {pages}
  {011014} (\bibinfo {year} {2013})}\BibitemShut {NoStop}%
\bibitem [{\citenamefont {Ninarello}\ \emph {et~al.}(2014)\citenamefont
  {Ninarello}, \citenamefont {Gnan},\ and\ \citenamefont
  {Sciortino}}]{doi:10.1063/1.4901526}%
  \BibitemOpen
  \bibfield  {author} {\bibinfo {author} {\bibfnamefont {Andrea~Saverio}\
  \bibnamefont {Ninarello}}, \bibinfo {author} {\bibfnamefont {Nicoletta}\
  \bibnamefont {Gnan}}, \ and\ \bibinfo {author} {\bibfnamefont {Francesco}\
  \bibnamefont {Sciortino}},\ }\bibfield  {title} {\enquote {\bibinfo {title}
  {Observable-dependence of the effective temperature in off-equilibrium
  diatomic molecular liquids},}\ }\href {\doibase 10.1063/1.4901526} {\bibfield
   {journal} {\bibinfo  {journal} {The Journal of Chemical Physics}\ }\textbf
  {\bibinfo {volume} {141}},\ \bibinfo {pages} {194507} (\bibinfo {year}
  {2014})}\BibitemShut {NoStop}%
\bibitem [{\citenamefont {Russo}\ and\ \citenamefont
  {Sciortino}(2010)}]{PhysRevLett.104.195701}%
  \BibitemOpen
  \bibfield  {author} {\bibinfo {author} {\bibfnamefont {John}\ \bibnamefont
  {Russo}}\ and\ \bibinfo {author} {\bibfnamefont {Francesco}\ \bibnamefont
  {Sciortino}},\ }\bibfield  {title} {\enquote {\bibinfo {title} {How do
  self-assembling polymers and gels age compared to glasses?}}\ }\href
  {\doibase 10.1103/PhysRevLett.104.195701} {\bibfield  {journal} {\bibinfo
  {journal} {Phys. Rev. Lett.}\ }\textbf {\bibinfo {volume} {104}},\ \bibinfo
  {pages} {195701} (\bibinfo {year} {2010})}\BibitemShut {NoStop}%
\bibitem [{\citenamefont {Gnan}\ \emph {et~al.}(2013)\citenamefont {Gnan},
  \citenamefont {Maggi}, \citenamefont {Parisi},\ and\ \citenamefont
  {Sciortino}}]{PhysRevLett.110.035701}%
  \BibitemOpen
  \bibfield  {author} {\bibinfo {author} {\bibfnamefont {Nicoletta}\
  \bibnamefont {Gnan}}, \bibinfo {author} {\bibfnamefont {Claudio}\
  \bibnamefont {Maggi}}, \bibinfo {author} {\bibfnamefont {Giorgio}\
  \bibnamefont {Parisi}}, \ and\ \bibinfo {author} {\bibfnamefont {Francesco}\
  \bibnamefont {Sciortino}},\ }\bibfield  {title} {\enquote {\bibinfo {title}
  {Generalized fluctuation-dissipation relation and effective temperature upon
  heating a deeply supercooled liquid},}\ }\href {\doibase
  10.1103/PhysRevLett.110.035701} {\bibfield  {journal} {\bibinfo  {journal}
  {Phys. Rev. Lett.}\ }\textbf {\bibinfo {volume} {110}},\ \bibinfo {pages}
  {035701} (\bibinfo {year} {2013})}\BibitemShut {NoStop}%
\bibitem [{\citenamefont {Nisoli}\ \emph {et~al.}(2013)\citenamefont {Nisoli},
  \citenamefont {Moessner},\ and\ \citenamefont
  {Schiffer}}]{nisoli-artificial2013}%
  \BibitemOpen
  \bibfield  {author} {\bibinfo {author} {\bibfnamefont {Cristiano}\
  \bibnamefont {Nisoli}}, \bibinfo {author} {\bibfnamefont {Roderich}\
  \bibnamefont {Moessner}}, \ and\ \bibinfo {author} {\bibfnamefont {Peter}\
  \bibnamefont {Schiffer}},\ }\bibfield  {title} {\enquote {\bibinfo {title}
  {Colloquium: Artificial spin ice: Designing and imaging magnetic
  frustration},}\ }\href {\doibase 10.1103/RevModPhys.85.1473} {\bibfield
  {journal} {\bibinfo  {journal} {Rev. Mod. Phys.}\ }\textbf {\bibinfo {volume}
  {85}},\ \bibinfo {pages} {1473--1490} (\bibinfo {year} {2013})}\BibitemShut
  {NoStop}%
\bibitem [{\citenamefont {Skj{\ae}rv{\o}}\ \emph {et~al.}(2020)\citenamefont
  {Skj{\ae}rv{\o}}, \citenamefont {Marrows}, \citenamefont {Stamps},\ and\
  \citenamefont {Heyderman}}]{skjaervo-artifical2020}%
  \BibitemOpen
  \bibfield  {author} {\bibinfo {author} {\bibfnamefont {Sandra~H.}\
  \bibnamefont {Skj{\ae}rv{\o}}}, \bibinfo {author} {\bibfnamefont
  {Christopher~H.}\ \bibnamefont {Marrows}}, \bibinfo {author} {\bibfnamefont
  {Robert~L.}\ \bibnamefont {Stamps}}, \ and\ \bibinfo {author} {\bibfnamefont
  {Laura~J.}\ \bibnamefont {Heyderman}},\ }\bibfield  {title} {\enquote
  {\bibinfo {title} {Advances in artificial spin ice},}\ }\href {\doibase
  10.1038/s42254-019-0118-3} {\bibfield  {journal} {\bibinfo  {journal} {Nature
  Reviews Physics}\ }\textbf {\bibinfo {volume} {2}},\ \bibinfo {pages}
  {13--28} (\bibinfo {year} {2020})}\BibitemShut {NoStop}%
\bibitem [{\citenamefont {Levis}\ and\ \citenamefont
  {Cugliandolo}(2012)}]{levis2012out}%
  \BibitemOpen
  \bibfield  {author} {\bibinfo {author} {\bibfnamefont {Demian}\ \bibnamefont
  {Levis}}\ and\ \bibinfo {author} {\bibfnamefont {Leticia~F}\ \bibnamefont
  {Cugliandolo}},\ }\bibfield  {title} {\enquote {\bibinfo {title}
  {Out-of-equilibrium dynamics in the bidimensional spin-ice model},}\
  }\href@noop {} {\bibfield  {journal} {\bibinfo  {journal} {EPL (Europhysics
  Letters)}\ }\textbf {\bibinfo {volume} {97}},\ \bibinfo {pages} {30002}
  (\bibinfo {year} {2012})}\BibitemShut {NoStop}%
\bibitem [{\citenamefont {Levis}\ and\ \citenamefont
  {Cugliandolo}(2013)}]{levis2013defects}%
  \BibitemOpen
  \bibfield  {author} {\bibinfo {author} {\bibfnamefont {Demian}\ \bibnamefont
  {Levis}}\ and\ \bibinfo {author} {\bibfnamefont {Leticia~F}\ \bibnamefont
  {Cugliandolo}},\ }\bibfield  {title} {\enquote {\bibinfo {title} {Defects
  dynamics following thermal quenches in square spin ice},}\ }\href@noop {}
  {\bibfield  {journal} {\bibinfo  {journal} {Physical Review B}\ }\textbf
  {\bibinfo {volume} {87}},\ \bibinfo {pages} {214302} (\bibinfo {year}
  {2013})}\BibitemShut {NoStop}%
\bibitem [{\citenamefont {Lib\'al}\ \emph {et~al.}(2020)\citenamefont
  {Lib\'al}, \citenamefont {del Campo}, \citenamefont {Nisoli}, \citenamefont
  {Reichhardt},\ and\ \citenamefont {Reichhardt}}]{Libal2020}%
  \BibitemOpen
  \bibfield  {author} {\bibinfo {author} {\bibfnamefont {A.}~\bibnamefont
  {Lib\'al}}, \bibinfo {author} {\bibfnamefont {A.}~\bibnamefont {del Campo}},
  \bibinfo {author} {\bibfnamefont {C.}~\bibnamefont {Nisoli}}, \bibinfo
  {author} {\bibfnamefont {C.}~\bibnamefont {Reichhardt}}, \ and\ \bibinfo
  {author} {\bibfnamefont {C.~J.~O.}\ \bibnamefont {Reichhardt}},\ }\bibfield
  {title} {\enquote {\bibinfo {title} {Quenched dynamics of artificial
  colloidal spin ice},}\ }\href {\doibase 10.1103/PhysRevResearch.2.033433}
  {\bibfield  {journal} {\bibinfo  {journal} {Phys. Rev. Research}\ }\textbf
  {\bibinfo {volume} {2}},\ \bibinfo {pages} {033433} (\bibinfo {year}
  {2020})}\BibitemShut {NoStop}%
\bibitem [{\citenamefont {Perrin}\ \emph {et~al.}(2016)\citenamefont {Perrin},
  \citenamefont {Canals},\ and\ \citenamefont {Rougemaille}}]{Perrin2016}%
  \BibitemOpen
  \bibfield  {author} {\bibinfo {author} {\bibfnamefont {Yann}\ \bibnamefont
  {Perrin}}, \bibinfo {author} {\bibfnamefont {Benjamin}\ \bibnamefont
  {Canals}}, \ and\ \bibinfo {author} {\bibfnamefont {Nicolas}\ \bibnamefont
  {Rougemaille}},\ }\bibfield  {title} {\enquote {\bibinfo {title} {Extensive
  degeneracy, coulomb phase and magnetic monopoles in artificial square ice},}\
  }\href {\doibase 10.1038/nature20155} {\bibfield  {journal} {\bibinfo
  {journal} {Nature}\ }\textbf {\bibinfo {volume} {540}},\ \bibinfo {pages}
  {410--413} (\bibinfo {year} {2016})}\BibitemShut {NoStop}%
\bibitem [{\citenamefont {{\"O}stman}\ \emph {et~al.}(2018)\citenamefont
  {{\"O}stman}, \citenamefont {Stopfel}, \citenamefont {Chioar}, \citenamefont
  {Arnalds}, \citenamefont {Stein}, \citenamefont {Kapaklis},\ and\
  \citenamefont {Hj{\"o}rvarsson}}]{Ostman2018}%
  \BibitemOpen
  \bibfield  {author} {\bibinfo {author} {\bibfnamefont {Erik}\ \bibnamefont
  {{\"O}stman}}, \bibinfo {author} {\bibfnamefont {Henry}\ \bibnamefont
  {Stopfel}}, \bibinfo {author} {\bibfnamefont {Ioan-Augustin}\ \bibnamefont
  {Chioar}}, \bibinfo {author} {\bibfnamefont {Unnar~B.}\ \bibnamefont
  {Arnalds}}, \bibinfo {author} {\bibfnamefont {Aaron}\ \bibnamefont {Stein}},
  \bibinfo {author} {\bibfnamefont {Vassilios}\ \bibnamefont {Kapaklis}}, \
  and\ \bibinfo {author} {\bibfnamefont {Bj{\"o}rgvin}\ \bibnamefont
  {Hj{\"o}rvarsson}},\ }\bibfield  {title} {\enquote {\bibinfo {title}
  {Interaction modifiers in artificial spin ices},}\ }\href {\doibase
  10.1038/s41567-017-0027-2} {\bibfield  {journal} {\bibinfo  {journal} {Nature
  Physics}\ }\textbf {\bibinfo {volume} {14}},\ \bibinfo {pages} {375--379}
  (\bibinfo {year} {2018})}\BibitemShut {NoStop}%
\bibitem [{\citenamefont {Macauley}\ \emph {et~al.}(2020)\citenamefont
  {Macauley}, \citenamefont {Paterson}, \citenamefont {Li}, \citenamefont
  {Mac\^edo}, \citenamefont {McVitie},\ and\ \citenamefont
  {Stamps}}]{Macauley-thermal2020}%
  \BibitemOpen
  \bibfield  {author} {\bibinfo {author} {\bibfnamefont {Gavin~M.}\
  \bibnamefont {Macauley}}, \bibinfo {author} {\bibfnamefont {Gary~W.}\
  \bibnamefont {Paterson}}, \bibinfo {author} {\bibfnamefont {Yue}\
  \bibnamefont {Li}}, \bibinfo {author} {\bibfnamefont {Rair}\ \bibnamefont
  {Mac\^edo}}, \bibinfo {author} {\bibfnamefont {Stephen}\ \bibnamefont
  {McVitie}}, \ and\ \bibinfo {author} {\bibfnamefont {Robert~L.}\ \bibnamefont
  {Stamps}},\ }\bibfield  {title} {\enquote {\bibinfo {title} {Tuning magnetic
  order with geometry: Thermalization and defects in two-dimensional artificial
  spin ices},}\ }\href {\doibase 10.1103/PhysRevB.101.144403} {\bibfield
  {journal} {\bibinfo  {journal} {Phys. Rev. B}\ }\textbf {\bibinfo {volume}
  {101}},\ \bibinfo {pages} {144403} (\bibinfo {year} {2020})}\BibitemShut
  {NoStop}%
\bibitem [{\citenamefont {Sendetskyi}\ \emph {et~al.}(2019)\citenamefont
  {Sendetskyi}, \citenamefont {Scagnoli}, \citenamefont {Leo}, \citenamefont
  {Anghinolfi}, \citenamefont {Alberca}, \citenamefont {L\"uning},
  \citenamefont {Staub}, \citenamefont {Derlet},\ and\ \citenamefont
  {Heyderman}}]{Sendetskyi2019}%
  \BibitemOpen
  \bibfield  {author} {\bibinfo {author} {\bibfnamefont {Oles}\ \bibnamefont
  {Sendetskyi}}, \bibinfo {author} {\bibfnamefont {Valerio}\ \bibnamefont
  {Scagnoli}}, \bibinfo {author} {\bibfnamefont {Na\"emi}\ \bibnamefont {Leo}},
  \bibinfo {author} {\bibfnamefont {Luca}\ \bibnamefont {Anghinolfi}}, \bibinfo
  {author} {\bibfnamefont {Aurora}\ \bibnamefont {Alberca}}, \bibinfo {author}
  {\bibfnamefont {Jan}\ \bibnamefont {L\"uning}}, \bibinfo {author}
  {\bibfnamefont {Urs}\ \bibnamefont {Staub}}, \bibinfo {author} {\bibfnamefont
  {Peter~Michael}\ \bibnamefont {Derlet}}, \ and\ \bibinfo {author}
  {\bibfnamefont {Laura~Jane}\ \bibnamefont {Heyderman}},\ }\bibfield  {title}
  {\enquote {\bibinfo {title} {Continuous magnetic phase transition in
  artificial square ice},}\ }\href {\doibase 10.1103/PhysRevB.99.214430}
  {\bibfield  {journal} {\bibinfo  {journal} {Phys. Rev. B}\ }\textbf {\bibinfo
  {volume} {99}},\ \bibinfo {pages} {214430} (\bibinfo {year}
  {2019})}\BibitemShut {NoStop}%
\bibitem [{\citenamefont {Cathelin}(2020)}]{Cathelin2020}%
  \BibitemOpen
  \bibfield  {author} {\bibinfo {author} {\bibfnamefont {Vadim}\ \bibnamefont
  {Cathelin}},\ }\emph {\bibinfo {title} {Les glaces de spins : monopoles
  magn\'etiques, relation fluctuation-dissipation et fragmentation}},\
  \href@noop {} {\bibinfo {type} {{Ph.D.~Thesis}}},\ \bibinfo  {school}
  {Universit\'e Grenoble Alpes} (\bibinfo {year} {2020})\BibitemShut {NoStop}%
\bibitem [{\citenamefont {Paulsen}\ \emph {et~al.}(2016)\citenamefont
  {Paulsen}, \citenamefont {Giblin}, \citenamefont {Lhotel}, \citenamefont
  {Prabhakaran}, \citenamefont {Balakrishnan}, \citenamefont {Matsuhira},\ and\
  \citenamefont {Bramwell}}]{paulsen2016wien}%
  \BibitemOpen
  \bibfield  {author} {\bibinfo {author} {\bibfnamefont {C.}~\bibnamefont
  {Paulsen}}, \bibinfo {author} {\bibfnamefont {S.~R.}\ \bibnamefont {Giblin}},
  \bibinfo {author} {\bibfnamefont {E.}~\bibnamefont {Lhotel}}, \bibinfo
  {author} {\bibfnamefont {D.}~\bibnamefont {Prabhakaran}}, \bibinfo {author}
  {\bibfnamefont {G.}~\bibnamefont {Balakrishnan}}, \bibinfo {author}
  {\bibfnamefont {K.}~\bibnamefont {Matsuhira}}, \ and\ \bibinfo {author}
  {\bibfnamefont {S.~T.}\ \bibnamefont {Bramwell}},\ }\bibfield  {title}
  {\enquote {\bibinfo {title} {Experimental signature of the attractive coulomb
  force between positive and negative magnetic monopoles in spin ice},}\ }\href
  {\doibase 10.1038/nphys3704} {\bibfield  {journal} {\bibinfo  {journal}
  {Nature Physics}\ }\textbf {\bibinfo {volume} {12}},\ \bibinfo {pages}
  {661--666} (\bibinfo {year} {2016})}\BibitemShut {NoStop}%
\bibitem [{\citenamefont {Paulsen}\ \emph {et~al.}(2019)\citenamefont
  {Paulsen}, \citenamefont {Giblin}, \citenamefont {Lhotel}, \citenamefont
  {Prabhakaran}, \citenamefont {Matsuhira}, \citenamefont {Balakrishnan},\ and\
  \citenamefont {Bramwell}}]{paulsen2019nuclear}%
  \BibitemOpen
  \bibfield  {author} {\bibinfo {author} {\bibfnamefont {C}~\bibnamefont
  {Paulsen}}, \bibinfo {author} {\bibfnamefont {SR}~\bibnamefont {Giblin}},
  \bibinfo {author} {\bibfnamefont {Elsa}\ \bibnamefont {Lhotel}}, \bibinfo
  {author} {\bibfnamefont {D}~\bibnamefont {Prabhakaran}}, \bibinfo {author}
  {\bibfnamefont {K}~\bibnamefont {Matsuhira}}, \bibinfo {author}
  {\bibfnamefont {G}~\bibnamefont {Balakrishnan}}, \ and\ \bibinfo {author}
  {\bibfnamefont {ST}~\bibnamefont {Bramwell}},\ }\bibfield  {title} {\enquote
  {\bibinfo {title} {Nuclear spin assisted quantum tunnelling of magnetic
  monopoles in spin ice},}\ }\href@noop {} {\bibfield  {journal} {\bibinfo
  {journal} {Nature communications}\ }\textbf {\bibinfo {volume} {10}},\
  \bibinfo {pages} {1--8} (\bibinfo {year} {2019})}\BibitemShut {NoStop}%
\end{thebibliography}%

\end{document}